\pdfoutput=1
\documentclass[cernpreprint, texlive=2023, UKenglish, texmf, orcidlogo]{atlasdoc}
\PreprintIdNumber{CERN-EP-2022-221}

\usepackage{atlaspackage}
\usepackage{atlasheavyion}
 
\usepackage{atlasbiblatex}
 
\usepackage{atlasphysics}
 
\graphicspath{{logos/}{figures/}}
 
\usepackage{ANA-HION-2018-17-PAPER-defs}
\usepackage{multirow}
\usepackage{afterpage}

\AtlasTitle{Charged-hadron production in $pp$, $p$+Pb, Pb+Pb, and Xe+Xe collisions at $\sqrt{s_{_\text{NN}}}=5$\,TeV with the ATLAS detector at the LHC}
 
\AtlasAbstract{
This paper presents measurements of charged-hadron spectra obtained in $pp$, $p$+Pb, and Pb+Pb collisions at $\sqrt{s}$ or $\sqrt{s_{_\text{NN}}}=5.02$\,TeV, and in Xe+Xe collisions at $\sqrt{s_{_\text{NN}}}=5.44$\,TeV. The data recorded by the ATLAS detector at the LHC have total integrated luminosities of 25\,pb${}^{-1}$, 28\,nb${}^{-1}$, 0.50\,nb${}^{-1}$, and 3\,$\mu$b${}^{-1}$, respectively.
The nuclear modification factors $R_{p\text{Pb}}$ and $R_\text{AA}$ are obtained by comparing the spectra in heavy-ion and $pp$ collisions in a wide range of charged-particle transverse momenta and pseudorapidity.
The nuclear modification factor $R_{p\text{Pb}}$ shows a moderate enhancement above unity with a maximum at $p_{\mathrm{T}} \approx 3$\,GeV; the enhancement is stronger in the Pb-going direction.
The nuclear modification factors in both Pb+Pb and Xe+Xe collisions feature a significant, centrality-dependent suppression.
They show a similar distinct $p_{\mathrm{T}}$-dependence with a local maximum at $p_{\mathrm{T}} \approx 2$\,GeV and a local minimum at $p_{\mathrm{T}} \approx 7$\,GeV.
This dependence is more distinguishable in more central collisions.
No significant $|\eta|$-dependence is found.
A comprehensive comparison with several theoretical predictions is also provided. They typically describe $R_\text{AA}$ better in central collisions and in the $p_{\mathrm{T}}$ range from about 10 to 100\,GeV.
}

\author{The ATLAS Collaboration}
 
\AtlasRefCode{HION-2018-17}
 
\AtlasJournal{JHEP}
 
\AtlasJournalRef{JHEP 07 (2023) 074}
 
\AtlasDOI{10.1007/JHEP07(2023)074}

\hypersetup{pdftitle={ATLAS document},pdfauthor={The ATLAS Collaboration}}
 
\begin{document}
 
\maketitle
 
\tableofcontents

\section{Introduction}
 
Collisions of heavy nuclei at high centre-of-mass energy produce a hot and dense state of matter in which quarks and gluons are no longer confined, the quark--gluon plasma (QGP)~\cite{ROLAND201470,Busza:2018rrf}.
In these collisions, soft (low-energy) particles are produced mainly by the thermalizing QGP~\cite{Berges:2020fwq}, while hard (high-energy) particles are mostly found in jets.

Jets originating from high transverse momentum (\pT) partons are modified because of interactions with the QGP medium~\cite{HION-2012-11,CMS-HIN-11-013}.
Previous measurements of jets performed by the ATLAS, ALICE, and CMS Collaborations in \PbPb collisions at $\sqn=5.02$\,\TeV revealed that the energy loss of high-\pT partons results in a lower jet yield at a fixed \pT~\cite{HION-2017-10,ALICE_jet_Raa,CMS-HIN-18-014}.
The jet yield is about one half to three quarters of what it would have been in the absence of the medium's effects.
The medium's effects also modify the jet fragmentation functions~\cite{HION-2017-04}.
The measurements show an enhancement of the charged-hadron yield inside jets at $\pT\lesssim 4$\,\GeV, suppression at $4 \lesssim \pT \lesssim 30$\,\GeV, and a small enhancement at $\pT \gtrsim 30$\,\GeV
independent of jet $\pT$ for jets up to $\pT^{jet} \approx 400$\,\GeV.
 
Jet $\pT$ spectra, jet fragmentation functions, and (inclusive) charged-hadron $\pT$ spectra are three complementary observables and it is thus expected that the production of charged hadrons will be modified by the presence of the QGP as well.
The results from the LHC experiments show that hadron production in \PbPb and \XeXe (hereinafter collectively referred to as \AA) collisions is suppressed by a factor of about seven at $\pT \approx 7$\,\GeV~\cite{HION-2011-03,CMS-HIN-15-015,ALICE_Raa_PbPb,CMS-HIN-18-004,ALICE_Raa_XeXe} in the most central collisions. The suppression diminishes towards higher \pT.
The results can test models of jet quenching in fine detail.
The models often incorporate radial flow~\cite{PhysRevC.82.064903,PhysRevLett.115.132301} as well.
 
The observed modifications in \AA collisions are interpreted as arising mainly from the modifications of the parton showering processes in the final stages of the collisions because of the presence of QGP\@.
However, initial-state effects arising from the presence of a large nucleus may also play a role.
For example, nucleons bound in a nucleus are expected to have a somewhat different structure than free nucleons~\cite{Salgado:2011wc}, or partons may lose energy in the nuclear environment before hard scattering~\cite{Kang:2015mta}.
Proton--ion collisions are used to distinguish between the initial- and final-state effects.
The results from the LHC experiments show no more than a mild enhancement of hadron production at $\pT\gtrsim 2$\,\GeV in \pPb collisions~\cite{HION-2012-14,CMS-HIN-15-015,ALICE_Raa_PbPb}, indicating that initial-state effects do not influence hadron production to any large extent.
The significant suppression of charged-hadron yields in \AA collisions is thus a result of the final-state effects.

The suppression of hadron production in heavy-ion (HI) collisions, encompassing both \AA and \pPb collisions, can be quantified using the nuclear modification factor, $\Rab$.
It is a ratio of the measured charged-particle production yield in HI collisions to the appropriately scaled yield in nucleon--nucleon collisions:
\begin{equation}
\label{eq:Rab}
\Rab = \frac{1}{\avgTab} \frac{(1/N_\mathrm{evt})\ \mathrm{d}^2 N_{\mathrm{ch}}/\mathrm{d}\pT \mathrm{d}\eta}{\mathrm{d}^2\sigma_{pp}/\mathrm{d}\pT \mathrm{d}\eta},
\end{equation}
where $N_\mathrm{evt}$ is the number of HI events, $\mathrm{d}^2 N_{\mathrm{ch}}/\mathrm{d}\pT \mathrm{d}\eta$ is the differential yield of charged hadrons in HI collisions,
$\mathrm{d}^2\sigma_{pp}/\mathrm{d}\pT \mathrm{d}\eta$ is the differential charged-hadron production cross-section measured in inelastic $\pp$ collisions,
and $\avgTab$ is the mean nuclear thickness function.
The nuclear modification factor also refers to quantities integrated over $\pT$ or $\eta$ variables,\footnote{ATLAS uses a right-handed coordinate system with its origin at the nominal interaction point (IP) in the centre of the detector and the $z$-axis along the beam pipe.
The $x$-axis points from the IP to the centre of the LHC ring, and the $y$-axis points upwards.
Cylindrical coordinates $(r,\phi)$ are used in the transverse plane, $\phi$ being the azimuthal angle around the $z$-axis.
The pseudorapidity is defined in terms of the polar angle $\theta$ as $\eta = -\ln \tan(\theta/2)$ and the rapidity of the components of the beam, $y$, are defined in terms of their energy, $E$, and longitudinal momentum, $p_z$, as $y=0.5 \ln[(E+p_z)/(E-p_z)]$.
Angular distance is measured in units of $\Delta R \equiv \sqrt{(\Delta\eta)^{2} + (\Delta\phi)^{2}}$.} or both.
The mean nuclear thickness function $\avgTab$ is estimated as the number of nucleon--nucleon collisions divided by their cross-section~\cite{glauber}.
The nucleon--nucleon hadron-production cross-section can be approximated by the corresponding \pp cross-section if isospin effects are neglected~\cite{isospin}.
To eliminate any centre-of-mass energy dependence of the charged-hadron cross-section, both the HI and \pp collisions are measured at the same $\sqn$.

The HI colliding systems considered in this study are \pPb, \PbPb, and \XeXe.
The nuclear modification factors measured for \pPb collisions are denoted by \Rppb and they use $\avgTpPb$; the nuclear modification factors measured for \PbPb and \XeXe collisions are collectively denoted by \Raa and they use their respective $\avgTaa$.
It is expected that in the absence of initial- and final-state effects, the nuclear modification factor will be unity in the region of \pT where hadron production is dominated by hard scattering processes.
All three nuclear modification factors are estimated as functions of $\pT$ and $\eta$, and in intervals of collision centrality.
In \pPb collisions, the rapidity in the nucleon--nucleon centre-of-mass frame, $\ystar$, is used instead of $\eta$~\cite{HION-2012-14};
the same applies for its \pp baseline in \RpPb.
Collision centrality is further discussed in Section~\ref{sec:centrality}.

This analysis uses \pp, \pPb, \PbPb, and \XeXe data recorded by the ATLAS detector to measure charged-hadron spectra and nuclear modification factors.
The \pp, \pPb, and \PbPb collisions have nucleon--nucleon centre-of-mass energy $\sqs=\sqn=5.02$\,\TeV while the \XeXe collisions have $\sqn=5.44$\,\TeV.
The \pp reference for \XeXe collisions is obtained by an extrapolation of the existing \pp data at lower $\sqs$.
The \pp data were recorded in 2015 with an integrated luminosity of 25\,\ipb;
the \pPb collisions were recorded in 2013 and they have an integrated luminosity of 28\,\inb;
the \PbPb collisions recorded in 2015 have 0.50\,\inb of data;
and the \XeXe collisions were recorded in 2017 with an integrated luminosity of 3\,\iub.
The measurements use tracks with $\pT>0.3$\,\GeV for all samples except for the \PbPb sample, where the $\pT$ threshold is increased to 0.5\,\GeV.
The measurements cover $|\eta|<2.5$ for all samples, apart from \pPb, which uses $-2.5<\ystar<2.0$.
Unlike the others, the \pPb collisions are asymmetric; protons with $E=4$\,\TeV collide with lead nuclei of energy $E=82\times 4$\,\TeV.
The resulting \pPb centre-of-mass reference frame is boosted relative to the laboratory frame with a rapidity shift of $\Delta y=\pm0.465$.

This paper is organized in nine sections:
Section~\ref{sec:detector} reviews the ATLAS detector;
Section~\ref{sec:events} introduces the data and simulation samples;
Section~\ref{sec:centrality} provides a brief overview of the centrality definition;
Section~\ref{sec:selection} outlines the event and track selection;
Section~\ref{sec:analysis} describes all the corrections applied to the data;
Section~\ref{sec:systematics} summarizes the systematic uncertainties of this measurement;
Section~\ref{sec:result} presents the results;
and finally Section~\ref{sec:conclusion} contains the conclusions.


\section{ATLAS detector}
\label{sec:detector}
 
The ATLAS detector~\cite{PERF-2007-01} at the LHC is a multipurpose particle detector with a forward/backward symmetric cylindrical geometry
that covers nearly the entire solid angle around the collision point.
It consists of an inner tracking detector surrounded by a thin superconducting solenoid, electromagnetic and hadronic calorimeters,
and a muon spectrometer incorporating three large superconducting toroid magnets.
 
The inner-detector system (ID) is immersed in a 2\,T axial magnetic field
and provides charged particle tracking in the range $|\eta| < 2.5$.
The high-granularity silicon pixel detector (Pixel) covers the vertex region and typically provides three or four measurements per track; the original three layers have been augmented by the insertable B-layer~\cite{ATLAS-TDR-2010-19,PIX-2018-001}, a new innermost layer operating as a part of the pixel detector since 2015.
It is surrounded by the silicon microstrip tracker (SCT) which usually provides four two-dimensional measurement points per track.
These silicon detectors are complemented by the transition radiation tracker, which enables radially extended track reconstruction up to $|\eta| = 2.0$.
 
The calorimeter system covers the pseudorapidity range $|\eta| < 4.9$.
In the region $|\eta|< 3.2$, electromagnetic calorimetry is provided by barrel and endcap high-granularity lead/liquid-argon (LAr) electromagnetic calorimeters,
with an additional thin LAr presampler covering $|\eta| < 1.8$ to correct for energy loss in material upstream of the calorimeters.
Hadronic calorimetry is provided by the steel/scintillator-tile calorimeter, segmented into three barrel structures within $|\eta| < 1.7$, and two copper/LAr hadronic endcap calorimeters.
The system is completed with forward copper/LAr and tungsten/LAr calorimeter modules (FCal) covering $3.1<|\eta|<4.9$,
optimized for electromagnetic and hadronic measurements respectively.
 
The muon spectrometer comprises separate trigger and high-precision tracking chambers.
A set of precision chambers covers the pseudorapidity range $|\eta| < 2.7$ with three layers of monitored drift tubes,
complemented by cathode-strip chambers in the forward region.
The muon trigger system covers the range $|\eta| < 2.4$ with resistive-plate chambers in the barrel, and thin-gap chambers in the endcap regions.
 
The minimum-bias trigger scintillator (MBTS) counters are located at $z = \pm3.56$\,m along the beamline from the centre of the ATLAS detector and cover $2.1<|\eta|<3.9$ on each side.
The number of scintillator pads forming each MBTS counter was reduced from the original 16 to 12 in 2015~\cite{TRIG-2016-01}.
 
The zero-degree calorimeter (ZDC) consists of two arms, positioned at $z = \pm140$\,m from the centre of the ATLAS detector, and detects neutrons and photons with $|\eta|>8.3$. Signals from the ZDC are used by the trigger systems. The ZDC thresholds were set to select events with even a single neutron reaching the ZDC\@. The ZDC was installed only during the \PbPb data-taking.
 
An extensive software suite~\cite{ATL-SOFT-PUB-2021-001} is used in data simulation, in the reconstruction and analysis of real and simulated data, in detector operations, and in the trigger and data acquisition systems of the experiment.
A two-level trigger system was used to select events~\cite{TRIG-2016-01}, with a first-level (L1) trigger implemented in hardware followed by a software-based high-level trigger (HLT) that reconstructs the event in a manner similar to the final offline reconstruction, accessing information from all ATLAS subdetectors. Events used for the analysis were selected using several carefully designed triggers.


\section{Data and Monte Carlo simulation samples}
\label{sec:events}
 
\subsection{Data samples}
 
A summary of all data samples used in the measurement, including triggers, luminosities, and years of data-taking is given in Table~\ref{tab:data_overview}.
 
The \pp events were required to either satisfy a minimum-bias (MB) trigger or one of several jet triggers with different $\ET$ thresholds.
The MB trigger required an event to be randomly selected at L1, requiring only filled bunch crossings, and a track to be reconstructed in the ID at the HLT.
Jet triggers used an \antikt jet algorithm~\cite{anti-kt,Fastjet,HION-2015-04} with a radius parameter of $R=0.4$ and one of seven different $\ET$ thresholds at the HLT\@;
the events were preselected in different ways at L1.
The jet trigger with an HLT threshold of $\ET=20$\,\GeV required an event to be randomly selected at L1.
The jet triggers with thresholds of $\ET=30$\,\GeV and 40\,\GeV required at L1 the total transverse energy (TE) measured in the calorimeter system to be greater than 5\,\GeV and 10\,\GeV, respectively.
The jet triggers with thresholds of $\ET=50$\,\GeV, 60\,\GeV, 75\,\GeV, and 85\,\GeV required the identification of a jet by the L1 jet trigger algorithm with thresholds of 12\,\GeV, 15\,\GeV, 20\,\GeV, and again 20\,\GeV, respectively.
 
The \pPb events were recorded in two data-taking periods with proton and lead beam directions swapped in the LHC rings: period A with `Pb+$p$' geometry and period B with `$p$+Pb' geometry.
Regardless of the period, the physics results of \pPb collisions are presented such that the $p$-going direction is towards positive rapidity; this is the convention used in previous ATLAS publications as well as in publications of other LHC experiments~\cite{HION-2013-08,CMS-HIN-13-001,ALICE:2014nqx}.
The events were required to either satisfy an MB trigger or one of several jet triggers with different thresholds.
The MB trigger required an event to have signals on both sides of the MBTS\@.
Jet triggers used an \antikt jet algorithm with a radius parameter of $R=0.4$ and six different thresholds at the HLT, while the events were preselected in different ways at L1~\cite{HION-2015-04}.
The jet triggers with HLT thresholds of $\ET=20$\,\GeV, 30\,\GeV, and 40\,\GeV required signals at L1 in both sides of the MBTS\@.
The jet trigger with the $\ET=50$\,\GeV threshold required the identification of a jet by the L1 jet trigger algorithm with a threshold of 10\,\GeV, and the triggers with 60 and 75\,\GeV thresholds required identification by the L1 jet trigger algorithm with a threshold of 15\,\GeV.
 
The \PbPb events were required to either satisfy one of the MB triggers or one of several jet triggers with different thresholds.
The first MB trigger required an event to have TE below 50\,\GeV, signals in both sides of the ZDC, and a track in the ID\@.
The second MB trigger was complementary and it required TE above 50\,\GeV.
Jet triggers used an \antikt jet algorithm with a radius parameter of $R=0.4$ and three different thresholds at the HLT, while the events were preselected in the same way at L1~\cite{HION-2017-04}.
The jet triggers had thresholds of $\ET=60$\,\GeV, 75\,\GeV, and 100\,\GeV.
They required the identification of a jet by the L1 jet trigger algorithm with a threshold of 50\,\GeV.
 
The \XeXe events were required to satisfy one of the MB triggers.
The first MB trigger required an event to have TE smaller than 4\,\GeV and have a track in the ID\@.
The second MB trigger was complementary and it required TE greater than 4\,\GeV.
No jet triggers were used for \XeXe collisions.
 
\begin{table}[tb]
\centering
\caption{Summary of the data samples used in the analysis. The entire luminosity is sampled only by the triggers indicated in bold font. For the MB triggers, `track' means a track must be  reconstructed in the ID; `MBTS' means a signal is required in both sides of the MBTS; `ZDC' means a signal is required in both sides of the ZDC; `TE50' and `TE4' means the total energy deposited in the calorimeter must be above 50\,\GeV and 4\,\GeV, respectively; `VTE50' and `VTE4' means a veto of the complementary trigger. More details are provided in the text.}
\begin{tabular}{l|lllll} \hline
Sample & Year & MB triggers & Jet trigger thresholds [$\GeV$] & Luminosity & $\sqs$, $\sqn$ \\ \hline
\pp   & 2015 & track & 20, 30, 40, 50, 60, 75, \textbf{85} & 25.0\,\ipb & 5.02\,\TeV \\
\pPb  & 2013 & MBTS & 20, 30, 40, 50, 60, \textbf{75} & 28.1\,\inb & 5.02\,\TeV \\
\PbPb & 2015 & TE50, VTE50+ZDC+track & 60, 75, \textbf{100} & 0.50\,\inb & 5.02\,\TeV \\
\XeXe & 2017 & \textbf{TE4}, VTE4+track & --- & \phantom{0.0}3\,\iub & 5.44\,\TeV \\ \hline
\end{tabular}
\label{tab:data_overview}
\end{table}
 
\subsection{Monte Carlo simulation samples}
 
Several Monte Carlo (MC) simulation samples are used to evaluate the impact of the detector performance on the obtained data and to estimate necessary corrections to derive particle-level results.
 
For \pp data, MC samples were generated at leading order using \Pythia[8.2]~\cite{pythia8} with \NNPDF[2.3lo]~\cite{nnpdf23lo} parton distribution functions (PDFs), the A14 set of tuned parameters~\cite{ATL-PHYS-PUB-2014-021}, and the same $\sqs$ as in the data.
Since the production of the MC samples requires significant resources, a special effort was made to optimize it.
The event production was split into 12 subsamples according to the $\pT$ of the leading charged particle in the range of $|\eta| < 2.6$.
The samples used $\pT$ ranges of 0--3\,\GeV, 3--7\,\GeV, 7--10\,\GeV, 10--15\,\GeV, 15--20\,\GeV, 20--30\,\GeV, 30--45\,\GeV, 45--65\,\GeV, 65--90\,\GeV, 90--140\,\GeV, 140--200\,\GeV, and above 200\,\GeV.
For each subsample, \Pythia was set up with high momentum transfer\footnote{Denoted by \enquote{PhaseSpace:pTHatMin} in \Pythia.} that still covered the production of all tracks within the given range;
if \Pythia produced an event with the leading charged-particle \pT too high for the given range, the event was discarded.
This procedure ensures the statistical significance of the MC sample is approximately uniform in the \pT range studied.
There are $4.8 \cdot 10^6$ simulated events in total.
 
Another three sets of \Pythia samples are required for the \pp data extrapolation as described in Section~\ref{sec:analysis}.
Each set is also split into 12 subsamples with the same $\pT$ ranges as those described above.
There are $3\times 19.2\cdot 10^6$ events in total.
The sets have $\sqs=5.44$\,\TeV, 5.02\,\TeV, and 2.76\,\TeV.
The first two sets are used for extrapolation of data with $\sqs=5.02$\,\TeV to $\sqs=5.44$\,\TeV as required for the calculation of \Raa for \XeXe collisions.
The last two sets are used to estimate the systematic uncertainty in this extrapolation.
For these three sets of \Pythia samples, only the generation was performed.
The simulation of the detector response and consequent reconstruction, described later in this section, were not performed.
 
For \pPb data, separate \Hijing~(version 1.38b)~\cite{hijing} and `data overlay' MC samples~\cite{SOFT-2010-01} were used.
There are separate MC samples for each data-taking period.
The samples using the \Hijing event generator consist of $9.7 \cdot 10^6$ minimum-bias events in total.
For the overlay samples, the \Pythia hard-scattering events, using the same parameter tune and PDFs as for \pp simulation, were embedded in real \pPb data events specifically recorded for this purpose with dedicated triggers.
Thus, data overlay events contain an underlying-event contribution identical to that in data.
Only one hard-scattering event is embedded in each \pPb event.
The events are split into five subsamples according to the generated jet $\pT$: 0--20\,\GeV, 20--80\,\GeV, 80--200\,\GeV, 200--500\,\GeV, and 500--1000\,\GeV.
The data overlay samples consist of $33.9 \cdot 10^6$ events in total.
 
The samples for \PbPb data also use \Hijing and `data overlay'.
The \Hijing sample has $1.5\cdot 10^6$ events.
The \Pythia events, split into 12 subsamples analogous to those of \pp, were embedded in real \PbPb data events, recorded for this purpose with dedicated triggers. The data overlay sample has $16\cdot 10^6$ events in total.
For \XeXe data, `\Hijing overlay' and `data overlay' MC samples are used.
The \Pythia events, split into 12 subsamples analogous to those of \pp, were embedded either in \Hijing events or in real \XeXe data events, recorded for this purpose with dedicated triggers.
There are $2.9\cdot 10^6$ \Hijing overlay events and $16\cdot 10^6$ data overlay events in total.
 
Once the MC events were generated, \Geant~\cite{geant} simulated the detector response.
It was configured with geometry and digitization parameters matching those of the data~\cite{SOFT-2010-01}.
The MC events were reconstructed using the same configuration as the corresponding data~\cite{PERF-2015-08}.
 
To achieve precise correspondence between data and MC samples, a reweighting of the MC samples is necessary.
This is done on an event-by-event basis as well as on a particle-by-particle basis.
Events in the data overlay MC samples are reweighted on an event-by-event basis such that they have the same distributions of total transverse energy deposited in the FCal (\FCalET) as in data.
Events in the \Hijing and \Hijing overlay MC samples are reweighted on an event-by-event basis such that they have the same charged-particle multiplicity distributions as those in data; \Hijing does not reproduce the correlation between \FCalET\ and multiplicity reliably.
No reweighting on an event-by-event basis is applied to \Pythia MC samples.
Particles in all MC samples are reweighted on a particle-by-particle basis such that the relative abundances of particle types match those measured in data by the ALICE Collaboration~\cite{alice_partcomp_pp_pbpb_2019,alice_partcomp_ppb}. The abundances in \XeXe collisions are approximated by those measured in \PbPb collisions in centrality intervals with the same $\langle \mathrm{d}N/\mathrm{d}\eta\rangle$~\cite{ALICE_Raa_XeXe}.
Since ALICE measured only pions, kaons, and protons in sufficiently wide $\pT$ ranges in all systems (except \XeXe), abundances of strange baryons are estimated with another procedure.
It is assumed that the number of particles follows the dependency described in Ref.~\cite{Grassi:2003dc}.
The unknown parameters that are independent of the particle species are calculated using known ratios of kaons to pions and protons to pions~\cite{alice_partcomp_pp_pbpb_2019,alice_partcomp_ppb}, while also accounting for differences between the model and the data~\cite{Andronic:2017pug}.
The estimate yields about 15\% strange baryons in the most central \PbPb collisions at around 3\,\GeV and less elsewhere.
This is significantly higher than the 6\% expected from \Pythia.
Their contribution is a result of the combination of several effects: the particle production cross-sections, kinematics and phase spaces of the decay modes, and momentum distributions of parent and child particles.


\section{Centrality}
\label{sec:centrality}

In HI collisions, the event centrality reflects the overlap of the two colliding nuclei.
The description of the centrality is based on the Monte Carlo Glauber model~\cite{glauber}.
ATLAS analyses determine centrality by a measurement of the event activity in the FCal.
In the case of \pPb collisions, only the Pb-going side of the FCal is used~\cite{HION-2013-08},
whereas in the case of \PbPb and \XeXe collisions, the sum of energies measured on both sides is used~\cite{HION-2016-04,HION-2019-03}.
The centrality intervals used in this analysis are defined according to successive percentiles of the \FCalET distribution obtained from MB-triggered HI events ordered from the most central (highest \FCalET, smallest impact parameter, centrality interval 0--5\%) to the most peripheral collisions (lowest \FCalET, larger impact parameter, centrality intervals 60--90\% or 60--80\%).
The hard-scattering rate is higher in more central collisions.
Such an enhancement is expected because the number of binary nucleon--nucleon collisions is greater in central than in peripheral collisions.
The bias in \FCalET from activity related to the triggering jet is negligible.
 
A Glauber model analysis of the \FCalET distribution is used to evaluate geometric quantities together with their systematic uncertainties \cite{HION-2013-08,HION-2016-04,HION-2019-03};
they are summarized in Table~\ref{tab:all_cent}.
The number of participants, $\Npart$, corresponds to the number of nucleons that interacted inelastically at least once.
The number of nucleon--nucleon collisions, $\Ncoll$, can be derived from their total inelastic cross-section and the nuclear thickness function:
\begin{equation*}
\Ncoll = \avgTab\ \sigma_{_\text{NN}}
\end{equation*}
 
\begin{table}[tb]
\def\arraystretch{1.3} 
\centering
\caption{Average values and the uncertainties of centrality parameters evaluated by a Glauber model analysis of the \FCalET distributions. Only values of the centrality intervals used in this analysis are mentioned. The calculations use the value of $\sigma_{_\text{NN}}=67.6 \pm 0.5$\,mb at 5.02\,\TeV~\cite{LoizidesXS} and $71\pm 3$\,mb at 5.44~\TeV. The uncertainties are discussed in Ref.~\cite{HION-2018-25}.}
\begin{tabular}{l|cc|cc|cc} \hline
\multirow{2}{*}{Centrality}         & \multicolumn{2}{c|}{\pPb}    & \multicolumn{2}{c|}{\PbPb}                    & \multicolumn{2}{c}{\XeXe} \\
& $\avgTpPb$ [$\mu$b${}^{-1}$] & \avgNpart  & \avgTaa[mb${}^{-1}$]  & \avgNpart  & \avgTaa[mb${}^{-1}$] & \avgNpart \\  \hline
\phantom{0}0--5\%\phantom{0} & $222^{+25}_{-6}$ & $16.5^{+1.9}_{-0.9}$ & $26.0\phantom{0} \pm 0.1\phantom{00}$ & $383.7 \pm 0.4$ & $13.9\phantom{0} \pm 0.1\phantom{00}$ & $237.4 \pm 0.6$ \\
\phantom{0}5--10\%  & $194^{+14}_{-4}$ & $14.6^{+1.2}_{-0.9}$  &  $20.4\phantom{0} \pm 0.1\phantom{00}$  & $331.4 \pm 0.9$  &  $10.9\phantom{0} \pm 0.1\phantom{00}$  & $208.6 \pm 1.2$ \\
10--20\% & $172^{+7\phantom{0}}_{-3}$  & $13.1^{+0.8}_{-0.7}$  &  $14.4\phantom{0} \pm 0.1\phantom{00}$  & $262.3 \pm 1.5$  &  $7.54 \pm 0.09$ & $166.6 \pm 1.5$ \\
20--30\% & $148^{+4\phantom{0}}_{-2}$  & $11.4^{+0.6}_{-0.6}$  &  $8.77 \pm 0.13$ & $188.2 \pm 1.9$  &  $4.48 \pm 0.10$ & $119.9 \pm 1.7$ \\
30--40\% & $126^{+3\phantom{0}}_{-4}$  &  $\phantom{0}9.8^{+0.6}_{-0.6}$  &  $5.09 \pm 0.12$ & $131.0 \pm 2.0$  &  $2.55 \pm 0.08$ &  $\phantom{0}83.6 \pm 1.7$ \\
40--50\% &                  &                       &  $2.75 \pm 0.09$ &  $\phantom{0}87.2 \pm 2.0$  &  $1.37 \pm 0.06$ &  $\phantom{0}55.6 \pm 1.6$ \\
40--60\% & $\phantom{0}92^{+4\phantom{0}}_{-6}$   & $\phantom{0}7.4^{+0.4}_{-0.5}$   &                  &                  &                  &                 \\
50--60\% &                  &                       &  $1.35 \pm 0.07$ &  $\phantom{0}54.4 \pm 1.8$  &  $0.69 \pm 0.04$ &  $\phantom{0}34.9 \pm 1.4$ \\
60--80\% &                  &                       &  $0.42 \pm 0.03$ &  $\phantom{0}23.4 \pm 1.2$  &  $0.23 \pm 0.02$ &  $\phantom{0}15.5 \pm 0.9$ \\
60--90\% & $\phantom{0}43^{+3\phantom{0}}_{-4}$   & $\phantom{0}4.0^{+0.2}_{-0.3}$   &                 &                  &                  &                 \\ \hline
\end{tabular}
\label{tab:all_cent}
\end{table}


\section{Event and track selection}
\label{sec:selection}
 
\subsection{Event selection}
 
Offline event selection always requires a vertex to be reconstructed less than 150\,mm from the centre of the detector in the $z$ direction~\cite{HION-2011-03}.
In \PbPb and \XeXe data-taking the instantaneous luminosities resulted in less than $2\times10^{-4}$ interactions per bunch crossing, and reconstruction of only one vertex is attempted because the probability of having pile-up vertices is very small.
In \pp and \pPb data, with instantaneous luminosity reaching 1.5 interactions and $2\times10^{-3}$ interactions per bunch crossing, respectively, multiple collision vertices are possible.
 
In \pp collisions, pile-up events are not rejected, as the measured observable is the total inelastic cross-section.
On the other hand, the measured observable in HI collisions is the per-event yield, and thus the pile-up events are rejected.
This is achieved by multiple means: by requiring correlations between \FCalET\ and the number of reconstructed tracks (in the case of \PbPb and \XeXe collisions), by requiring signals in  both sides of the ZDC (in \PbPb collisions), by requiring energy deposits in both the positive and negative intervals of $2.9 < |\eta| < 5.0$ (in \pPb, \PbPb, and \XeXe collisions) and by requiring that there is only one reconstructed vertex (in \pPb collisions).
Beam--gas events are also rejected by this procedure.
 
\subsection{Track selection}
\label{sec:tracks}
 
This analysis of the charged-particle spectra refers to primary charged particles with a mean lifetime greater than $0.3 \times 10^{-10}$\,s, directly produced in the nucleus--nucleus interactions or long-lived charged particles created by subsequent decays of particles with shorter lifetimes~\cite{STDM-2010-06}.
All other particles are considered secondary.
Tracks produced by primary and secondary particles are referred to as primary and secondary tracks, respectively. Reconstructed tracks arising from spuriously associated detector-layer hits that originate from different particles are considered to be fake tracks.
Tracks are reconstructed in the pseudorapidity region $|\eta|<2.5$ and over the full azimuth~\cite{STDM-2015-02,PERF-2015-08}.
Tracks with $\pT>0.3$\,\GeV from \pp, \pPb, and \XeXe collisions are reconstructed; tracks with $\pT>0.5$\,\GeV from \PbPb collisions are reconstructed.
 
In the MC simulation, the categorization of tracks relies on their matching to generated particles.
This is based on the hits produced in the detector layers by the generated particles.
A reconstructed track is matched to a generated particle that contributed the most to the hits used to build this track.
The matching procedure is explained in more detail in Ref.~\cite{ATL-PHYS-PUB-2015-051}.
 
To enhance the fraction of primary tracks and suppress secondary and fake tracks, all reconstructed tracks are required to satisfy quality criteria.
These criteria are the same for tracks in both the data and MC samples.
In \pp collisions, tracks are required not to miss any hits in the Pixel detector and no more than two hits in the SCT detector if such hits are expected from the track trajectory.
Tracks are allowed to share either one hit in the Pixel detector or two hits in the SCT detector with other tracks.
Tracks are also required to have at least one hit in one of the two innermost layers of the Pixel detector if the reconstructed track passed through an active sensor.
Tracks with $|\eta|<1.65$ ($|\eta|>1.65$) are required to have at least 9 (11) hits in the Pixel and the SCT detectors combined.
Since primary tracks are more likely than secondary or fake tracks to originate from the collision vertex, a \pT-dependent requirement on \dz, the distance of the closest approach to the vertex in the transverse direction, is imposed on tracks.
There is no requirement on $z_0$, the longitudinal separation between the vertex and the point where \dz\ is measured.
 
In \pPb collisions, tracks must have at least two hits in the Pixel detector and at least six hits in the SCT detector.
Tracks are not allowed to have a missing hit in the innermost layer of the Pixel detector if such a hit is expected from the track trajectory.
Tracks must not to share any hits with other tracks.
To ensure a good match to the vertex, the significances $|\dz|/\edz$ and $|\zz|/\ezz$ may not exceed 3 where $\edz$ and $\ezz$ are the uncertainties in $\dz$ and $\zz$, respectively,
and $\theta$ is the polar angle of the track.
 
In \PbPb and \XeXe collisions, requirements on tracks are similar to those in \pp collisions but a match to the vertex is required as for \pPb collisions.
Tracks are required not to miss any hits in the Pixel or SCT detectors if such hits are expected from the track trajectory;
no restrictions on shared hits are imposed.

These track selection requirements produce a sufficiently clean track sample up to $\pT\lesssim 50$\,\GeV.
Above that, a significant number of tracks originate from lower-momentum primary particles whose track \pT is mismeasured.
The probability for this to occur increases with \pT, rapidity, and the detector occupancy and was studied in the simulated samples in detail.
To eliminate the problem, high-\pT tracks are counted only in the vicinity of jets, within an angular distance $\Delta R = 0.4$.
The jets are reconstructed in the calorimeter with the \antikt algorithm using a radius parameter of $R=0.4$;
the jet reconstruction and underlying-event subtraction is further described in Refs.~\cite{HION-2013-08,HION-2017-10}.
The relative difference of track \pT and $\pT^{\text{jet}}$ may not exceed three times their measurement uncertainties combined.
The limitation arising from a low jet reconstruction efficiency at low $\pT^{\text{jet}}$ is also taken into account.
The described requirement is not imposed if the track may originate from a jet whose reconstruction efficiency is less than 99\%.

Leptons are excluded from the measured charged-hadron spectra. Tracks forming part of reconstructed muons are identified, and their contribution is subtracted with a weight that also accounts for tracks of electrons. The probability of missing a track from a muon is very small ($\lesssim 1\%$).


\section{Analysis procedure}
\label{sec:analysis}
 
Events in \pp, \pPb, and \PbPb collisions are recorded by MB and jet triggers.
The spectra obtained with different triggers are combined to construct the spectra equivalent to the MB spectra, which extend to the highest \pT reachable with their luminosities~\cite{HION-2011-03}.
 
With $N$ denoting the number of jet triggers used in the data-taking for each analysed system as listed in Table~\ref{tab:data_overview},
the jet triggers are numbered $n=0, ...\,,N-1$ and to each one an \textit{interval of full efficiency} $[p^{\mathrm{jet}}_{\mathrm{T},n},\ p^{\mathrm{jet}}_{\mathrm{T},n+1}]$ is assigned.
The values of $p^{\mathrm{jet}}_{\mathrm{T},n}$ are defined as the momentum at which the $n$-th jet trigger is fully efficient.
Due to differences between the online and offline estimates of $p^{\mathrm{jet}}_{\mathrm{T}}$, they are somewhat higher than the nominal momenta of the jet triggers listed in Table~\ref{tab:data_overview}.
These values form an incremental sequence: $p^{\mathrm{jet}}_{\mathrm{T},0},\ p^{\mathrm{jet}}_{\mathrm{T},1}, ...\,, p^{\mathrm{jet}}_{\mathrm{T},N-1}$.
The interval $[0,p^{\mathrm{jet}}_{\mathrm{T},0}]$ is assigned to the MB trigger.
The highest-threshold triggers have no upper limit, i.e.~$p^{\mathrm{jet}}_{\mathrm{T},N}=\infty$.
 
In events acquired with the MB trigger, tracks that do not match to any jet are also selected.
In events acquired with the jet triggers, tracks are required to be matched only to the jets whose $\pT^{\mathrm{jet}}$ fall in the interval of full efficiency corresponding to the given jet trigger.
 
\begin{figure}[ht]
\centering
\includegraphics[width=0.48\textwidth]{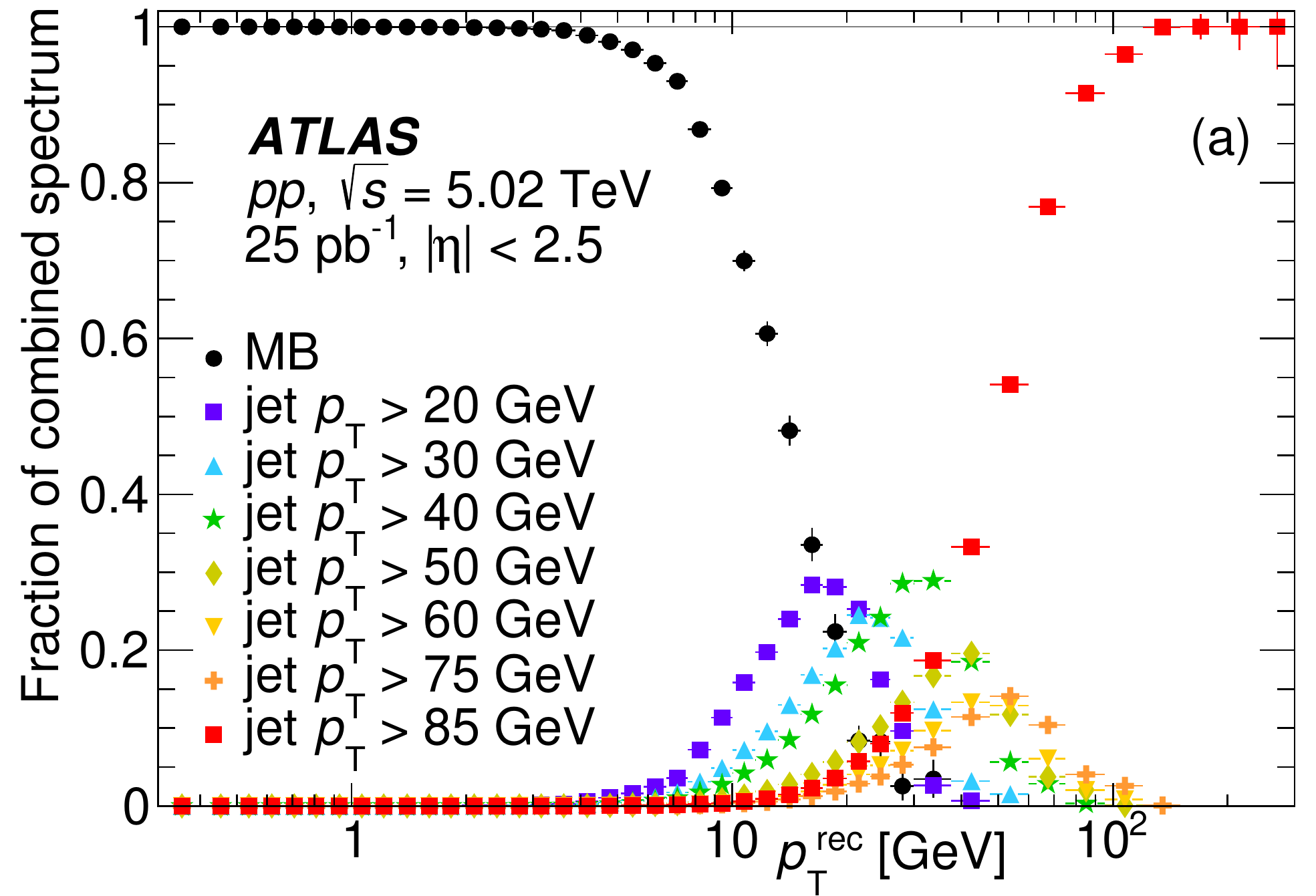} \includegraphics[width=0.48\textwidth]{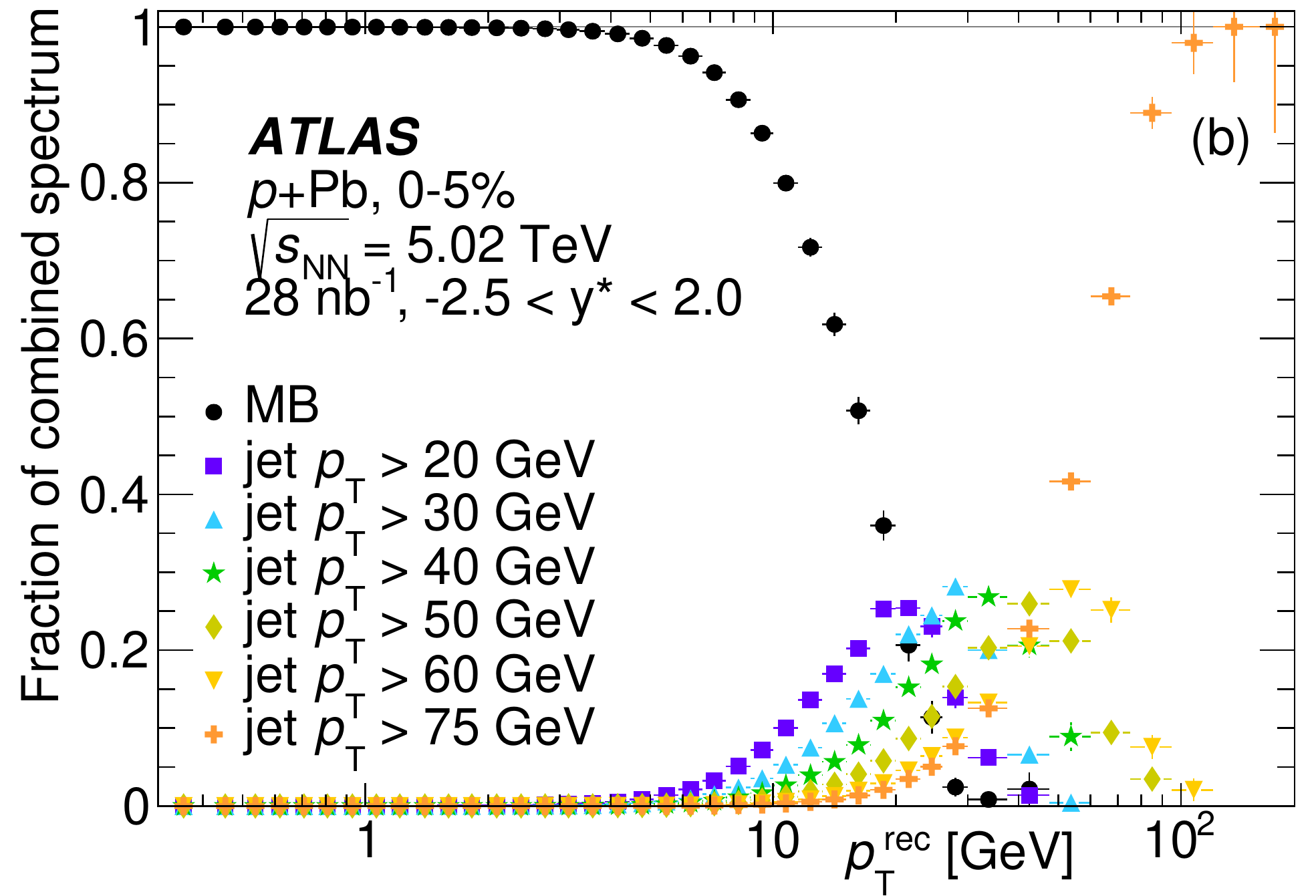} \\
\includegraphics[width=0.48\textwidth]{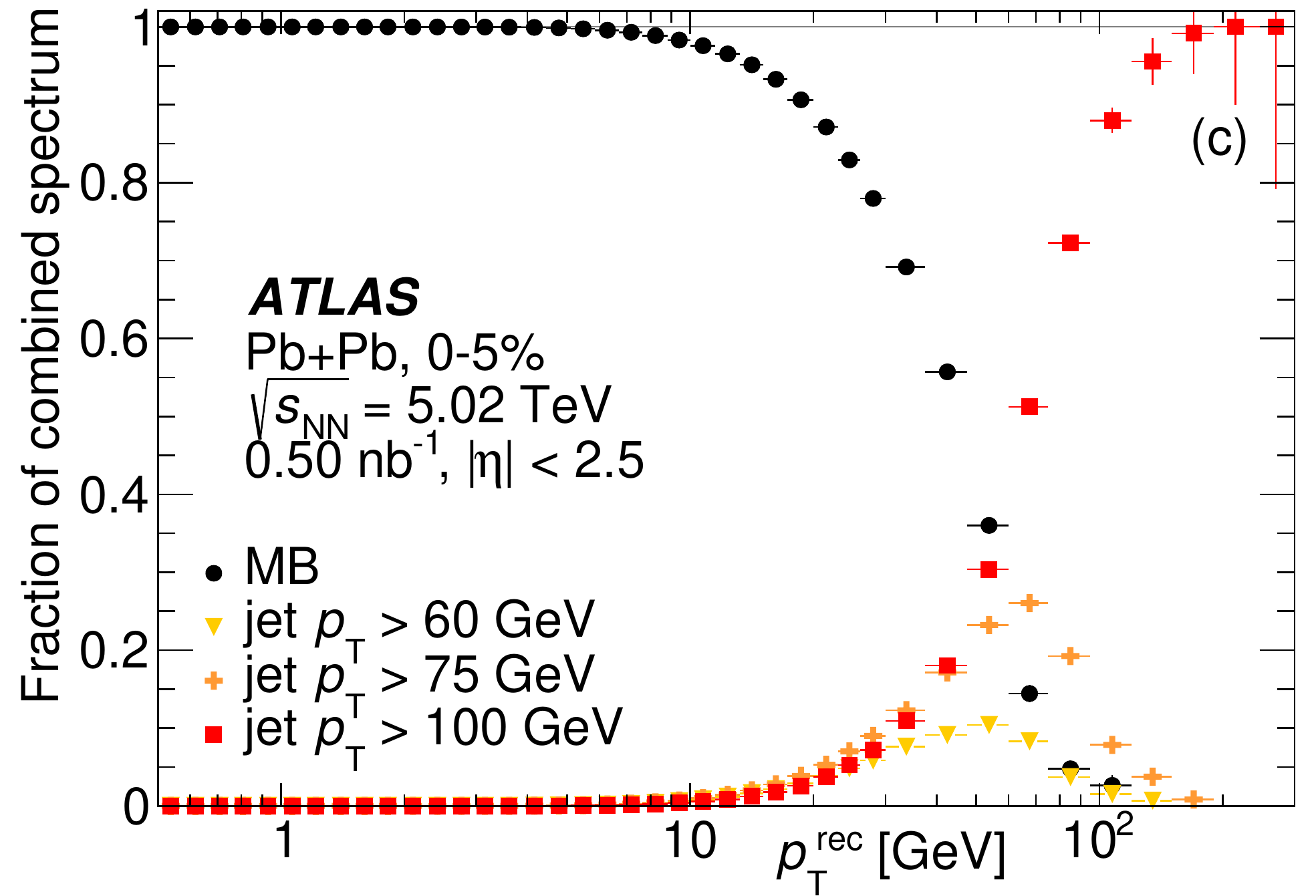}
\caption{The relative contributions of different triggers to the combined reconstructed track $\pT$ spectra in (a) \pp, (b) \pPb, and (c) \PbPb collisions. The \pPb and \PbPb results are shown for the 0--5\% centrality interval. The statistical uncertainties are shown with vertical lines.}
\label{fig:trigger_merge}
\end{figure}

After scaling the spectrum of each trigger by its appropriate effective integrated luminosity, spectra from all triggers are summed.
The procedure is illustrated in Figure~\ref{fig:trigger_merge}.
The low-\pT part of the combined spectrum is covered by the MB trigger with negligible contributions from other triggers.
With increasing \pT, the importance of the MB trigger diminishes and that of jet triggers increases.
At the highest \pT measured, all tracks are taken from the jet trigger with the highest $p^{\mathrm{jet}}_{\mathrm{T}}$.
 
After the summation, several subsequent corrections are applied in this order: for fake and secondary tracks, for $\pT$ resolution, for $\eta$ (or $\ystar$) resolution, and for track reconstruction efficiency. Moreover, an extrapolation to $\sqs=5.44$\,\TeV is applied as a final correction to the \pp distributions used as a baseline for the $\Raa$ values measured in $\XeXe$ collisions. Corrections accounting for the generated distribution of particle masses are applied to \pPb distributions and their \pp baseline.
All corrections together modify the measured distributions by several tens of percent.

The corrections are estimated by using MC samples.
In the case of \pp collisions, all corrections use \Pythia samples.
In the case of HI collisions, the corrections for fake and secondary tracks are derived using \Hijing samples, while the other corrections are derived using data overlay samples.
 
The first applied correction is for fake and secondary tracks.
The merged spectra are multiplied by the fraction of primary tracks, which is defined as the number of primary tracks divided by the total number of tracks, i.e.~including fake and secondary tracks.
The fractions are estimated in the MC simulation as a function of $\pT$, $\eta$ (or $\ystar$), collision system, and centrality.
 
\begin{figure}[ht]
\centering
\includegraphics[width=0.48\textwidth]{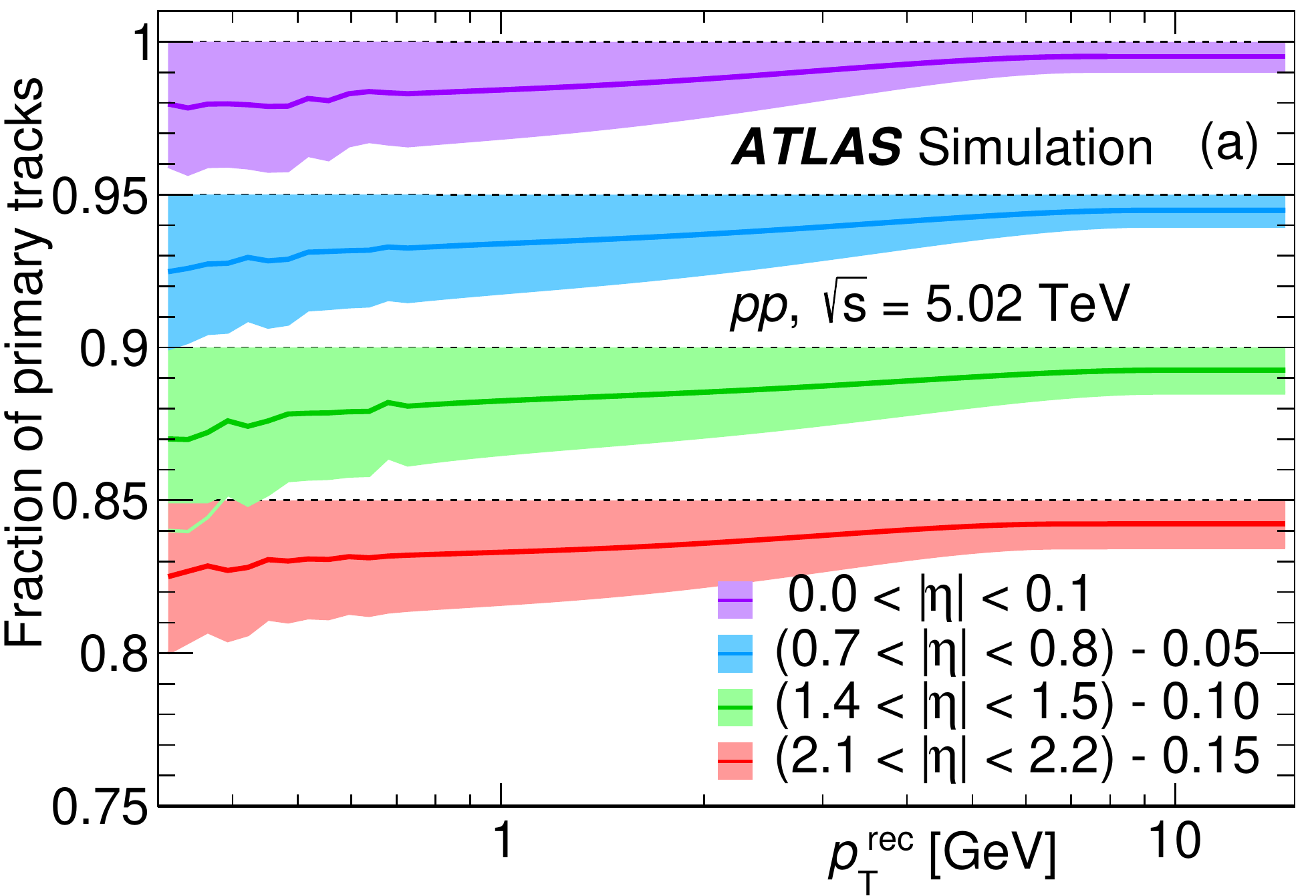} \includegraphics[width=0.48\textwidth]{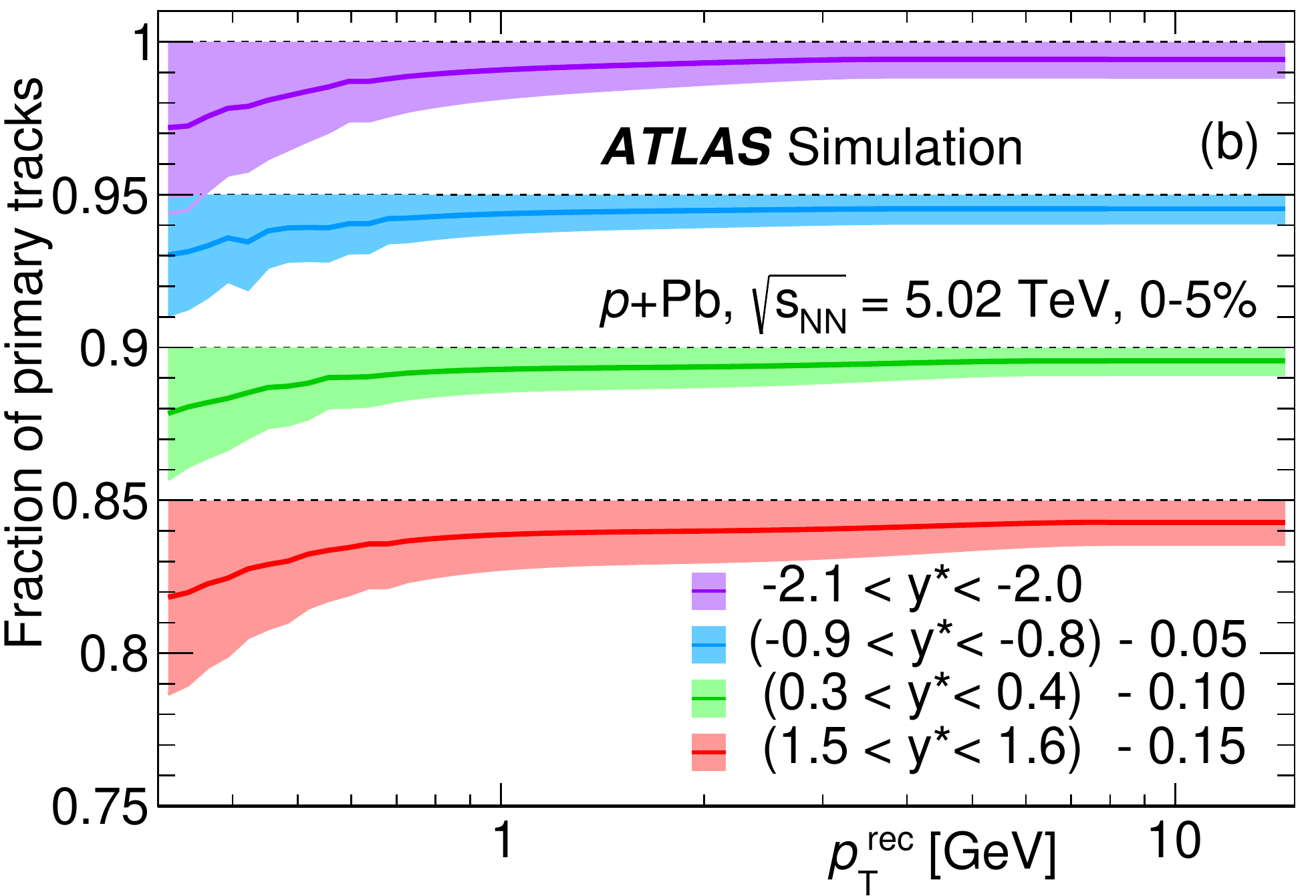}  \\
\includegraphics[width=0.48\textwidth]{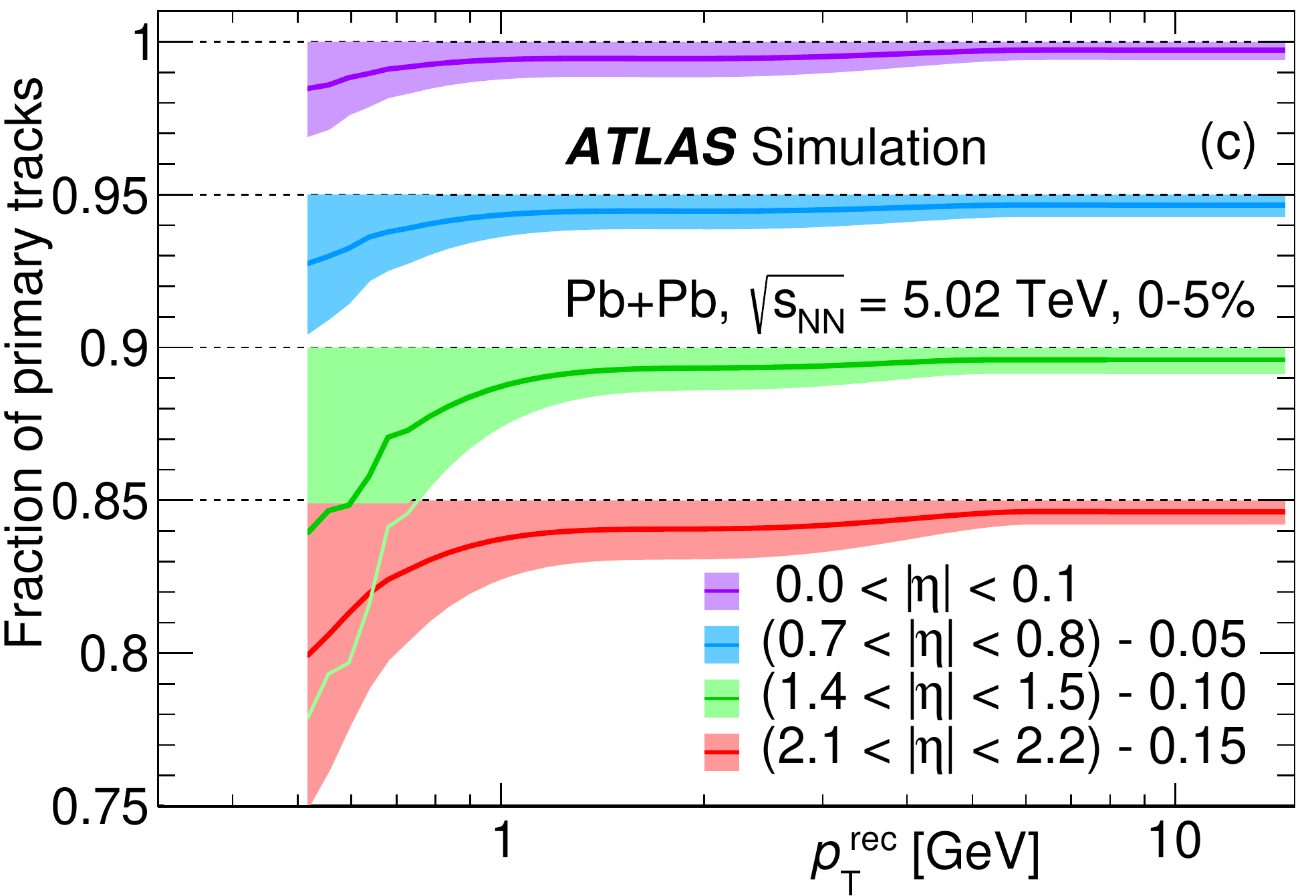} \includegraphics[width=0.48\textwidth]{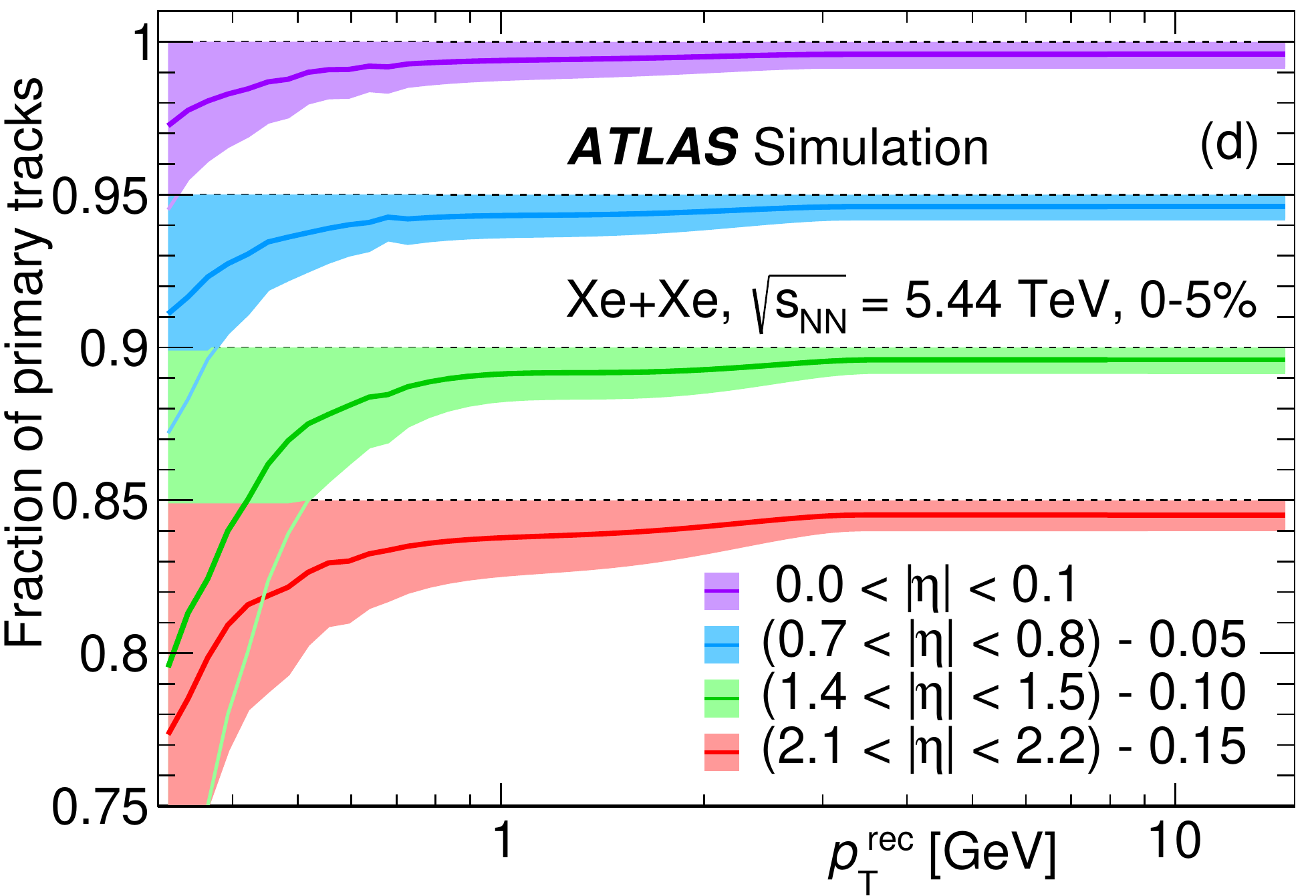}
\caption{The fraction of reconstructed tracks matched to primary particles as a function of reconstructed $\pT$ for (a) \pp, (b) \pPb, (c) \PbPb, and (d) \XeXe collisions and for four ranges of (pseudo)rapidity. The widths of the bands represent the systematic uncertainties; see Section~\ref{sec:systematics} for further details. The values are vertically offset for clarity.}
\label{fig:prim}
\end{figure}

Figure~\ref{fig:prim} shows examples of the fraction of primary tracks in several $\eta$ (or $\ystar$) ranges.
For HI collisions, only the most central collisions are shown.
The decrease at low $\pT$ is a feature common to all centrality intervals, although it is milder for more peripheral collisions.
 
To correct for the $\pT$ resolution, iterative Bayesian unfolding~\cite{bayes_unf} is used with migration matrices describing the relationship of the reconstructed $\pT$ of the tracks ($\pT^{\mathrm{rec}}$) to the generated $\pT$ of the matched particles.
 
The unfolding accuracy is limited by the number of MC events through statistical fluctuations within the migration matrices.
Therefore, a procedure to `smooth' the migration matrices via fitting is implemented.
 
First, the response $r$ defined as:
\begin{equation*}
r = \pT/\pT^{\mathrm{rec}} - 1,
\end{equation*}
is estimated for each collision system and centrality interval, and in small ranges of generated $\pT$ and $|\eta|$ (or $\ystar$) that coincide with the binning of the final distributions.
Every response distribution is fitted; the central part ($|r|\lesssim 0.2$) is described by two Gaussian distributions, while the positive ($r\gtrsim 0.2$) and negative ($r\lesssim -0.2$) tails are described by two independent exponential functions.
 
\begin{figure}[tbh]
\centering
\includegraphics[width=0.48\textwidth]{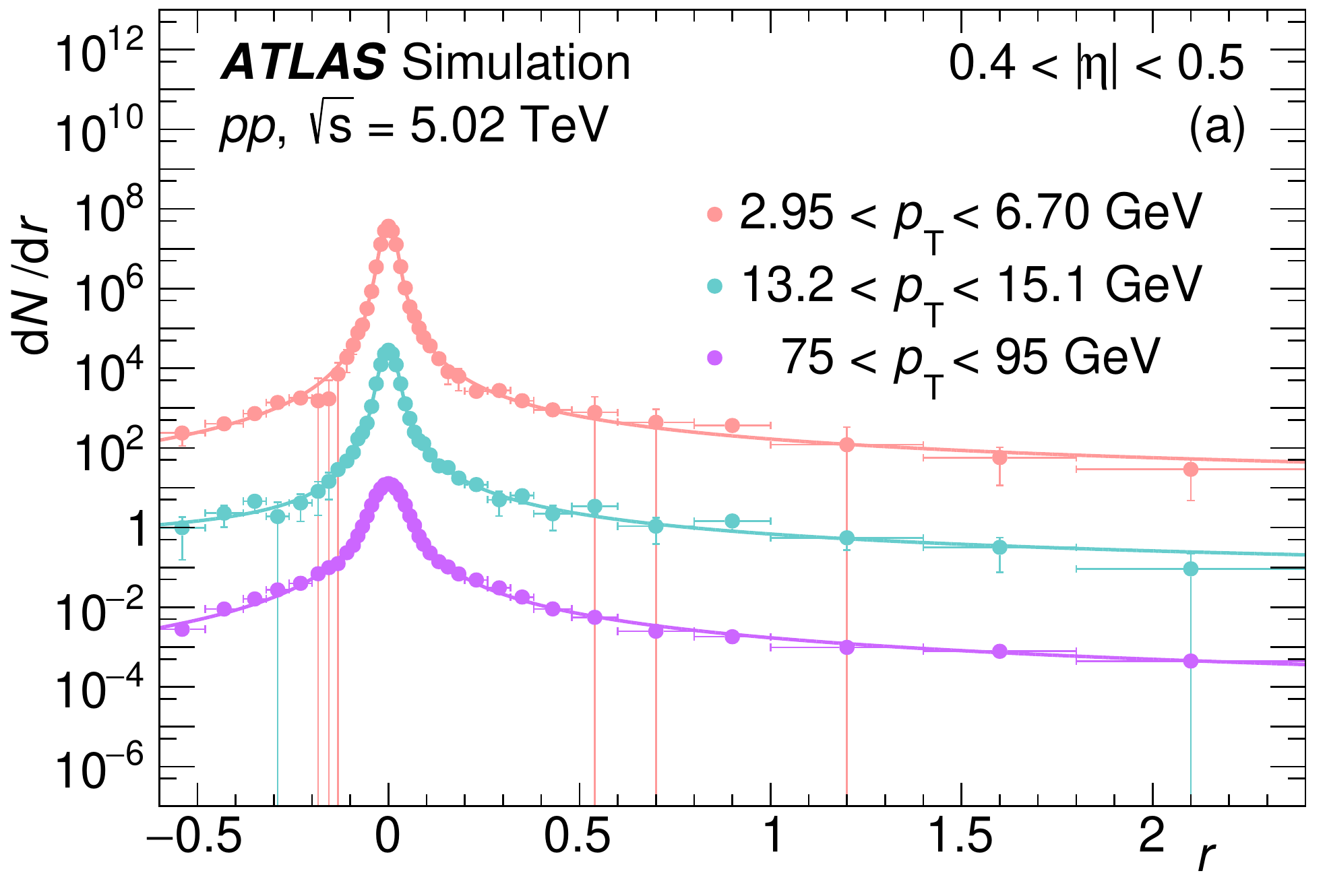} \includegraphics[width=0.48\textwidth]{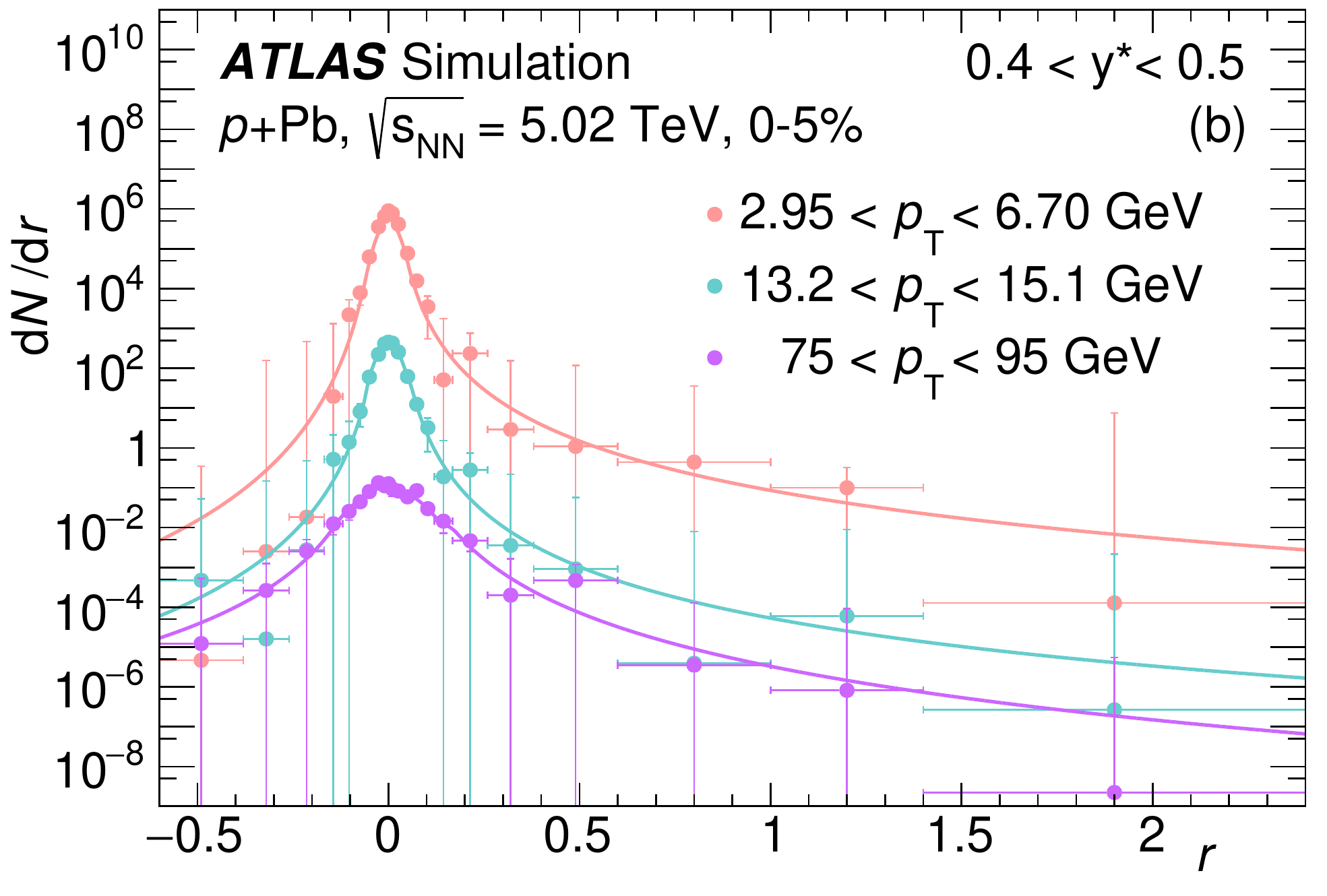}  \\
\includegraphics[width=0.48\textwidth]{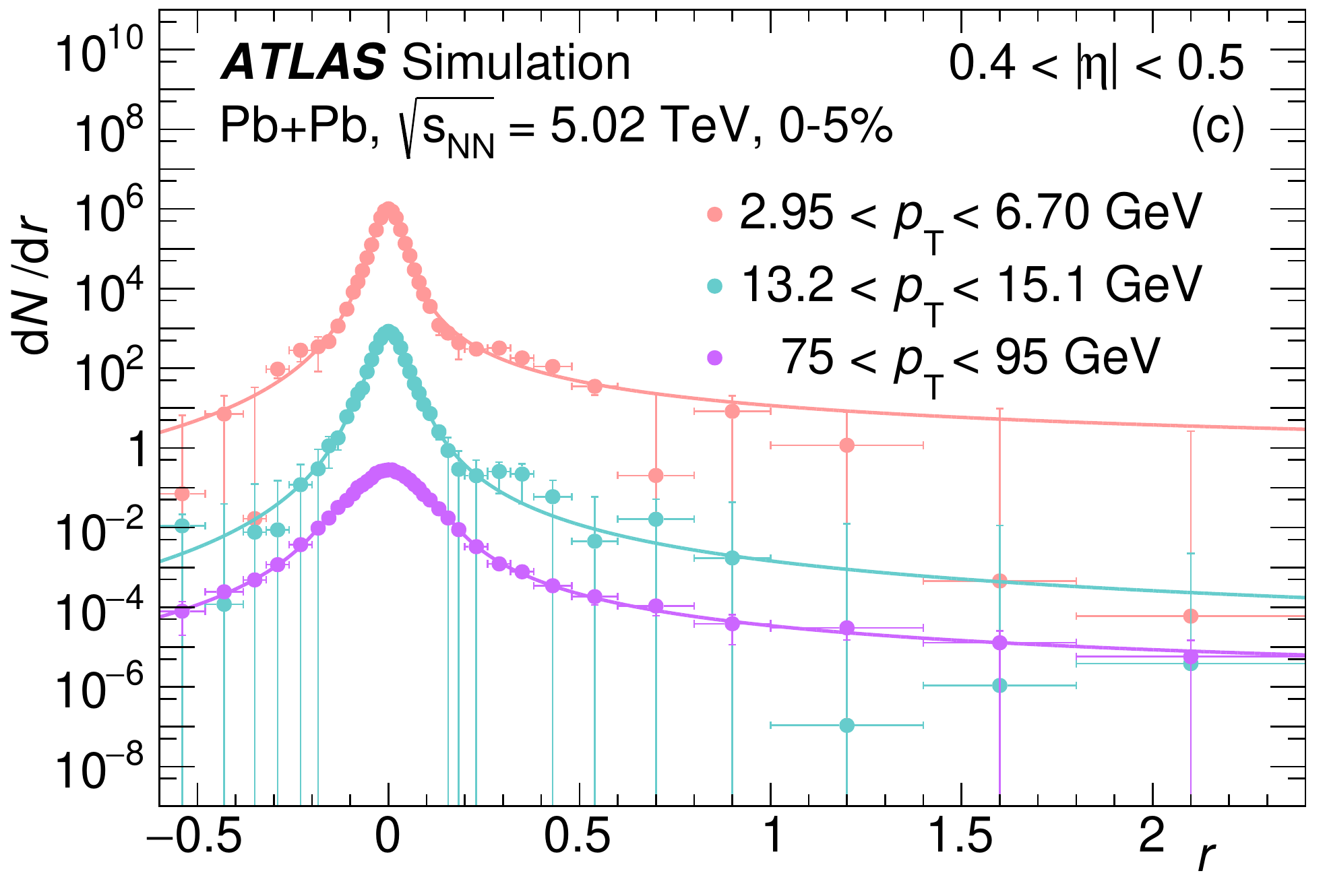} \includegraphics[width=0.48\textwidth]{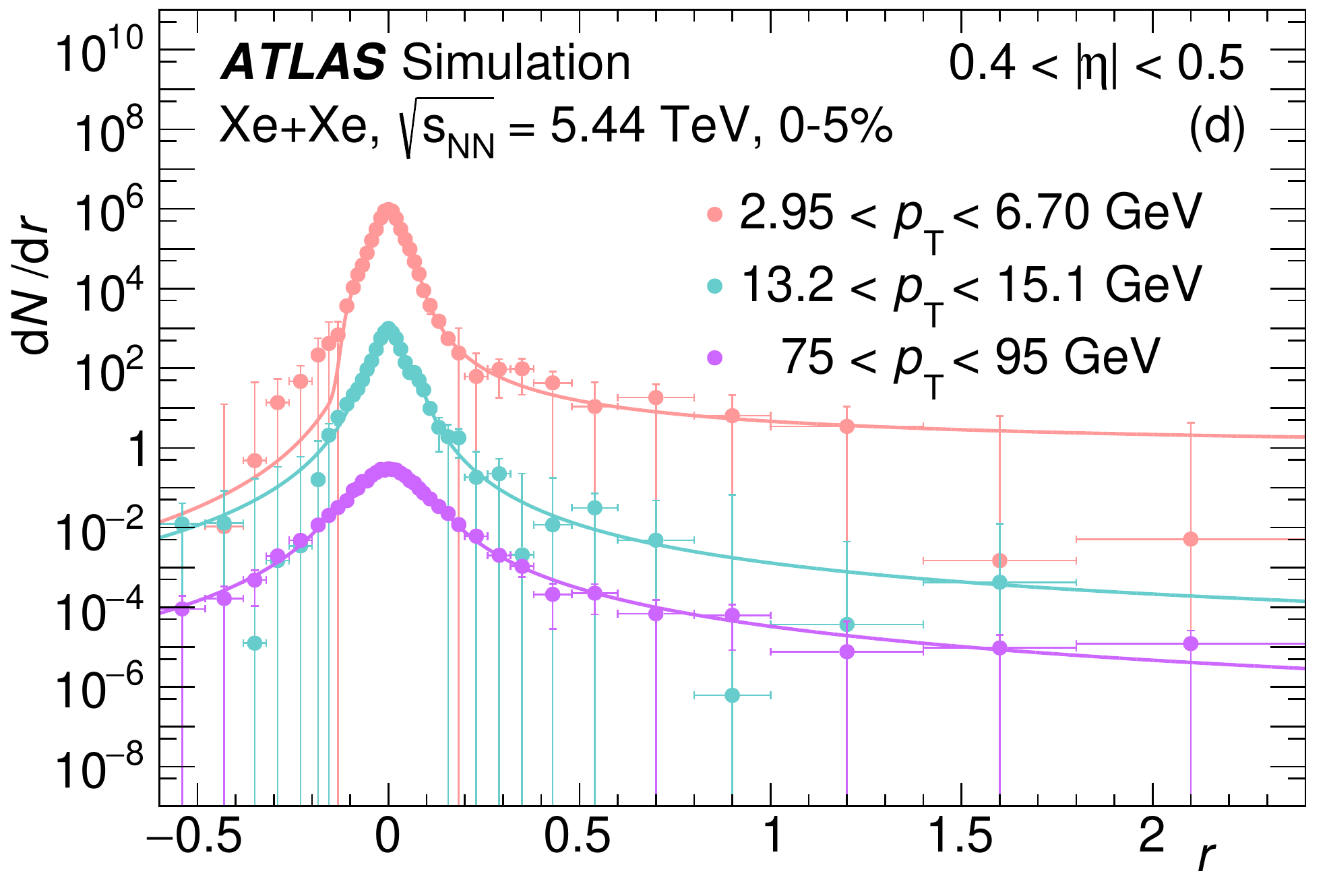}
\caption{The track \pT response as a function of $r$ for (a) \pp, (b) \pPb, (c) \PbPb, and (d) \XeXe collisions and for three ranges of generated $\pT$. The vertical bars represent the statistical uncertainties. The curves represent the corresponding fits.}
\label{fig:res}
\end{figure}

\begin{figure}[tbh]
\centering
\includegraphics[width=0.48\textwidth]{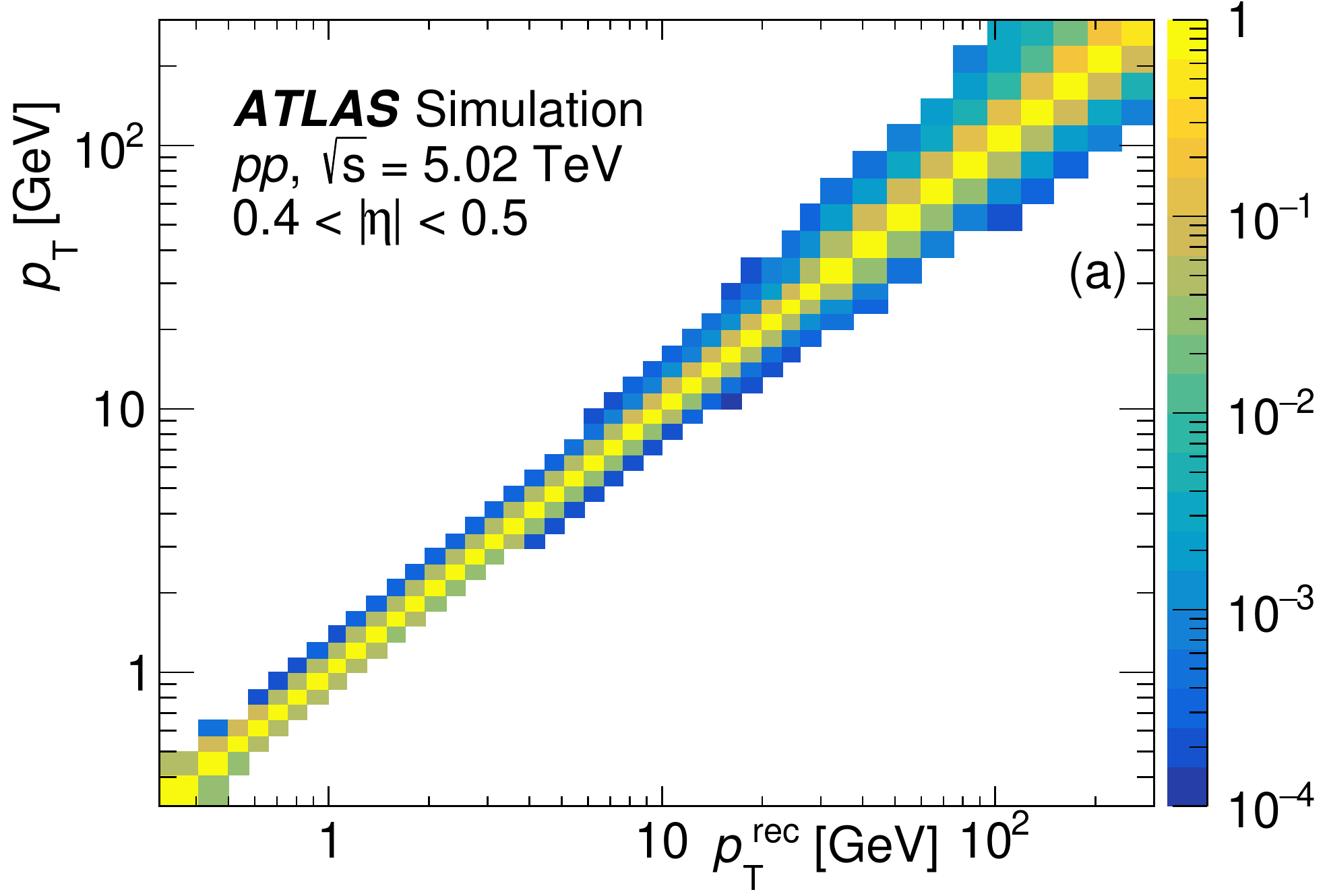} \includegraphics[width=0.48\textwidth]{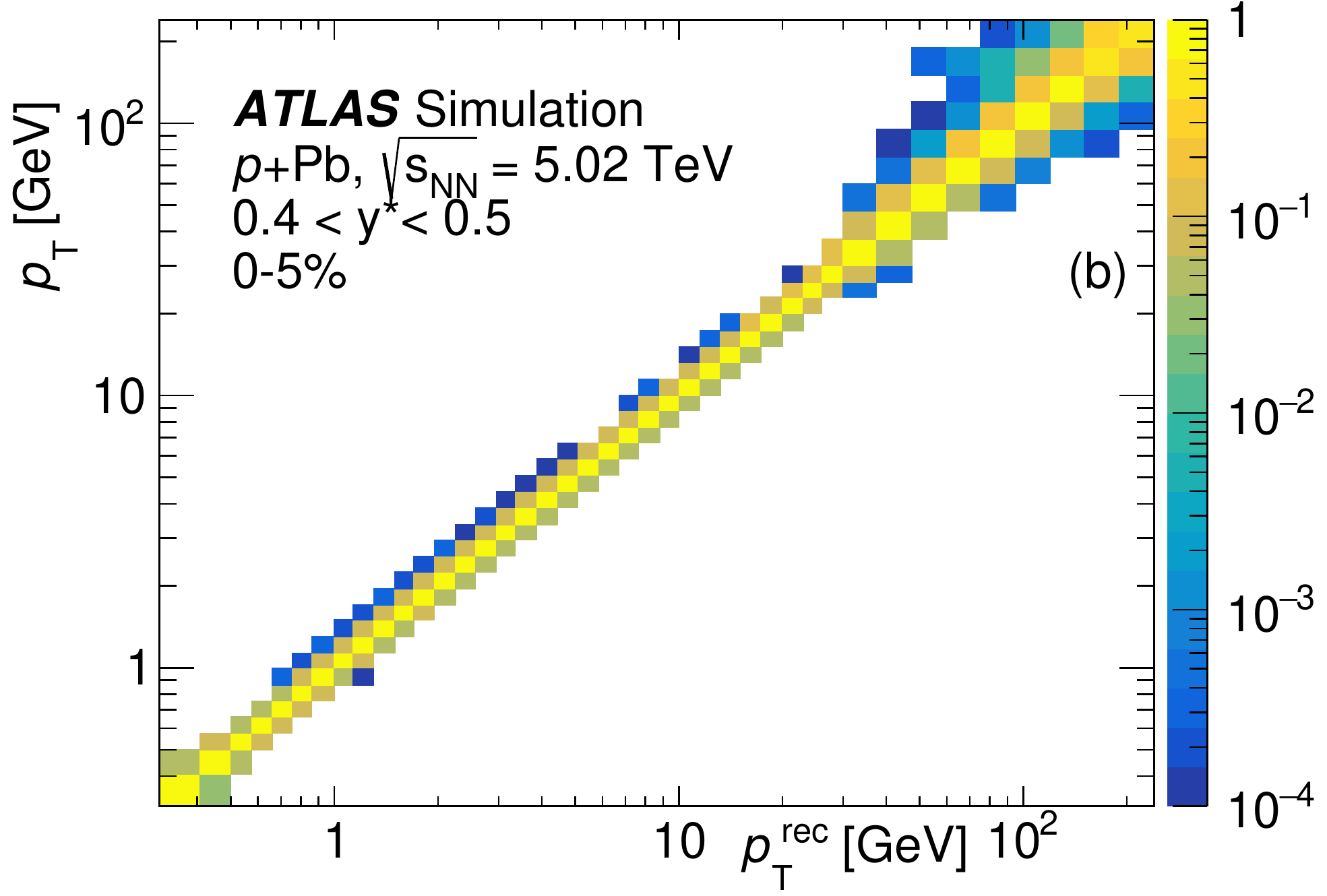}  \\
\includegraphics[width=0.48\textwidth]{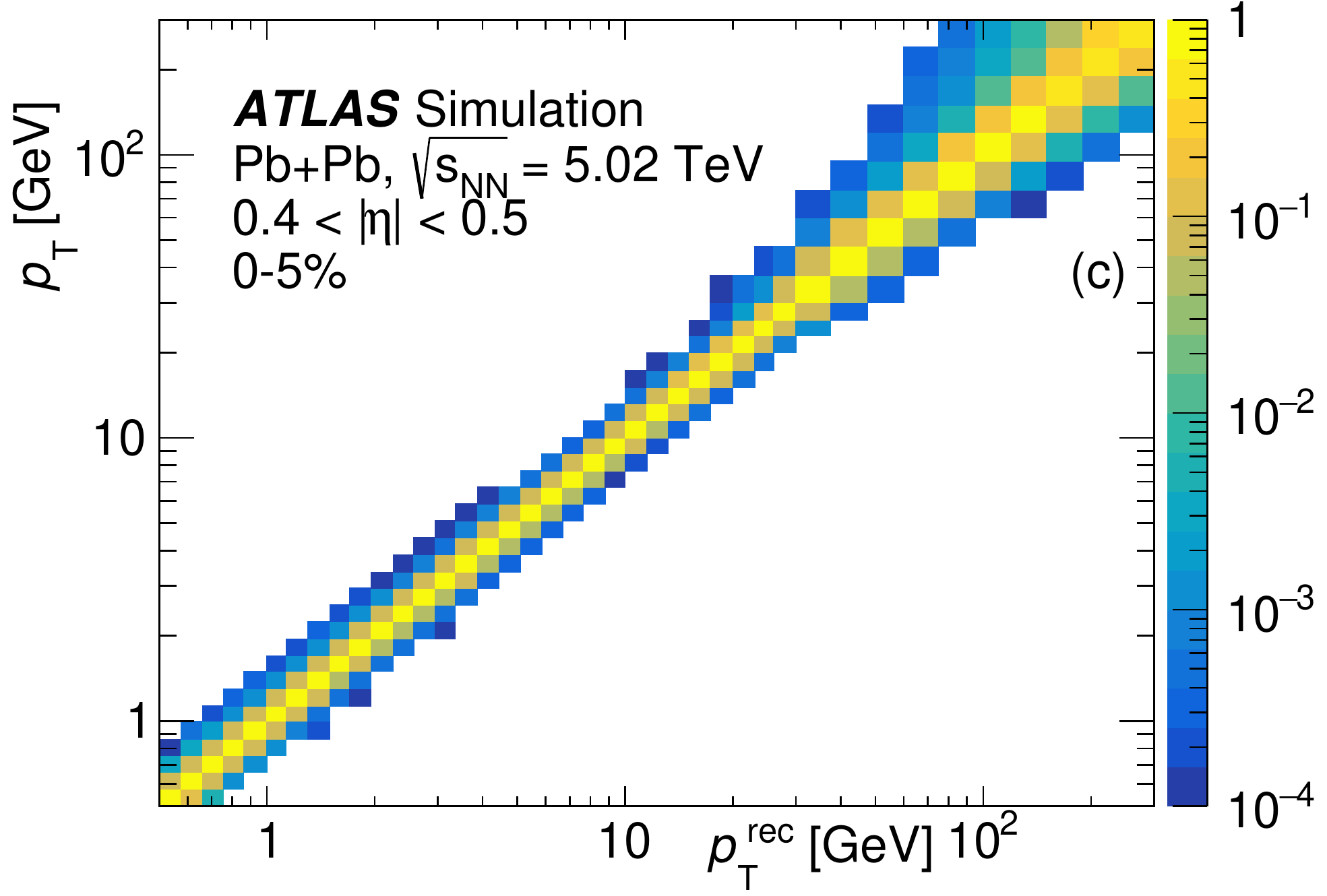} \includegraphics[width=0.48\textwidth]{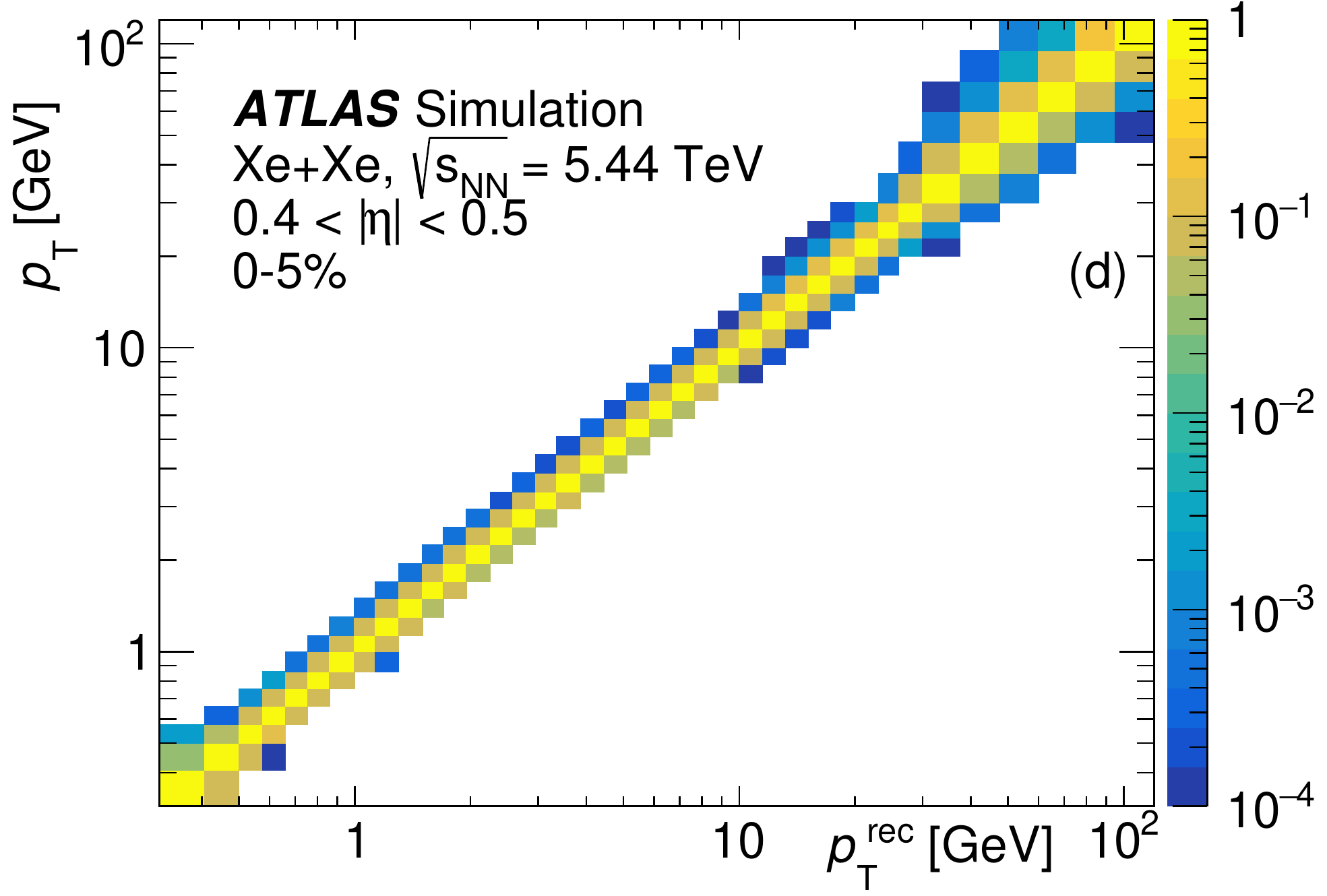}
\caption{The $\pT$ migration matrices for (a) \pp, (b) \pPb, (c) \PbPb, and (d) \XeXe collisions. The integral of the distribution in each row is normalized to unity.}
\label{fig:mm}
\end{figure}

Figure~\ref{fig:res} shows examples of the response distributions and their fits in a few $\pT$ ranges.
Migration matrices are built separately for each collision system, centrality interval, and small range of $|\eta|$ (or $\ystar$).
The matrices have fixed generated $\pT$ in rows and fixed reconstructed $\pT^{\mathrm{rec}}$ in columns.
To fill one bin of the matrix, an integral of the response fit with appropriate ranges has to be calculated.
The same fit is used for all bins in the same row, i.e.~for the same generated $\pT$.
Examples of the migration matrices after smoothing are shown in Figure~\ref{fig:mm} with the same centrality intervals and $|\eta|$ (or $\ystar$) ranges as the response fits in Figure~\ref{fig:res}.
The migration matrices are used in the iterative Bayesian unfolding.
The procedure converges within two iterations.
 
Iterative Bayesian unfolding is also used to correct for the $\eta$ (or $\ystar$) resolution.
In contrast to the $\pT$ resolution, the $\eta$ (and $\ystar$) resolution is better, the migration matrices are more diagonal, and therefore no smoothing procedure is required for the migration matrices used in unfolding.
Migration matrices are estimated separately for each collision system and centrality interval, and in small $\pT$ ranges.
The procedure converges within two iterations.
 
The unfolded spectra are corrected for the reconstruction efficiency, defined as the number of primary particles matched to reconstructed tracks divided by the total number of generated primary particles.
The reconstruction efficiency is calculated as a function of $\pT$, $\eta$ (or $\ystar$), collision system, and centrality.
To remove statistical fluctuations, the efficiency is fitted by a polynomial of the sixth order in $\log \left( \pT \right)$.
 
\begin{figure}[tbh]
\centering
\includegraphics[width=0.48\textwidth]{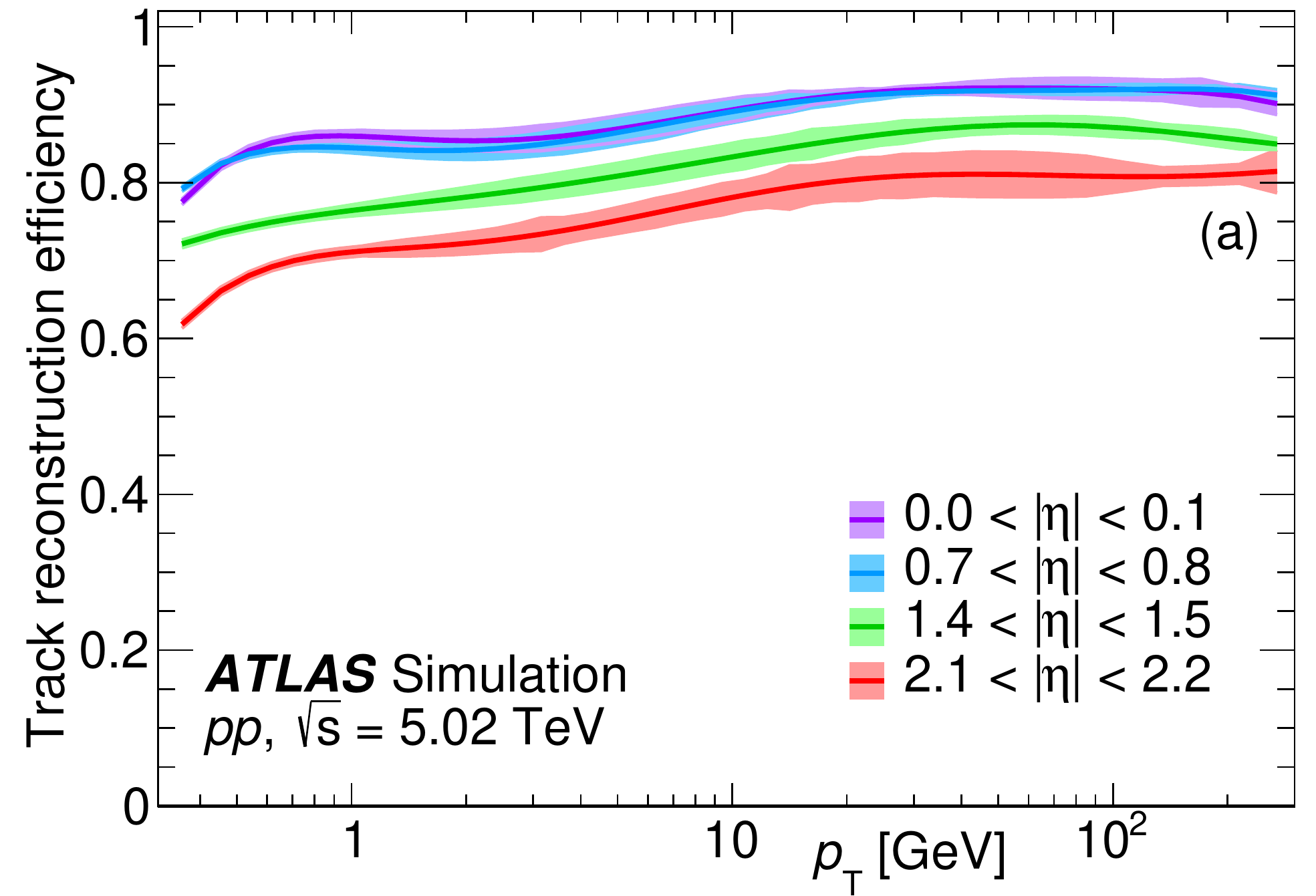} \includegraphics[width=0.48\textwidth]{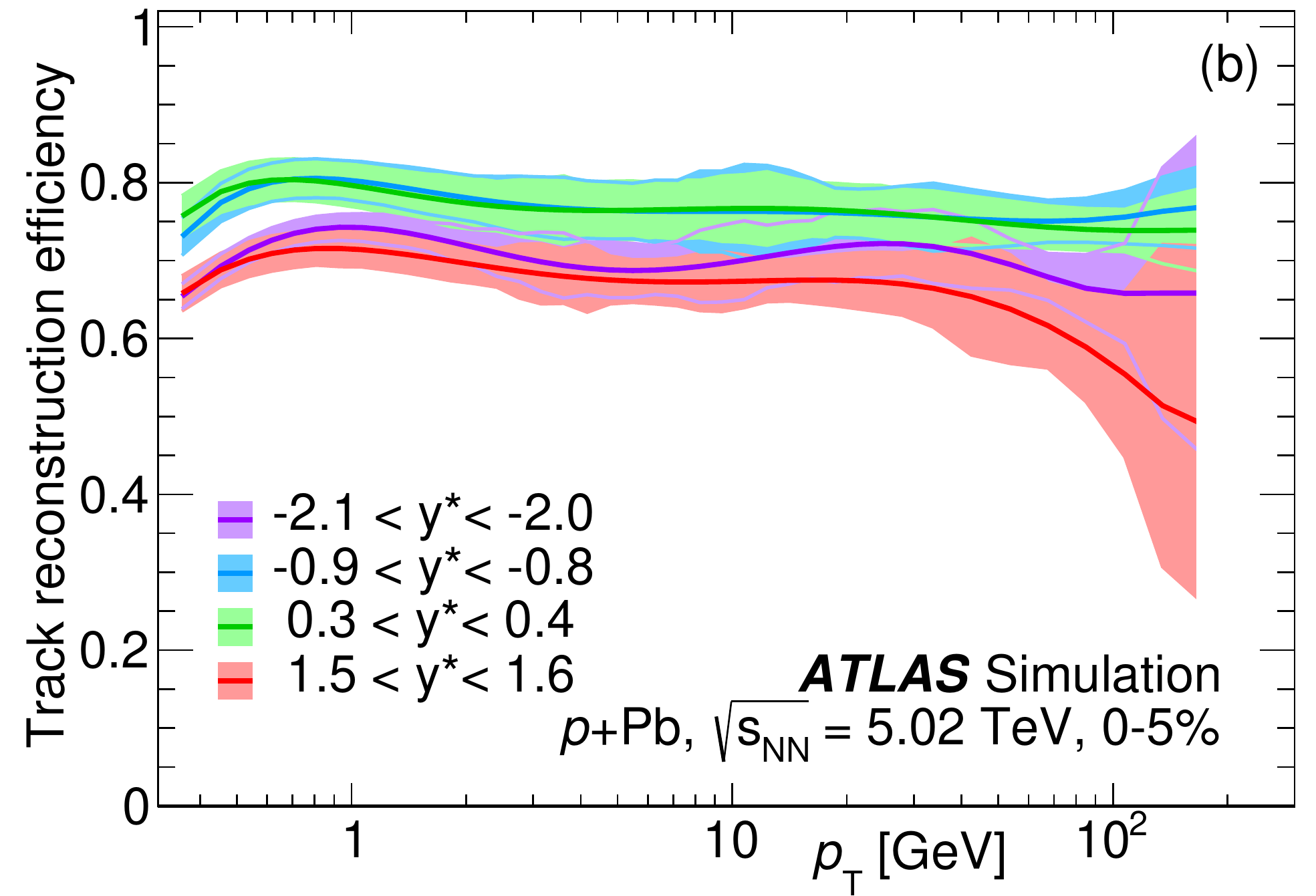} \\
\includegraphics[width=0.48\textwidth]{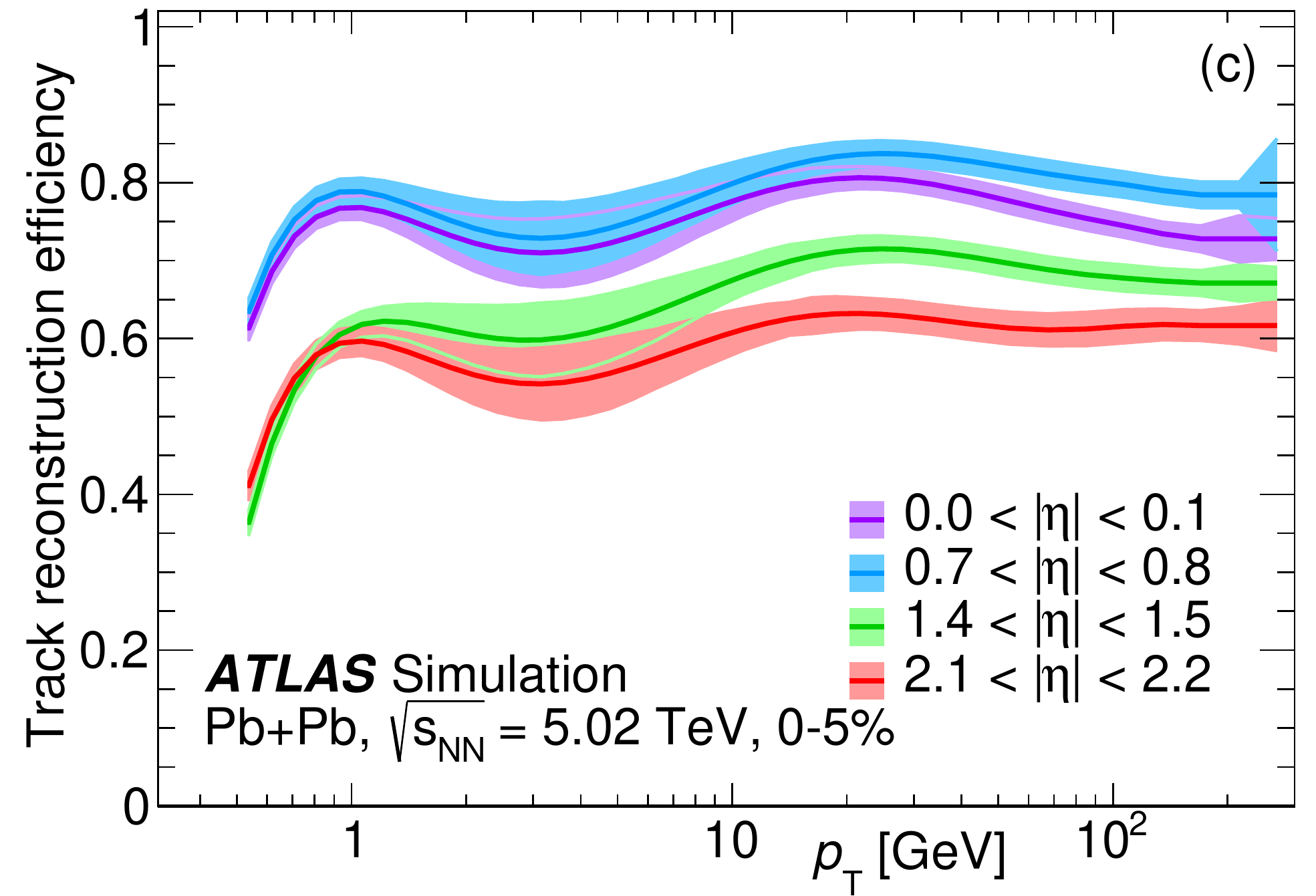} \includegraphics[width=0.48\textwidth]{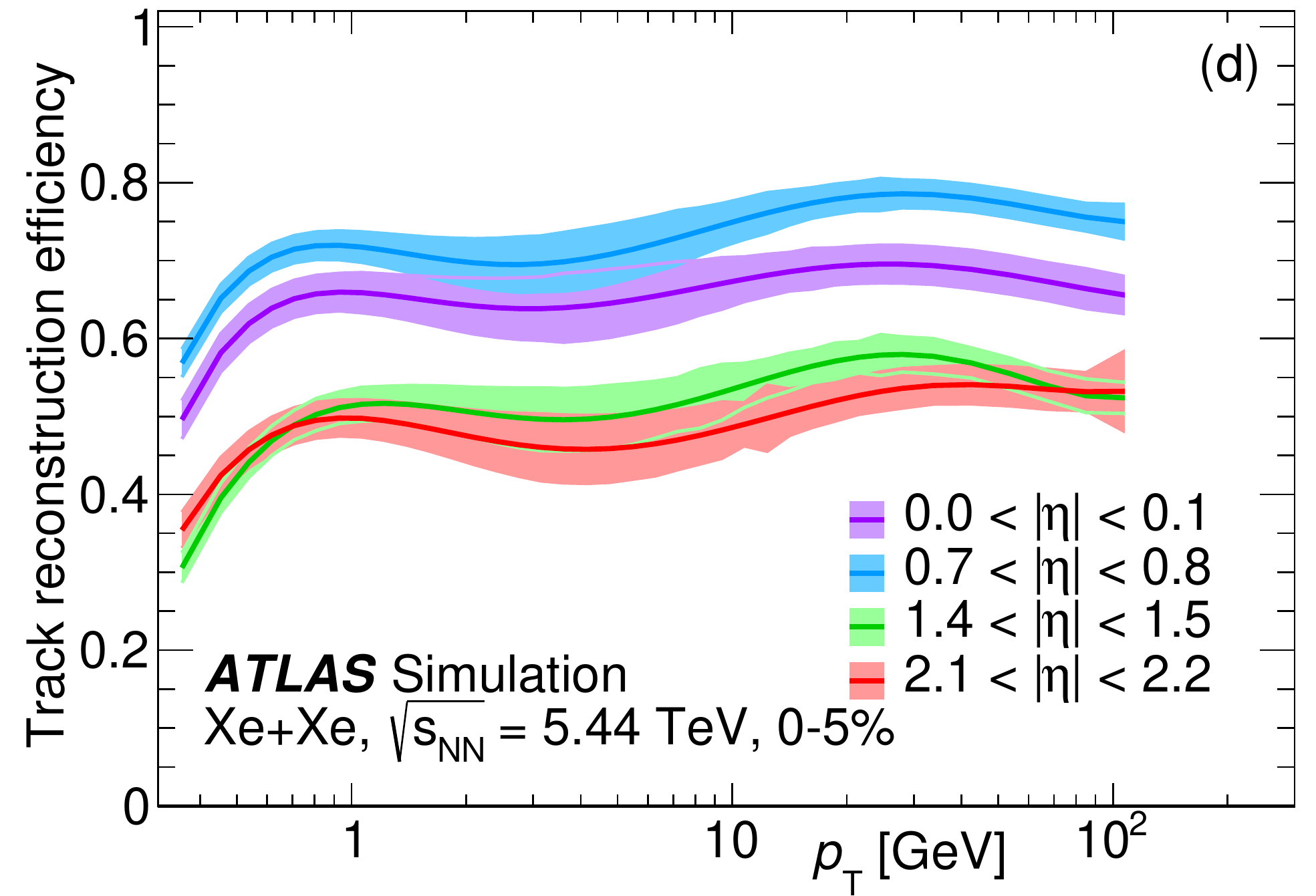}
\caption{Track reconstruction efficiency as a function of $\pT$ for (a) \pp, (b) \pPb, (c) \PbPb, and (d) \XeXe collisions and for four ranges of (pseudo)rapidity. The widths of the bands represent the systematic uncertainties; see Section~\ref{sec:systematics} for further details.}
\label{fig:effic}
\end{figure}

Figure~\ref{fig:effic} shows examples of the track reconstruction efficiency in several $\eta$ (or $\ystar$) ranges.
For HI collisions, only the most central collisions are shown.
The efficiency is typically higher for more peripheral collisions.
The efficiencies for different collision systems also reflect changing detector conditions and reconstruction software, which are implemented in the simulation.
The decrease at low $\pT$ is a feature common to all centrality intervals.
The decrease in reconstruction efficiency around $\pT$ of 3\,\GeV is due to a large number of strange baryons, which according to the definition in Section~\ref{sec:tracks} are considered primary particles.
At this $\pT$, they are still unlikely to fully traverse the ID -- with the probability growing at higher momentum.
The reconstruction efficiency is expected to deteriorate at high $\pT$ due to the higher likelihood of mismeasuring tracks in the dense core of high-$\pT$ jets~\cite{HION-2017-04}.
 
To construct the charged-hadron \Raa in the \XeXe system at $\sqs=5.44$\,\TeV the reference \pp spectrum is extrapolated from the measured \pp spectrum at $\sqs=5.02$\,\TeV.
A multiplicative extrapolation factor is estimated as a ratio of the generated spectrum at $\sqs=5.44$\,\TeV to that at $\sqs=5.02$\,\TeV.
To remove statistical fluctuations, the extrapolation factor is fitted by a polynomial of the third order in $\log \left( \pT \right)$.
The extrapolation factor is estimated as a function of \pT and $\eta$.
 
\begin{figure}[tbh]
\centering
\includegraphics[width=0.48\textwidth]{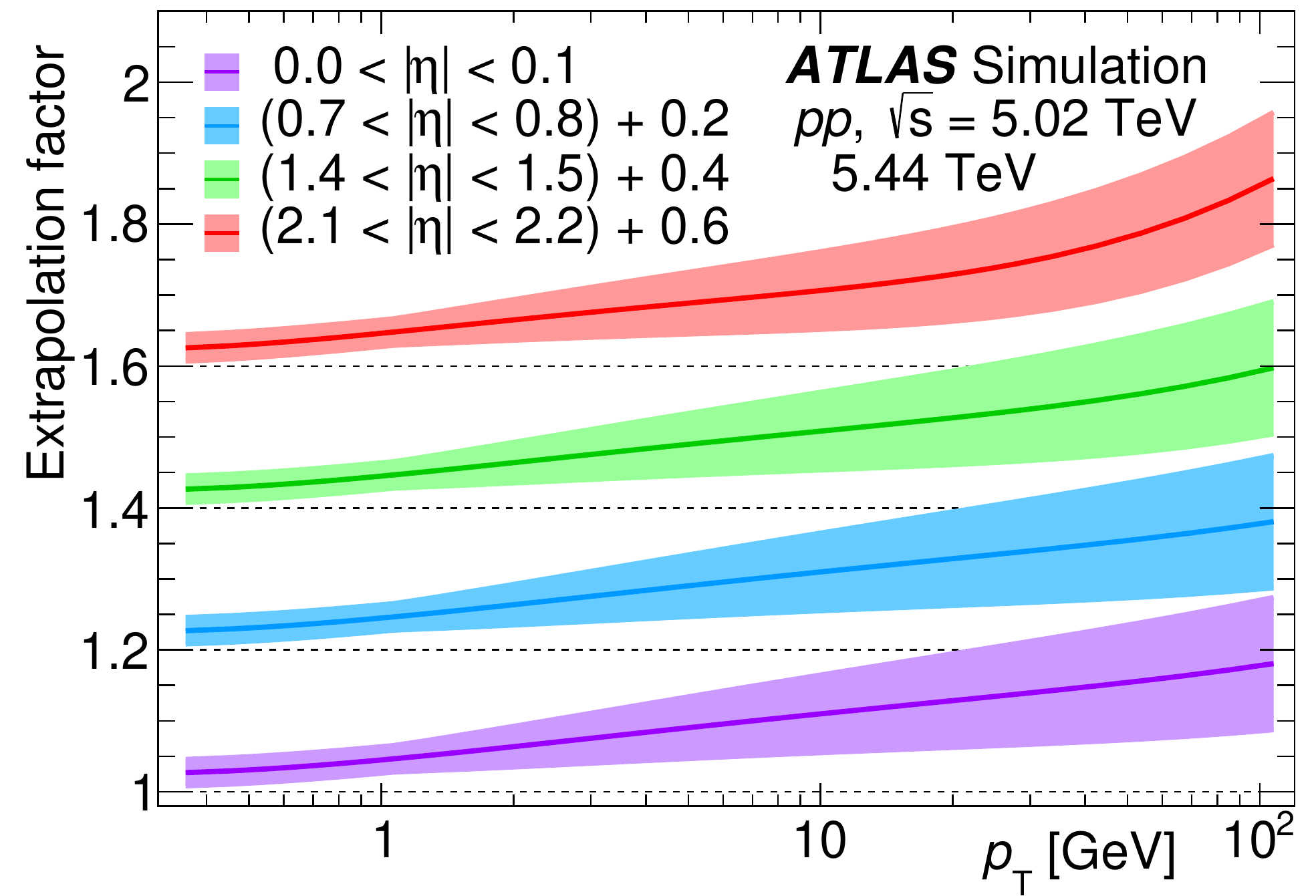}
\caption{Extrapolation factors as a function of $\pT$ for four ranges of pseudorapidity. The widths of the bands represent the systematic uncertainties; see Section~\ref{sec:systematics} for further details. The values are vertically offset for clarity.}
\label{fig:extrap}
\end{figure}

Figure~\ref{fig:extrap} shows the extrapolation factor in several $\eta$ ranges.
At low $\pT$, the factor is almost independent of $|\eta|$.
At high $\pT$, the factor is higher for high $|\eta|$ ranges.
 
For the measurements of \pPb collisions as well as of the corresponding \pp baseline, all the corrections are estimated as a function of $\ystar$, assuming the pion mass for all particles.
The same assumption is made for the tracks from the data samples.
To account for the generated mass of the particles, a correction similar to the one in Ref.~\cite{HION-2012-14} is applied.
It is defined as a ratio of the generated particle spectrum using the correct mass to the spectrum using the mass of the pion.
It is estimated as a function of $\pT$, $\ystar$, and centrality.
It has an effect only at low $\pT$ and high $|\ystar|$, where it corrects for migration in $\ystar$; there the correction factor can fall to about 0.85,
from values closer to unity elsewhere.

\FloatBarrier


\section{Systematic uncertainties}
\label{sec:systematics}
 
The systematic uncertainties of the measurements arise from several independent sources.
The contribution of each source is estimated by varying the corresponding parameters of the analysis and propagating the impact of the source to the final result by performing the full analysis chain.
 
The total systematic uncertainties are determined by adding the contributions from all relevant sources in quadrature. All systematic uncertainties are summarized in Table~\ref{tab:syst}.
 
\begin{description}
\item[Track selection.]
In addition to the track selection requirements explained in Section~\ref{sec:tracks}, two other sets of requirements are considered.
The more restrictive criteria, with more required hits per track and stricter pointing-to-vertex requirements, provide a higher fraction of primary tracks but with lower efficiency.
On the other hand, the relaxed criteria, with fewer required hits per track and looser pointing-to-vertex requirements, offer higher track reconstruction efficiency at the expense of more fake and secondary tracks.
The average of the absolute values of the deviations from the default selection defines the systematic uncertainty.
The uncertainty is higher in more central collisions and at higher $\pT$.
\item[Momentum bias.]
A residual detector misalignment present in \PbPb data causes a momentum bias that is absent in the MC simulations.
An additional correction is introduced to compensate for this effect.
Its uncertainty is propagated to the measured distributions.
No such misalignment is present in other data samples.
The uncertainty is higher at high $\pT$ and in more peripheral collisions.
\item[Track-to-particle matching quality.]
The efficiency corrections used in the analysis rely on matching the reconstructed tracks to generated particles.
To account for ambiguities in the matching procedure, the matching probability, defined in Ref.~\cite{ATL-PHYS-PUB-2015-051}, is varied to assess the systematic uncertainty.
This uncertainty is most pronounced at low \pT in the most central collisions.
\item[Particle composition.]
The systematic uncertainty of the measurements of particle compositions~\cite{alice_partcomp_pp_pbpb_2019,alice_partcomp_ppb}, as well as the full difference between the model and the data~\cite{Andronic:2017pug,ALICE:2013xmt}, are taken into account when estimating particle weights for the analysis as described in Section~\ref{sec:events}.
The resulting systematic uncertainty reaches its maximum at $3\lesssim \pT \lesssim 5$\,\GeV.
\item[Correction for fake and secondary tracks.]
The systematic uncertainty due to the fake and secondary tracks is estimated to be 100\% of the secondary and fake rate.
This is independent of the track selection criteria.
\item[Resolution correction procedures in \pT and $\eta$.]
To test the stability of \pT and $\eta$ resolution corrections, the content of each bin in the \pT resolution distributions or in the $\eta$ migration matrices is individually changed following a Poisson distribution, reflecting the statistical uncertainties of a bin.
The statistical uncertainties for \pT resolutions are shown in Figure~\ref{fig:res}.
\item[Unfolding procedure.]
In the analysis, the measured spectra are first unfolded to correct for \pT resolution and then to correct for $\eta$ (or $\ystar$) resolution.
To check the impact of this arbitrary choice, the order of these two steps is swapped.
\item[Track reconstruction efficiency.]
The track reconstruction efficiency is fitted to produce a smooth correction.
The difference between the fit and the data points is used to estimate an uncertainty in the track reconstruction efficiency; the difference is driven by the limited number of events in the MC samples.
\item[Detector material.]
This systematic uncertainty is associated with possible mis-modelling of the detector material~\cite{ATL-PHYS-PUB-2015-051}, effectively lowering the track reconstruction efficiency.
\item[Luminosity, $\avgTpPb$, and $\avgTaa$.]
The integrated luminosity determined for the \pp data is calibrated using data from dedicated beam separation scans, using procedures described in Ref.~\cite{DARP-2013-01}.
The systematic uncertainties in the mean nuclear thickness functions arise from the geometric modelling uncertainties and from the uncertainties in the fraction of selected inelastic HI collisions.
The uncertainties are adapted from Refs.~\cite{HION-2013-08,HION-2016-04,HION-2019-03} and are listed in Table \ref{tab:all_cent}.
\item[Extrapolation of the \textit{pp} baseline to $\sqs=5.44$\,\TeV.]
The measured \pp spectrum at $\sqs=2.76$\,\TeV~\cite{HION-2011-03} is extrapolated to $\sqs=5.02$\,\TeV in a way analogous to that described in Section~\ref{sec:analysis}, using generated spectra at $\sqs=2.76$\,\TeV and 5.02\,\TeV.
The extrapolated spectrum at $\sqs=5.02$\,\TeV is compared with the \pp spectrum at $\sqs=5.02$\,\TeV measured by this analysis.
At $\pT\approx 100$\,\GeV, the cross-section increase in the extrapolation is about 40\% less than it should be; it is more accurate at lower $\pT$.
It is assumed that the cross-section increase is underestimated by the same percentage, with its associated uncertainty, when extrapolating the \pp spectrum measured at $\sqs=5.02$\,\TeV to $\sqs=5.44$\,\TeV.
\end{description}
 
\begin{table}[tb]
\centering
\caption{The summary of systematic uncertainties. The first listed numbers are typical values and the second numbers are maximum values of the uncertainties. All values are in percent. More details are given in the text. }
\begin{tabular}{l|cccc|ccc} \hline
\multirow{2}{*}{Source} & \multicolumn{4}{c|}{Spectra [\%]} &     \multicolumn{3}{c}{Ratios [\%]}  \\
&  \pp        &   \pPb       &   \PbPb      &  \XeXe       & \pPb/\kern -0.1em \pp & \PbPb/\kern -0.1em \pp & \XeXe/\kern -0.1em \pp \\ \hline
Track selection         & \pz 2 / 12  & \pl 5 / 33   & \pl 2 / 13   & \pl 2 / 9\pz & \pl 5 / 32   & \pl 4 / 18  & \pl 3 / 12     \\
Momentum bias           &  ---        & \pl ---  \pz & \pl 2 / 28   & \pl ---  \pz & \pl ---  \pz & \pl 2 / 28  &  \pl ---  \pz  \\
Track-to-part.~match.   & $<$1 / $<$1 & \pl 1 / 3\pz & \pl 1 / 6\pz & \pl 1 / 5\pz & \pl 1 / 3\pz & \pl 1 / 6\pz &  \pl 1 / 5\pz \\
Particle composition    &    1 / 2    & \pl 2 / 8\pz & \pl 2 / 8\pz & \pl 2 / 4\pz & \pl 1 / 8\pz & \pl 1 / 7\pz &  \pl 1 / 2\pz \\
Fake \& second.~tracks  &    1 / 2    & \pl 1 / 2\pz & \pl 1 / 4\pz & \pl 1 / 6\pz & $<$1 / 1\pz  & $<$1 / 2\pz  &  \pl 1 / 3\pz \\
$\pT$ resolution correct. & $<$1 / $<$1 & $<$1 / 2\pz & $<$1 / 2\pz & $<$1 / 1\pz  & $<$1 / 3\pz  & $<$1 / 2\pz  &  $<$1 / 1\pz  \\
$\eta$ resolution correct. & $<$1 / $<$1 & $<$1 / 6\pz & $<$1 / 1\pz & $<$1 / 1\pz & $<$1 / 6\pz  & $<$1 / 1\pz  &  $<$1 / 1\pz  \\
Unfolding procedure     & $<$1 / $<$1 & $<$1 / 1\pz  &  $<$1 / 1\pz & $<$1 / 2\pz  & $<$1 / 2\pz  & $<$1 / 1\pz  &  $<$1 / 2\pz  \\
Track reconstr.~effic.  & $<$1 / $<$1 & \pl 1 / 2\pz & \pl 2 / 2\pz & \pl 2 / 2\pz & \pl 1 / 2\pz & \pl 2 / 2\pz & \pl 2 / 2\pz \\
Detector material       &    2 / 4    & \pl 3 / 6\pz & \pl 2 / 4\pz & \pl 2 / 4\pz & \pl 3 / 7\pz & \pl ---  \pz & \pl ---  \pz \\
Luminosity              &    2 / 2    & \pl ---  \pz & \pl ---  \pz & \pl ---  \pz & \pl 2 / 2\pz & \pl 2 / 2\pz & \pl 2 / 2\pz \\
$\avgTpPb$ or $\avgTaa$ &    ---      & \pz 6 / 11   & \pl 4 / 7\pz & \pl 4 / 8\pz & \pl 6 / 11   & \pl 4 / 7\pz & \pl 4 / 8\pz \\
Extrapolation of \pp    &    ---      & \pl ---  \pz & \pl ---  \pz & \pl ---  \pz & \pl ---  \pz & \pl ---  \pz & \pl 5 / 9\pz \\ \hline
Total                   & \pz 4 / 12  & \pz 8 / 38   & \pz 6 / 33   & \pz 6 / 11   & \pl 9 / 37   & \pl 7 / 35   & \pl 8 / 18   \\ \hline
\end{tabular}
\label{tab:syst}
\end{table}
 
Some sources of systematic uncertainty are considered correlated between the HI and $\pp$ systems and their variations can lead to their impacts partially cancelling out when the ratios $\Rppb$ and $\Raa$ are calculated.
This is the case for the uncertainties in track selection, track-to-particle matching quality, particle composition, the correction for fake and secondary tracks, and the unfolding procedure.
In these cases, the parameters are varied simultaneously for the HI and \pp collisions, so that the contributions properly reflect the correlated changes in the numerator and denominator of the ratio.
 
On the other hand, some sources are considered uncorrelated; this is the case for the uncertainties in resolution correction procedures in \pT and $\eta$, and the track reconstruction efficiency.
 
The uncertainties in detector material are considered correlated when comparing two samples from \RunTwo, which is the case for \PbPb/\kern -0.1em\pp and \XeXe/\kern -0.1em\pp.
The uncertainties from \RunOne and \RunTwo are not correlated by more than half of their combined values.
Since it is not feasible to work out the exact fractions, uncertainties are taken as uncorrelated when comparing a sample from \RunOne with a sample from \RunTwo, which is the case for \pPb/\kern -0.1em\pp.
 
The rest of the sources contribute only to either the numerator or the denominator.
 
The systematic uncertainties in the track quality selection, the momentum bias in \PbPb collisions, the extrapolation of the \pp baseline, and $\avgTpPb$ or $\avgTaa$ are the largest contributions to the total systematic uncertainty.
The uncertainties associated with other individual sources typically do not exceed 3\%.


\section{Results}
\label{sec:result}

This section presents the results for all observables measured in this analysis. It also compares them between centralities and different collision systems as well as with theoretical models and other experiments.
 
\begin{figure}[htb]
\centering
\includegraphics[width=0.60\textwidth]{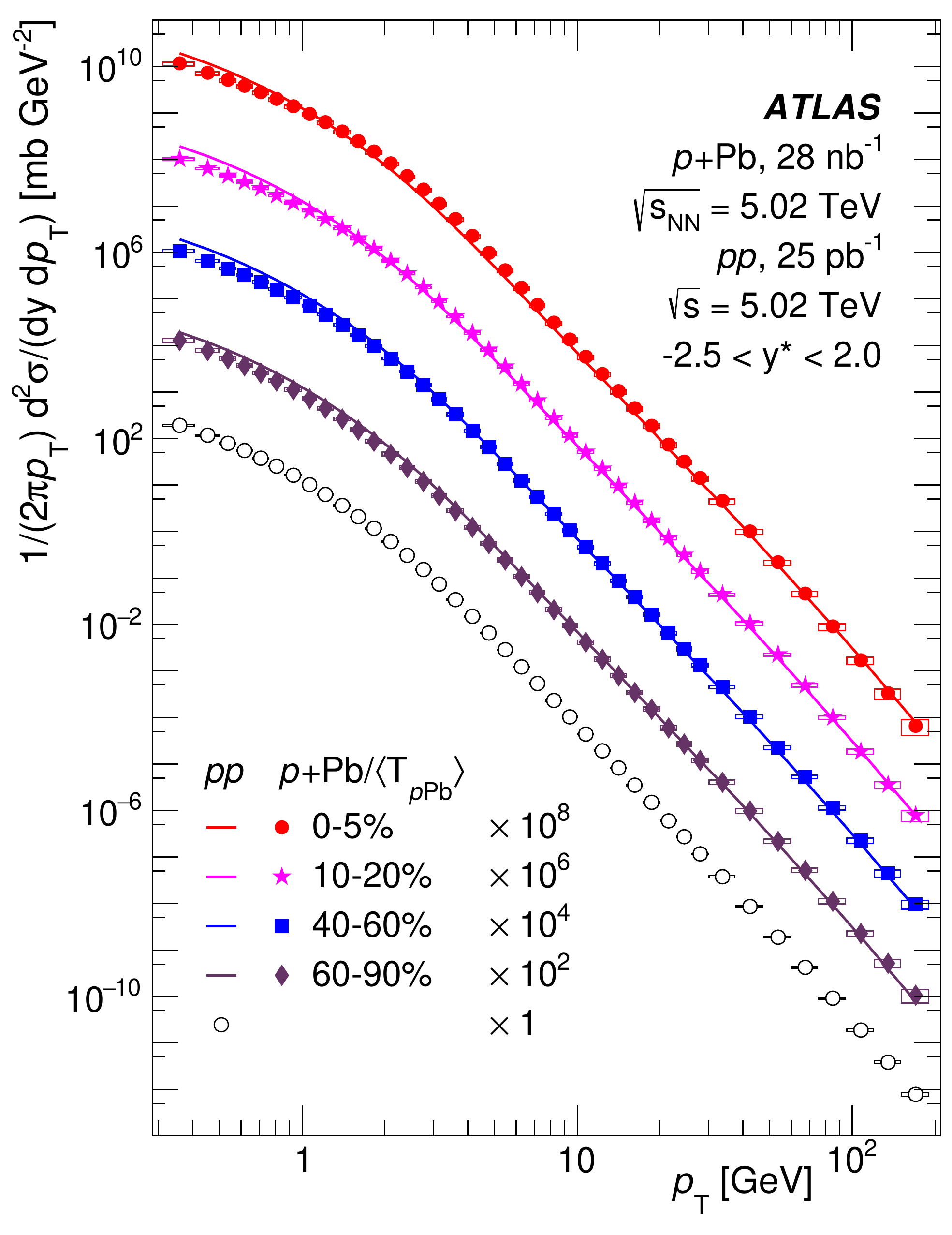}
\caption{Charged-hadron spectra divided by the nuclear thickness function in \pPb collisions and the cross-section in \pp collisions measured in the rapidity range $-2.5<\ystar<2.0$ at $\sqn=\sqs=5.02$\,\TeV. Measurements in centrality intervals are scaled by powers of ten for clarity in the plot. Systematic uncertainties are shown with boxes; statistical uncertainties are smaller than the marker size. All the lines represent the same \pp data, scaled by the same factor as the corresponding centrality interval.}
\label{fig:spectra_ppb}
\end{figure}

The corrected charged-hadron spectra from \pPb collisions at $\sqn=5.02$\,\TeV measured as a function of \pT for the rapidity range $-2.5<\ystar<2.0$ and for four centrality intervals are shown in Figure~\ref{fig:spectra_ppb}.
The \pPb spectra are divided by the $\avgTpPb$ of the corresponding centrality intervals and are compared with the charged-hadron production cross-section measured in the \pp collisions at the same nucleon--nucleon centre-of-mass energy and in the same rapidity range.
 
\begin{figure}[htb]
\centering
\includegraphics[width=0.60\textwidth]{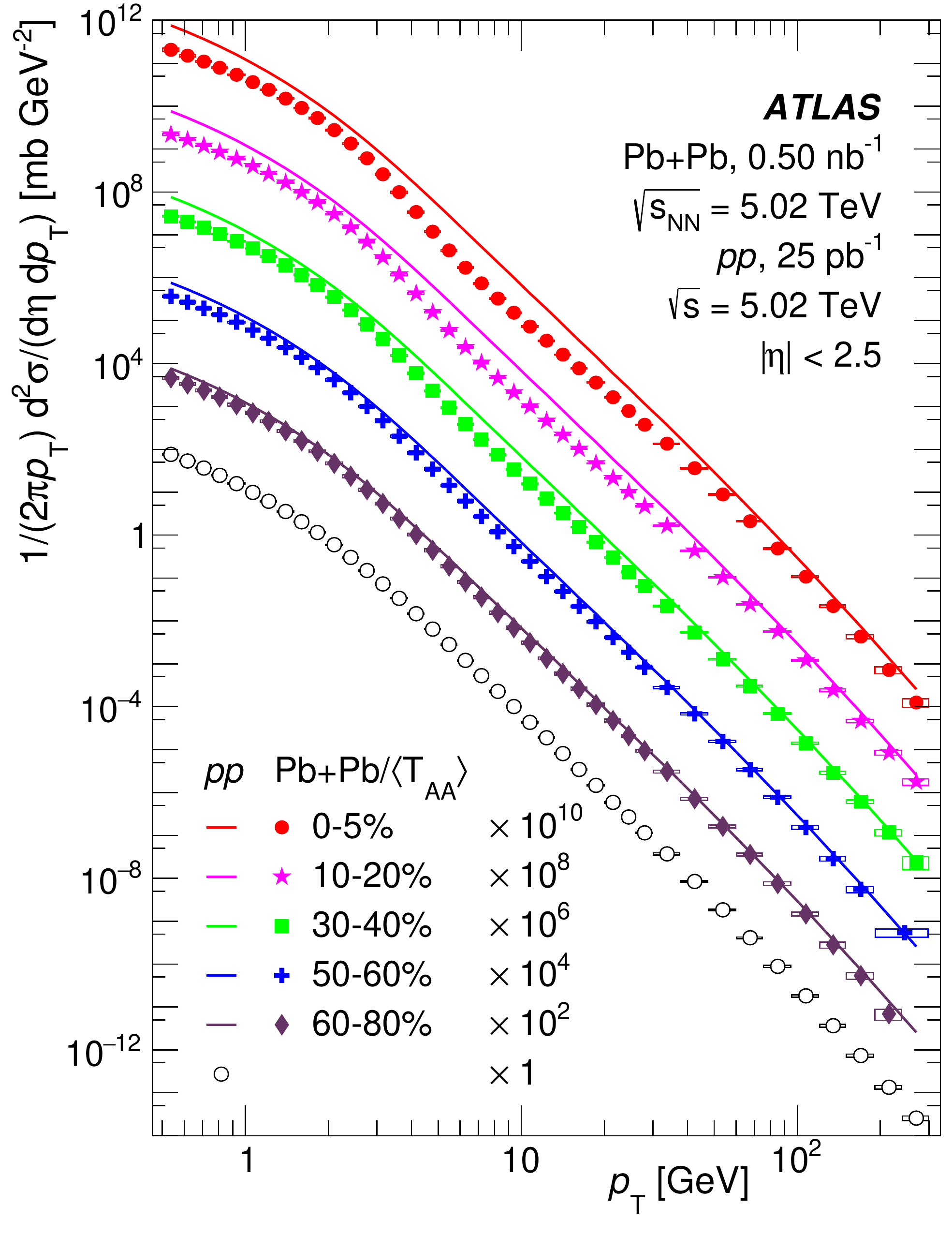}
\caption{Charged-hadron spectra divided by the nuclear thickness function in \PbPb collisions and the cross-section in \pp collisions measured in the pseudorapidity range $|\eta|<2.5$ at $\sqn=\sqs=5.02$\,\TeV. Measurements in centrality intervals are scaled by powers of ten for clarity in the plot. Systematic uncertainties are shown with boxes; statistical uncertainties are smaller than the marker size. Any data point with more than 30\% statistical uncertainty is not shown. All the lines represent the same \pp data, scaled by the same factor as the corresponding centrality interval.}
\label{fig:spectra_pbpb}
\end{figure}

\begin{figure}[htb]
\centering
\includegraphics[width=0.60\textwidth]{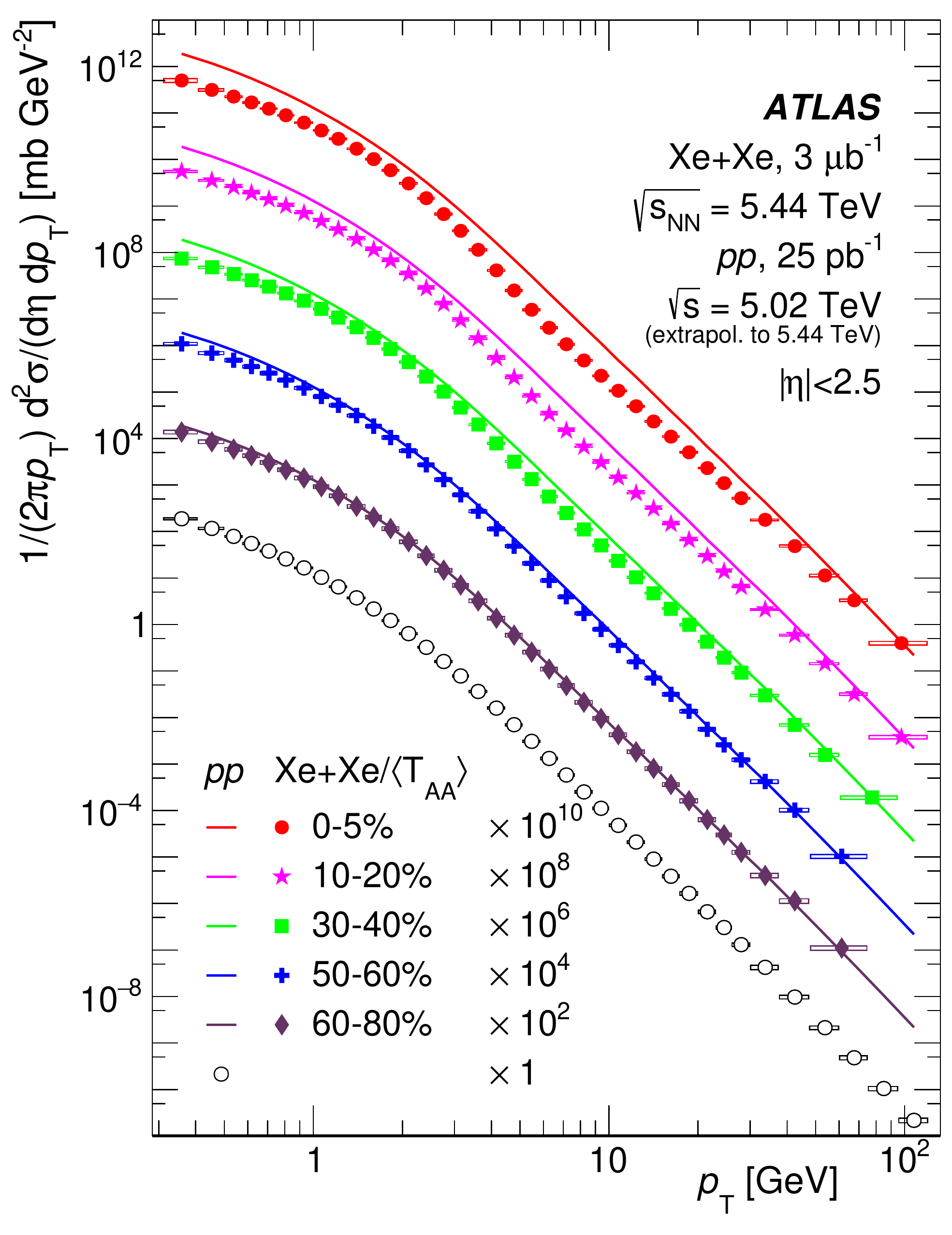}
\caption{Charged-hadron spectra divided by the nuclear thickness function in \XeXe collisions measured in the pseudorapidity range $|\eta| < 2.5$ at $\sqn=5.44$\,\TeV. Measurements in centrality intervals are scaled by powers of ten for clarity in the plot. Also shown is the cross-section measured in \pp collisions in the same pseudorapidity range and extrapolated to the same nucleon--nucleon centre-of-mass energy. Systematic uncertainties are shown with boxes; statistical uncertainties are smaller than the marker size. Any data point with more than 30\% statistical uncertainty is not shown. All the lines represent the same \pp data, scaled by the same factor as the corresponding centrality interval.}
\label{fig:spectra_xexe}
\end{figure}

The corrected charged-hadron spectra from \PbPb and \XeXe collisions at $\sqn=5.02$\,\TeV and 5.44\,\TeV, respectively, measured as a function of \pT for the pseudorapidity range $|\eta|<2.5$ and for five centrality intervals are shown in Figures~\ref{fig:spectra_pbpb} and~\ref{fig:spectra_xexe}, respectively.
Both sets of \AA spectra are divided by the $\avgTaa$ of the corresponding centrality intervals and are compared with the charged-hadron production \pp cross-section in the same pseudorapidity range and at the same centre-of-mass energy as the corresponding \AA collisions.

\begin{figure}[ptbh]
\centering
\includegraphics[width=0.60\textwidth]{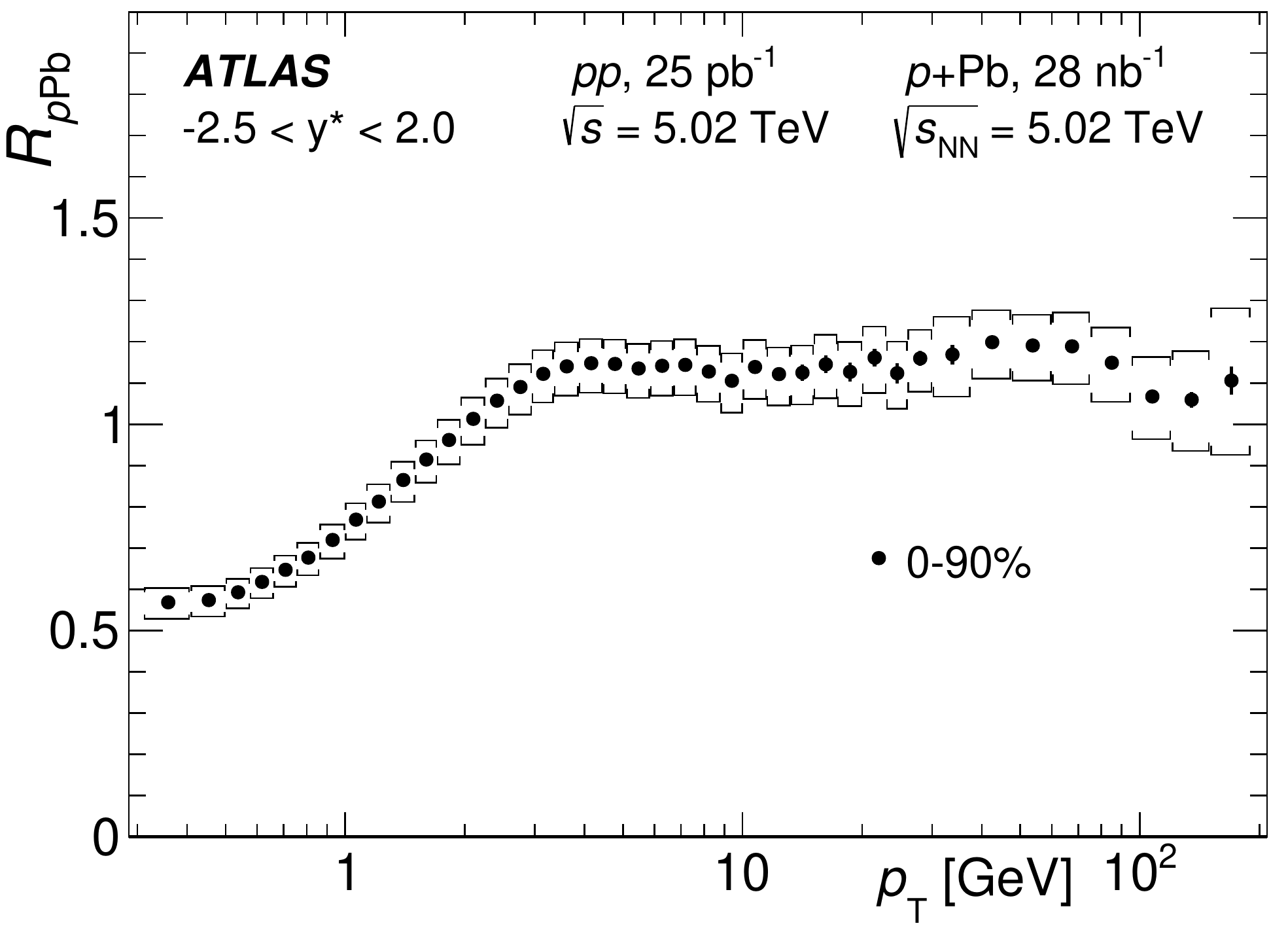}
\caption{Charged-hadron nuclear modification factor \RpPb in the rapidity region $-2.5 < \ystar < 2.0$. The markers show \RpPb for centrality interval 0--90\%. Systematic uncertainties are shown with brackets; statistical uncertainties are shown with vertical lines.}
\label{fig:raa_ppbMB}
\end{figure}

\begin{figure}[ptbh]
\centering
\includegraphics[width=0.60\textwidth]{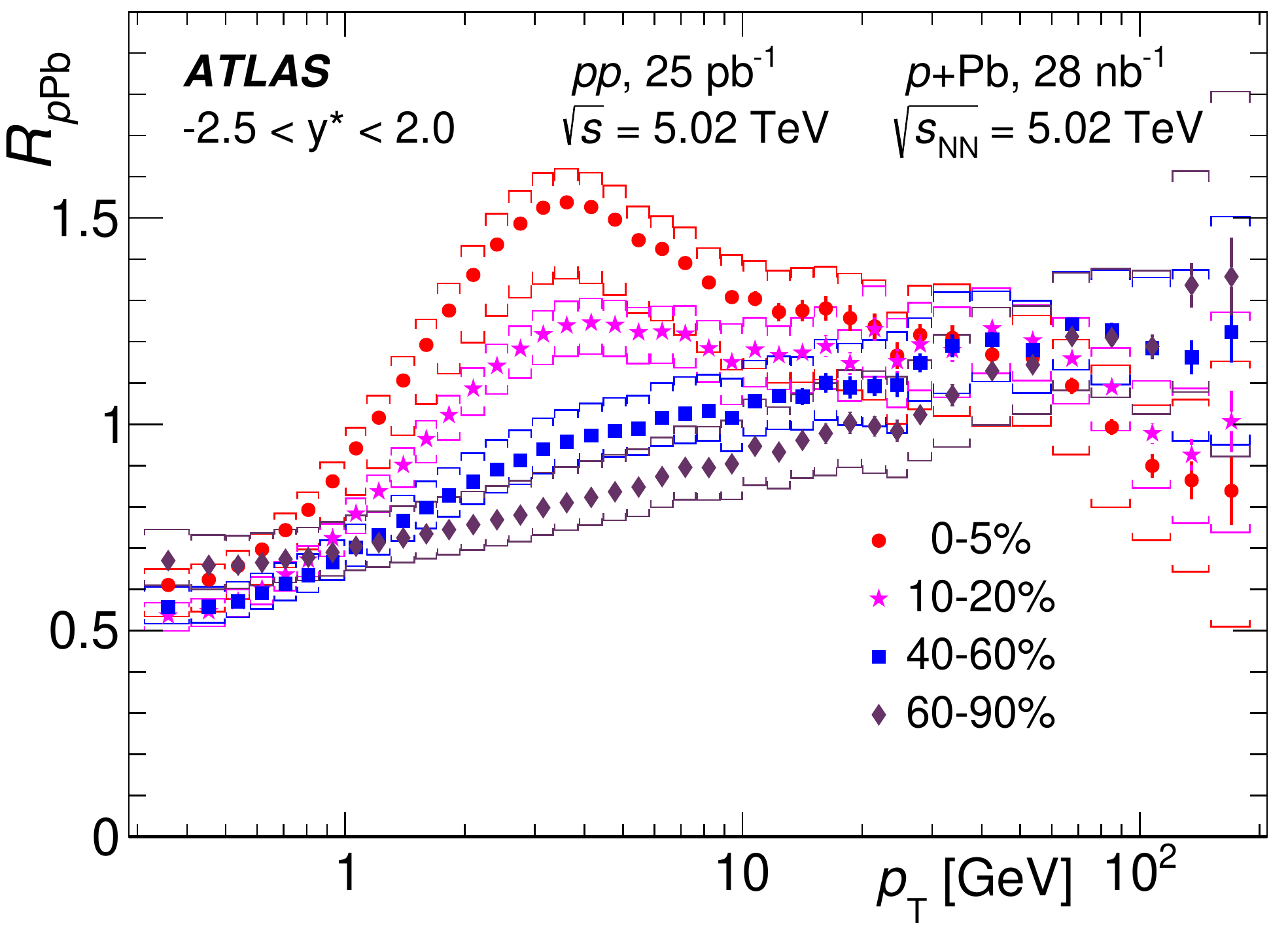}
\caption{Charged-hadron nuclear modification factor \RpPb in the rapidity region $-2.5 < \ystar < 2.0$. The full markers with different shapes show \RpPb for four different centralities. Systematic uncertainties are shown with brackets; statistical uncertainties are shown with vertical lines.}
\label{fig:raa_ppb}
\end{figure}

The nuclear modification factor, defined by Eq.~(\ref{eq:Rab}), quantifies the difference between HI and \pp collisions.
The difference is visible in all centrality intervals.
Figure~\ref{fig:raa_ppbMB} shows the nuclear modification factor $\Rppb$ in the centrality interval 0--90\%.
The \Rppb values exhibit only weak \pT dependence at $\pT > 4$\,\GeV and level off at a value $\Rppb = 1.14{}^{+0.06}_{-0.08}$ (syst.); this is obtained by fitting a constant to the data points in this $\pT$ range.
Figure~\ref{fig:raa_ppb} shows the nuclear modification factors $\Rppb$ for the same centrality intervals and $\ystar$ ranges as in the case of the charged-hadron spectra.
In the central collisions, $\Rppb$ rises up to $\pT\approx 3$\,\GeV where it has a local maximum.
In the more peripheral collisions, the maximum vanishes. Above 3\,\GeV, the values of $\Rppb$ in different centrality intervals converge and have very similar values at $\pT\approx 40$\,\GeV.
At higher momenta, \Rppb values from different centrality intervals follow different trends, but their values are consistent within the uncertainties.

\begin{figure}[ptbh]
\centering
\includegraphics[width=0.60\textwidth]{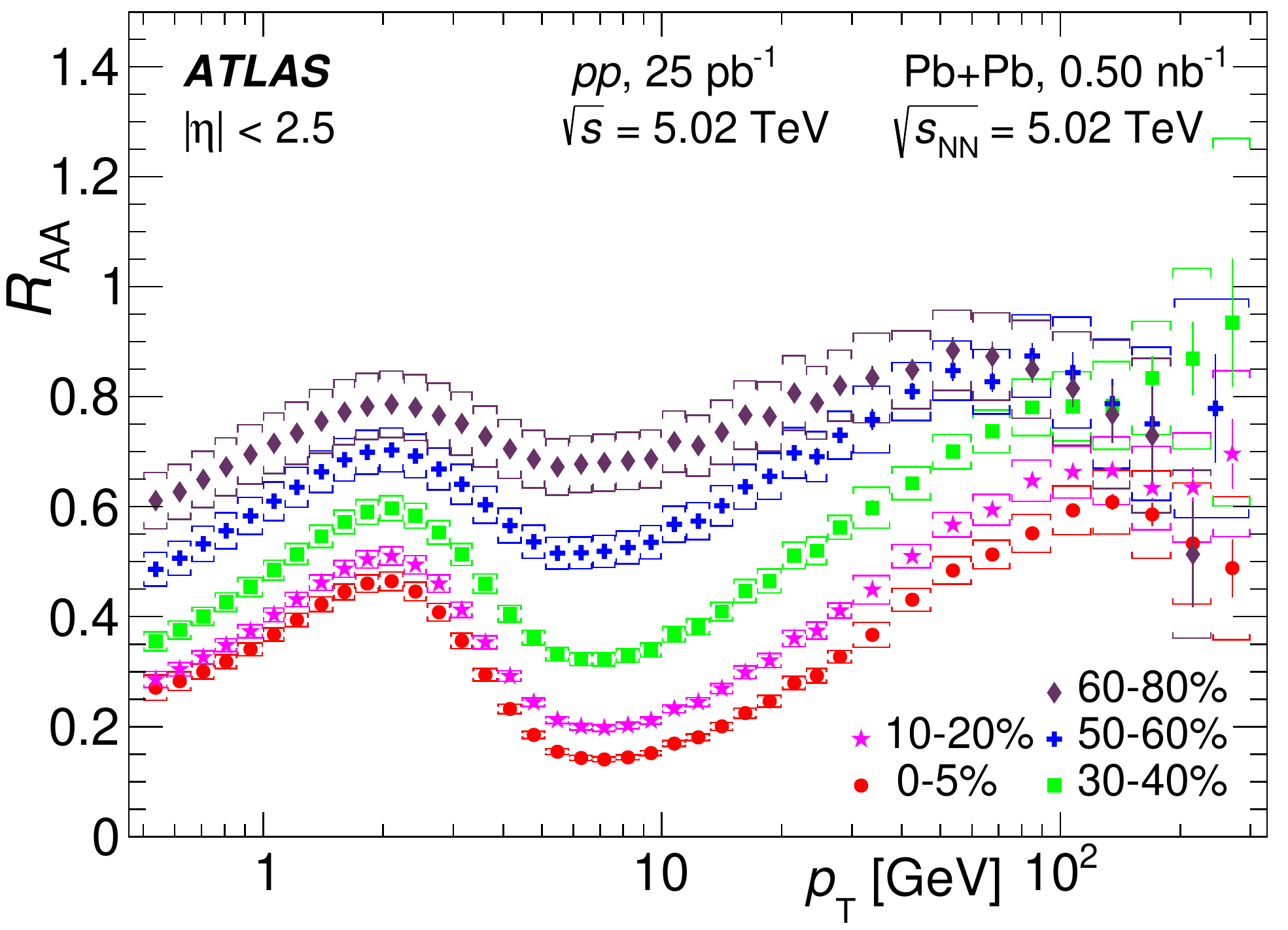}
\caption{Charged-hadron nuclear modification factor \Raa for \PbPb collisions in the pseudorapidity region $|\eta| < 2.5$. Full markers with different shapes show \Raa for five different centralities. Systematic uncertainties are shown with brackets; statistical uncertainties are shown with vertical lines. Any data point with more than 30\% statistical uncertainty is not shown.}
\label{fig:raa_pbpb}
\end{figure}

\begin{figure}[ptbh]
\centering
\includegraphics[width=0.60\textwidth]{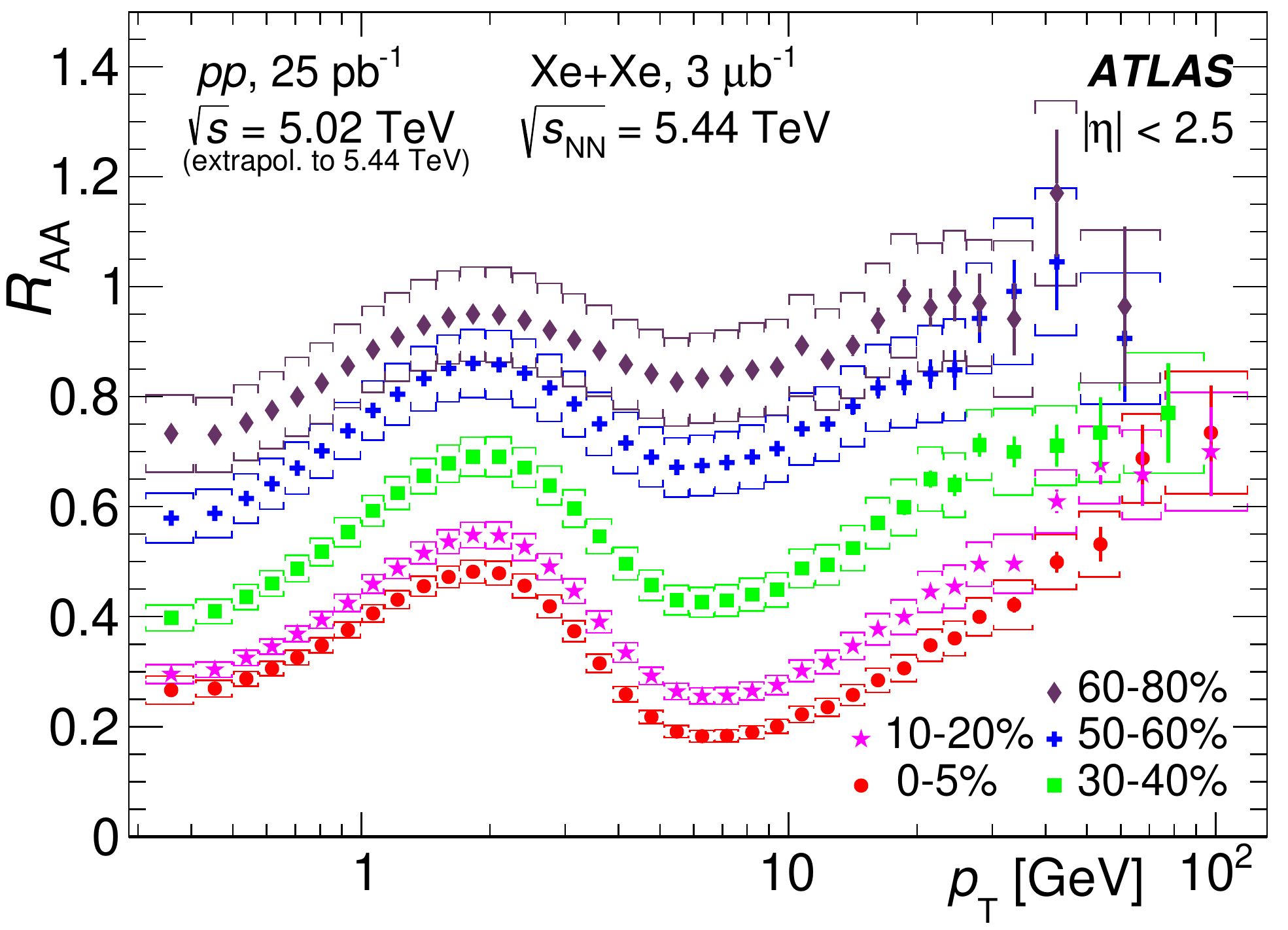}
\caption{Charged-hadron nuclear modification factor \Raa for \XeXe collisions in the pseudorapidity region $|\eta| < 2.5$. Full markers with different shapes show \Raa for five different centralities. Systematic uncertainties are shown with brackets; statistical uncertainties are shown with vertical lines. Any data point with more than 30\% statistical uncertainty is not shown.}
\label{fig:raa_xexe}
\end{figure}

Figures~\ref{fig:raa_pbpb} and \ref{fig:raa_xexe} show the nuclear modification factors $\Raa$ in \PbPb and \XeXe collisions, respectively, for the same centrality intervals and $|\eta|$ ranges as in the case of the charged-hadron spectra.
In both cases, the \Raa distributions have a characteristic shape seen previously~\cite{HION-2012-14,CMS-HIN-10-005,ALICE_Raa_PbPb_2p76, CMS-HIN-15-015,ALICE_Raa_PbPb, CMS-HIN-18-004,ALICE_Raa_XeXe}; the modification is stronger in central collisions.
First, \Raa increases to a local maximum at around 2\,\GeV, then it decreases to a local minimum at around 7\,\GeV, and then rises again.
In \PbPb collisions at $\pT\approx 100$\,\GeV, the slope of the distributions becomes smaller. At higher \pT, \Raa is consistent with a plateau with a magnitude between 0.6 for central collisions and 0.8 for peripheral collisions.
No similar conclusion can be drawn for the \XeXe data above $\pT\approx 100$\,\GeV due to their lower integrated luminosity and, therefore, higher statistical uncertainty.
 
\begin{figure}[p]
\centering
\includegraphics[width=0.60\textwidth]{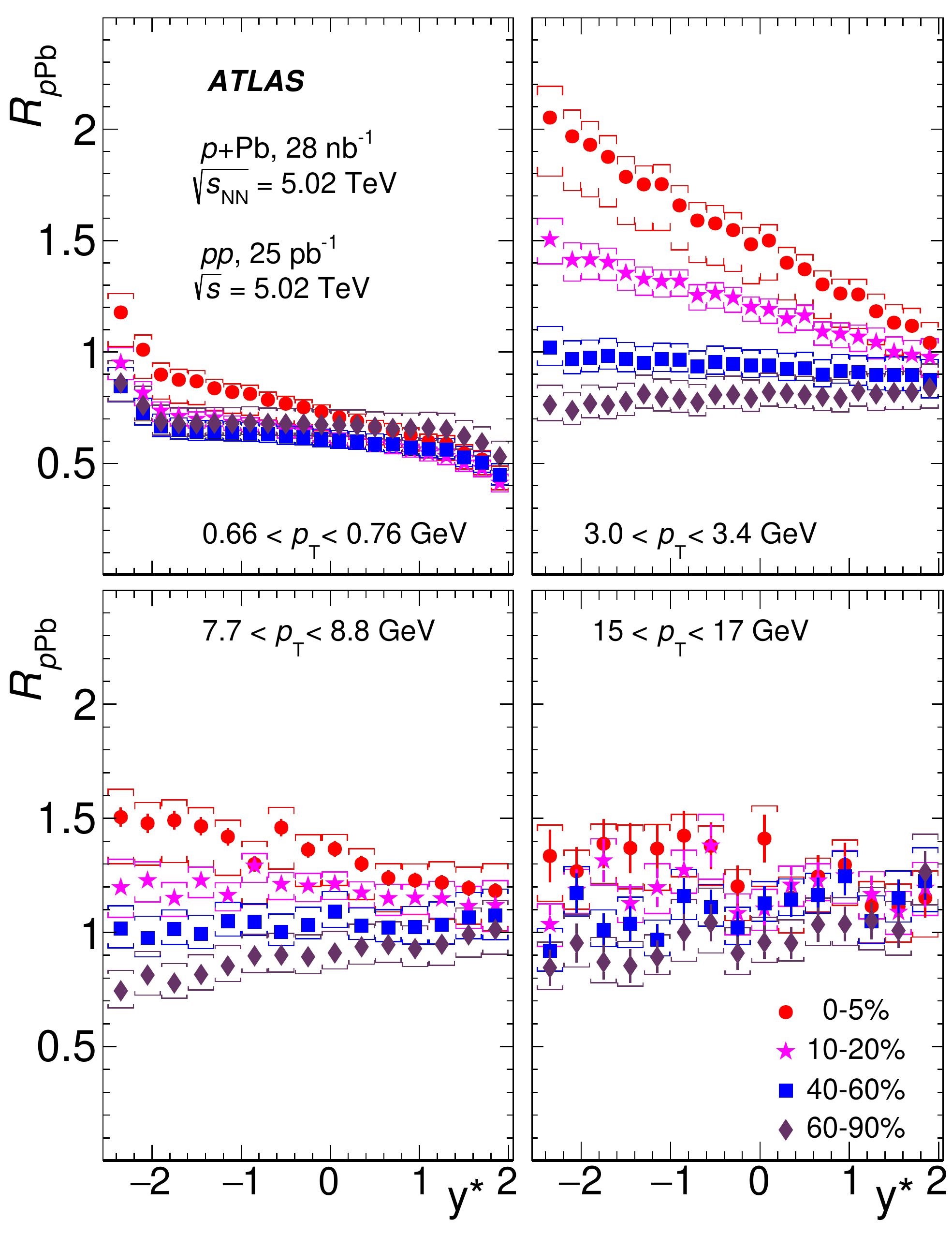}
\caption{Charged-hadron nuclear modification factor \RpPb for \pPb collisions in four \pT regions. Full markers with different shapes show \RpPb for four different centralities. Systematic uncertainties are shown with brackets; statistical uncertainties are shown with vertical lines.}
\label{fig:raaeta_ppb}
\end{figure}

Figure~\ref{fig:raaeta_ppb} shows \RpPb as a function of rapidity in four \pT ranges. They correspond to low $\pT$ where $\RpPb$ is below unity ($0.66<\pT<0.76$\,\GeV), the local maximum for central collisions ($3.0<\pT<3.4$\,\GeV), and two ranges where $\RpPb$ decreases towards unity ($7.7<\pT<8.8$\,\GeV and $15<\pT<17$\,\GeV).
The asymmetry is present in the central collisions even for lower \pT ranges, but it only reveals its full magnitude at about 3\,\GeV.
On the proton-going side, it is consistent with, or close to, unity; it shows an excess of up to a factor of ${\sim}2$ on the Pb-going side, with fragmentation of Pb nuclei.
With increasing centrality percentile and increasing $\pT$, the asymmetry diminishes.
This is in agreement with previous observations~\cite{HION-2012-14}.
 
\begin{figure}[p]
\centering
\includegraphics[width=0.60\textwidth]{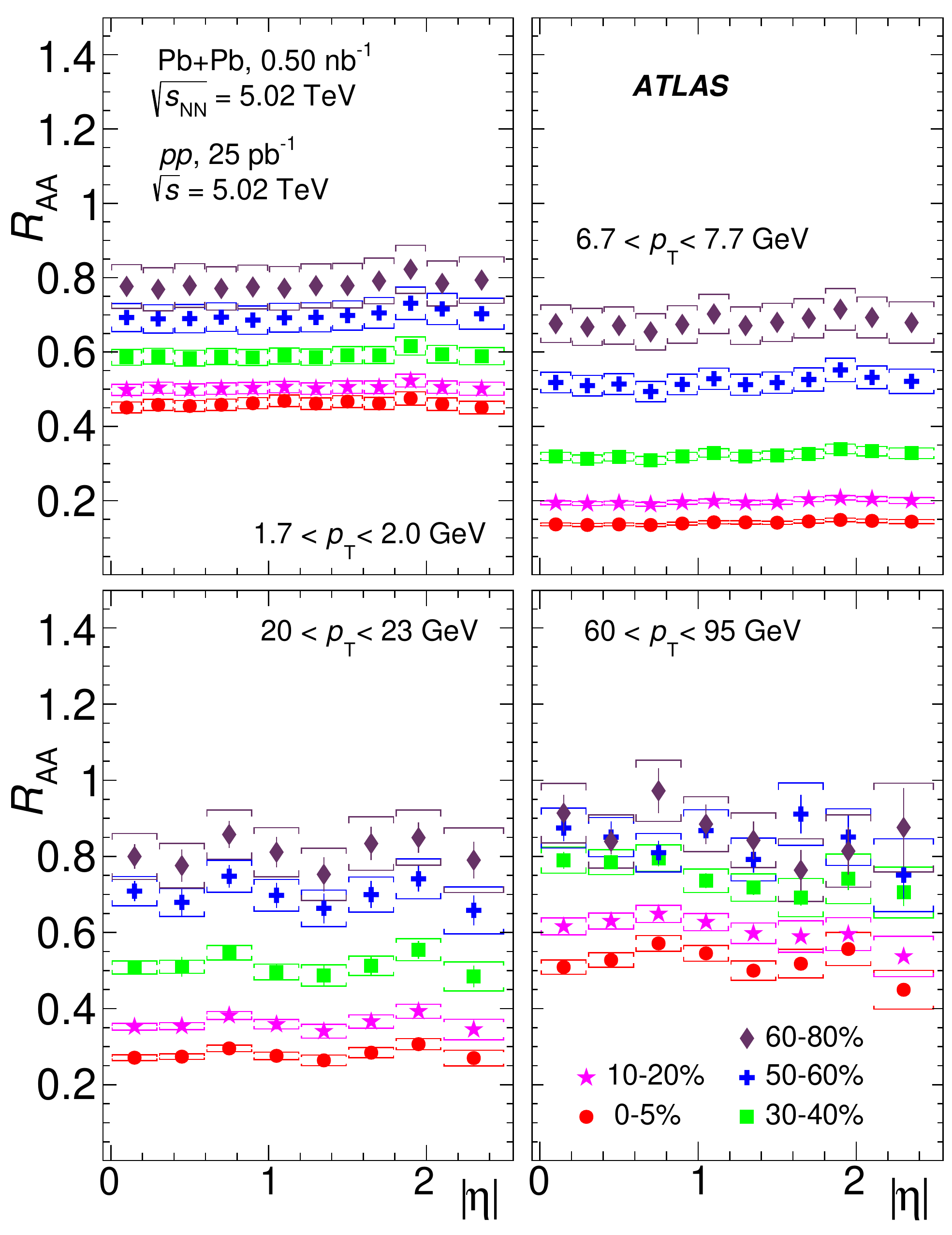}
\caption{Charged-hadron nuclear modification factor \Raa for \PbPb collisions in four $\pT$ regions. Full markers with different shapes show \Raa for five different centralities. Systematic uncertainties are shown with brackets; statistical uncertainties are shown with vertical lines.}
\label{fig:raaeta_pbpb}
\end{figure}
 
\begin{figure}[p]
\centering
\includegraphics[width=0.60\textwidth]{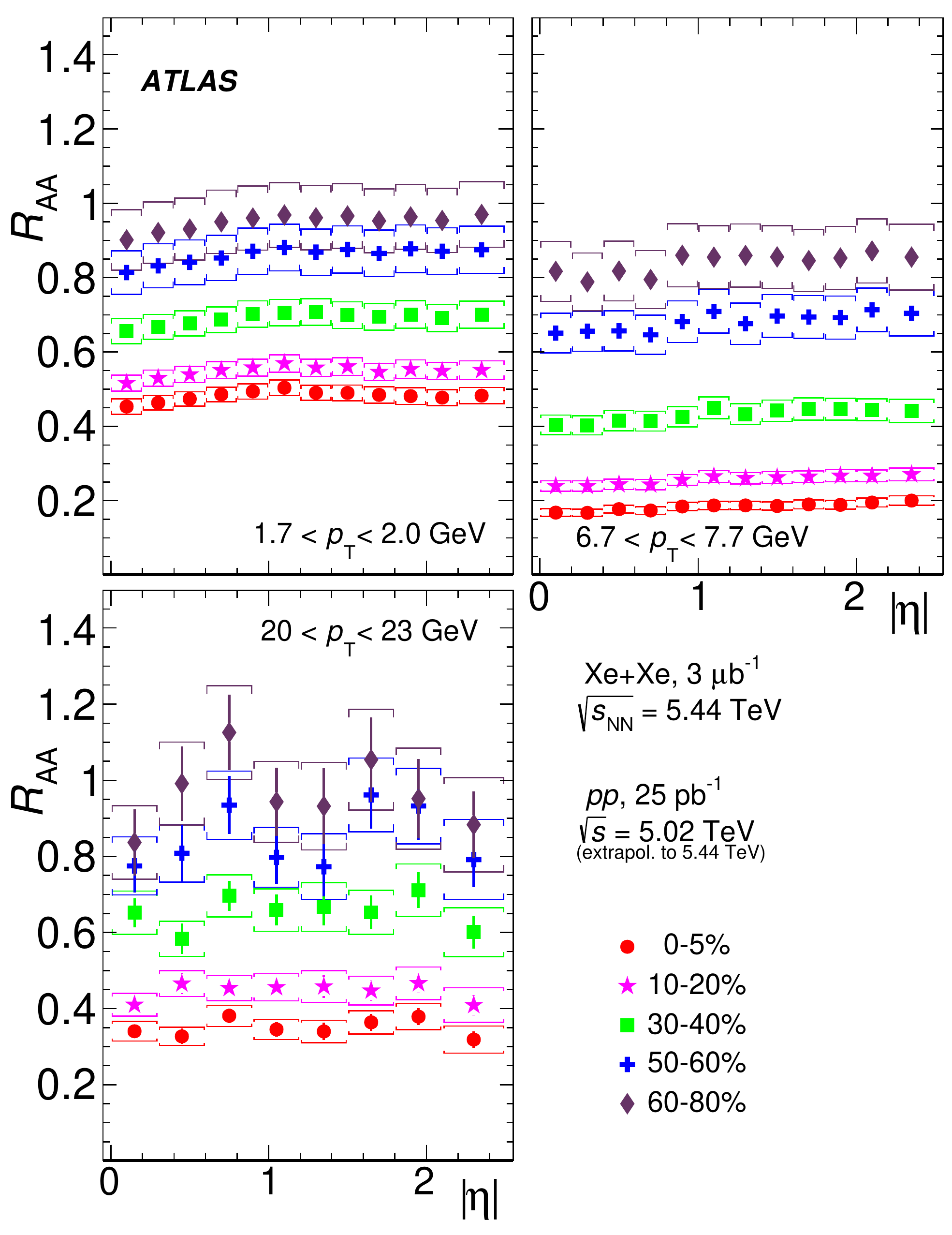}
\caption{Charged-hadron nuclear modification factor \Raa for \XeXe collisions in three \pT regions. Full markers with different shapes show \Raa for five different centralities. Systematic uncertainties are shown with brackets; statistical uncertainties are shown with vertical lines.}
\label{fig:raaeta_xexe}
\end{figure}
 
Figures~\ref{fig:raaeta_pbpb} and \ref{fig:raaeta_xexe} show \Raa in \PbPb and \XeXe collisions, respectively, as a function of pseudorapidity in several \pT ranges corresponding to the local maximum ($1.7<\pT<2.0$\,\GeV), the minimum ($6.7<\pT<7.7$\,\GeV), the region of increasing $\Raa$ ($20<\pT<23$\,\GeV), and the plateau ($60<\pT<95$\,\GeV; only for \PbPb).
The values of \Raa show no strong dependence on $\eta$.

\begin{figure}[ptbh]
\centering
\includegraphics[width=0.60\textwidth]{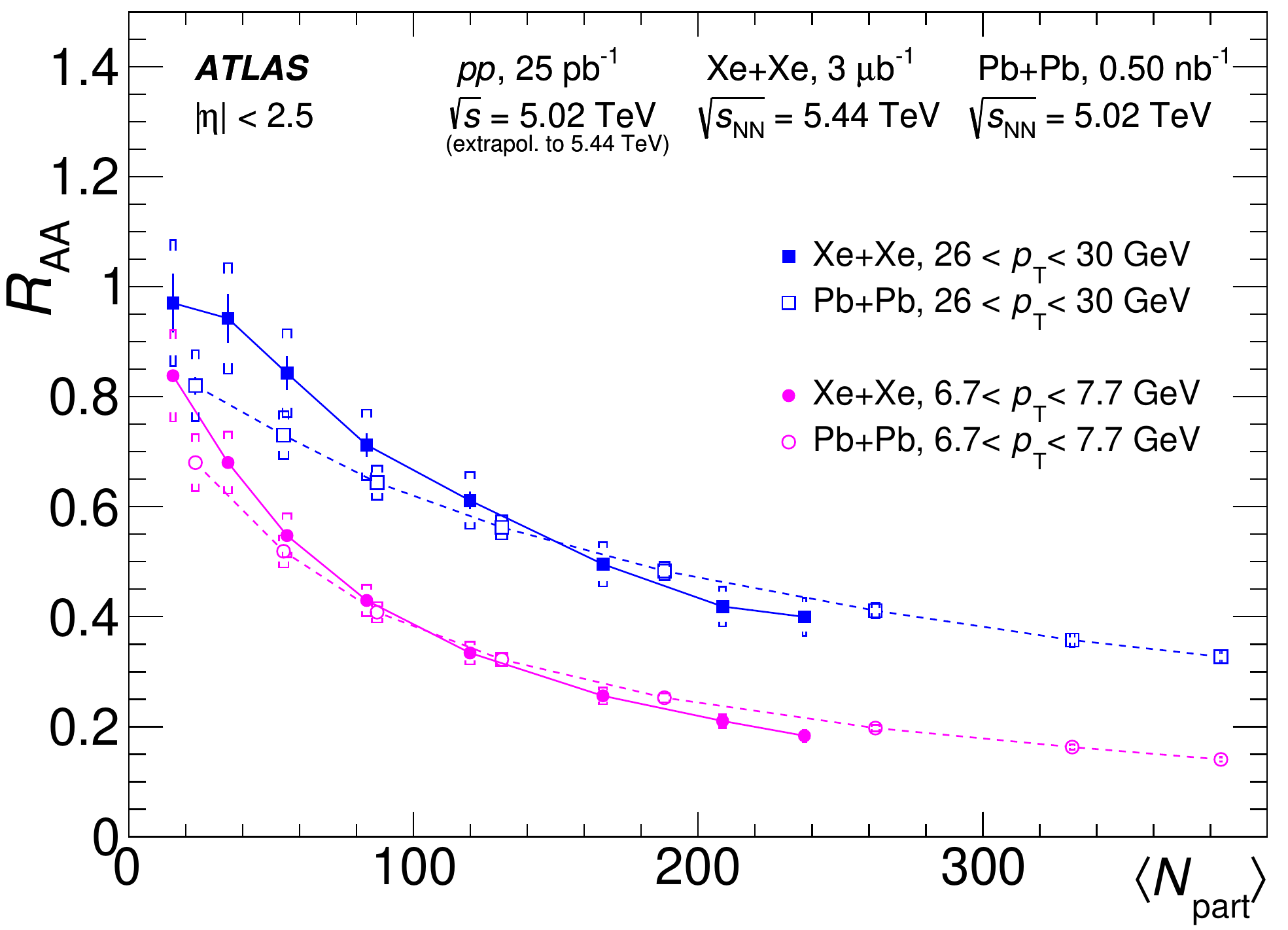}
\caption{Nuclear modification factor $\Raa$ as a function of $\avgNpart$ for selected ranges of $\pT$ measured in \XeXe collisions (full markers) and in \PbPb collisions (open markers). Systematic uncertainties are shown with brackets; statistical uncertainties are shown with vertical lines. The horizontal widths of the brackets represent the systematic uncertainties in $\avgNpart$. The lines are to help guide the eye. }
\label{fig:raa_npart}
\end{figure}

\begin{figure}[ptbh]
\centering
\includegraphics[width=0.60\textwidth]{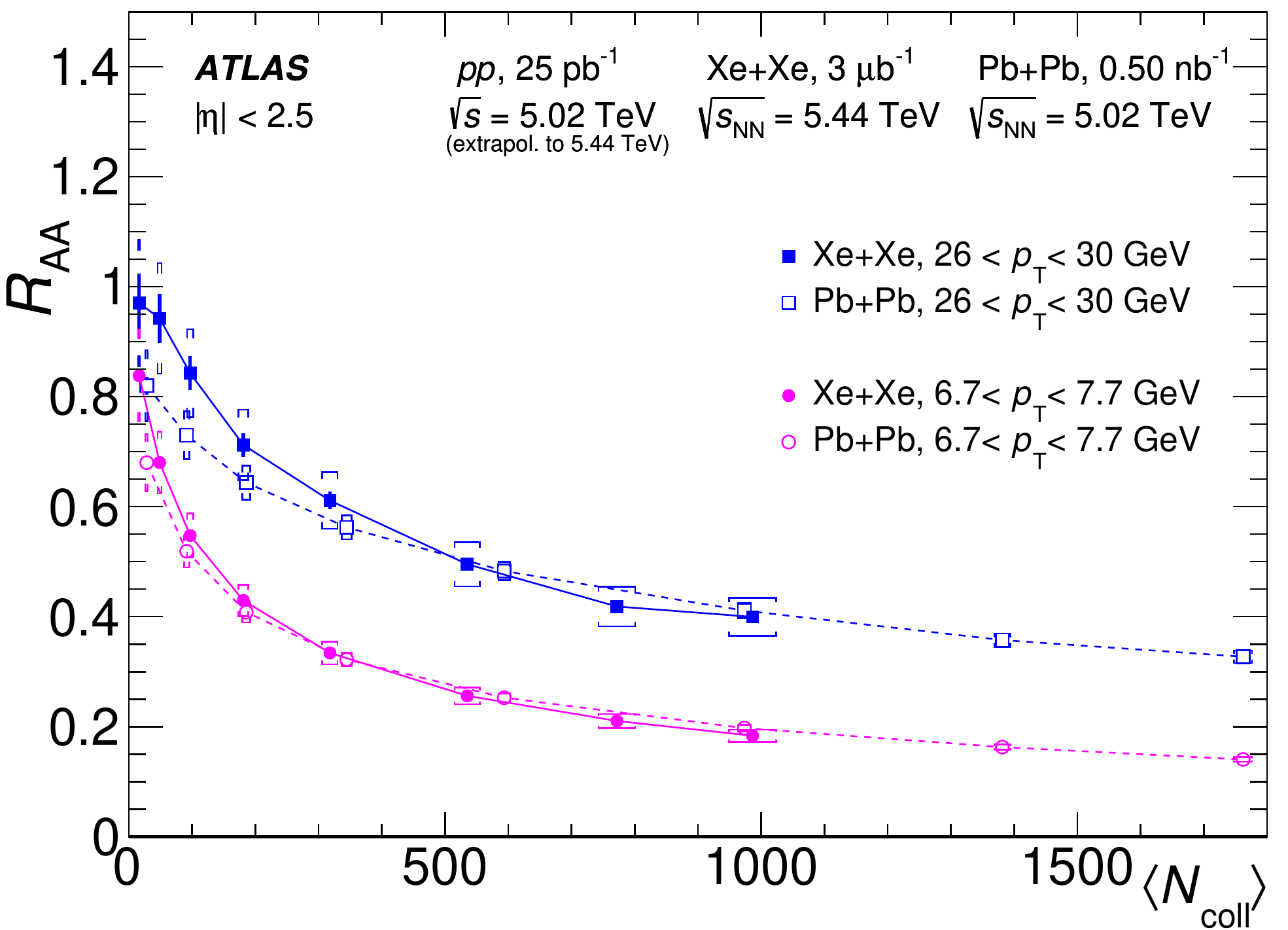}
\caption{Nuclear modification factor $\Raa$ as a function of $\avgNcoll$ for selected ranges of $\pT$ measured in \XeXe collisions (full markers) and in \PbPb collisions (open markers). Systematic uncertainties are shown with brackets; statistical uncertainties are shown with vertical lines. The horizontal widths of the brackets represent the systematic uncertainties in $\avgNcoll$. The lines are to help guide the eye. }
\label{fig:raa_ncoll}
\end{figure}

The nuclear modification factor $\Raa$ measured in \PbPb and \XeXe collisions as a function of the mean number of participants $\avgNpart$ and the mean number of collisions $\avgNcoll$ in two momentum intervals is shown in Figures~\ref{fig:raa_npart} and \ref{fig:raa_ncoll}, respectively.
These $\pT$ intervals correspond to the local minimum in the region $6.7<\pT<7.7$\,\GeV, and to the interval where $\Raa$ has an intermediate value, in the region $26<\pT<30$\,\GeV.
In both momentum intervals, $\Raa$ decreases in more central collisions; this decrease is stronger for the $\pT$ range corresponding to the local minimum.
Neither of the comparisons remove the dependency of $\Raa$ on the collision system.
 
\begin{figure}[ptbh]
\centering
\includegraphics[width=0.60\textwidth]{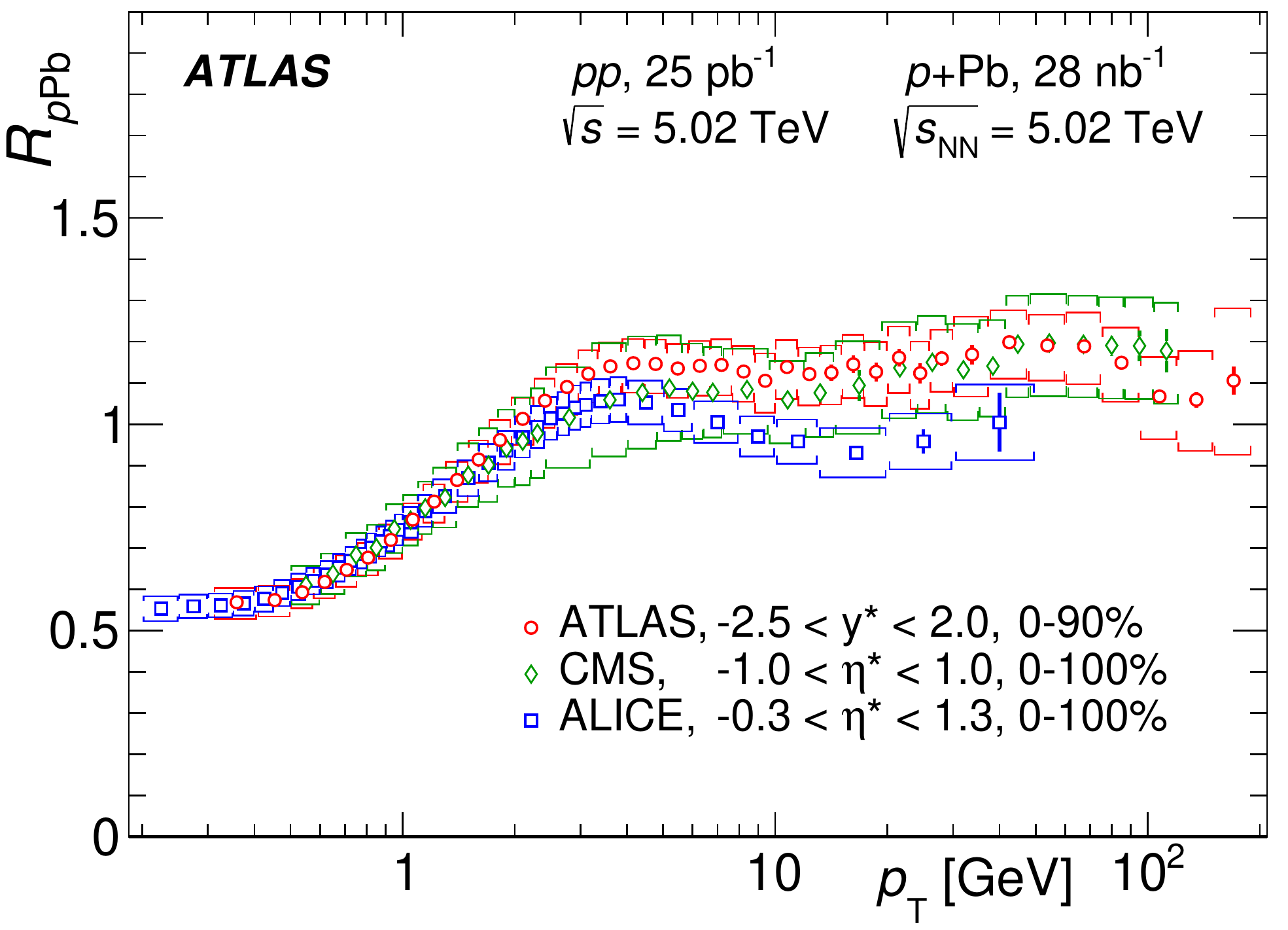}
\caption{Charged-hadron nuclear modification factor $\RpPb$ for \pPb collisions in the rapidity region $-2.5<\ystar < 2.0$ measured by ATLAS compared with results from CMS~\cite{CMS-HIN-15-015} and ALICE~\cite{ALICE_Raa_PbPb}. The centrality and (pseudo)rapidity intervals for CMS and ALICE data are different. Systematic uncertainties are shown with brackets; statistical uncertainties are shown with vertical lines.}
\label{fig:raa_compareto_ppb}
\end{figure}

\begin{figure}[ptbh]
\centering
\includegraphics[width=0.60\textwidth]{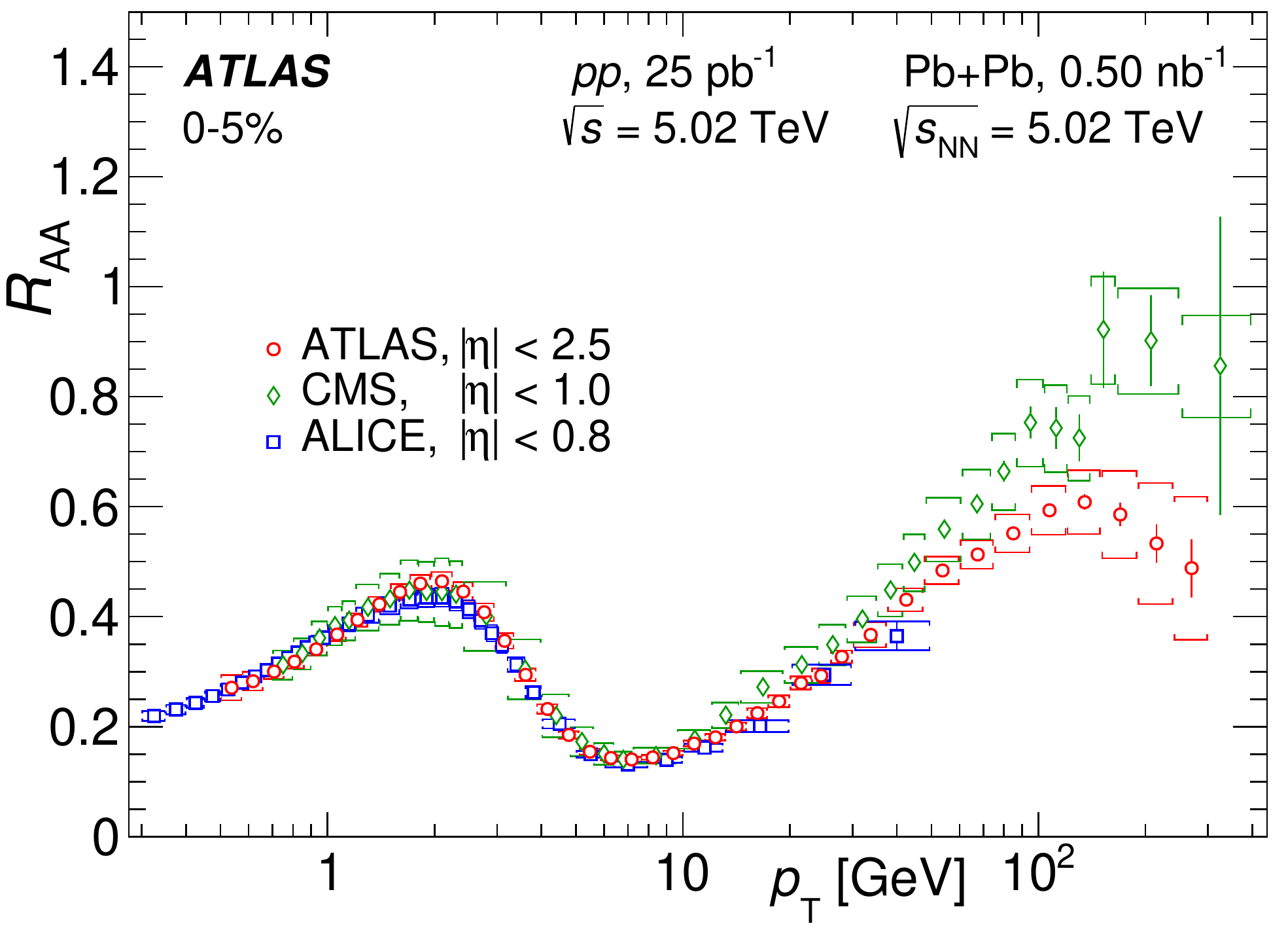}
\caption{Charged-hadron nuclear modification factor $\Raa$ for \PbPb collisions in the pseudorapidity region $|\eta| < 2.5$ measured by ATLAS compared with results from CMS~\cite{CMS-HIN-15-015} and ALICE~\cite{ALICE_Raa_PbPb}. Systematic uncertainties are shown with brackets; statistical uncertainties are shown with vertical lines.}
\label{fig:raa_compareto_pbpb}
\end{figure}

\begin{figure}[ptbh]
\centering
\includegraphics[width=0.60\textwidth]{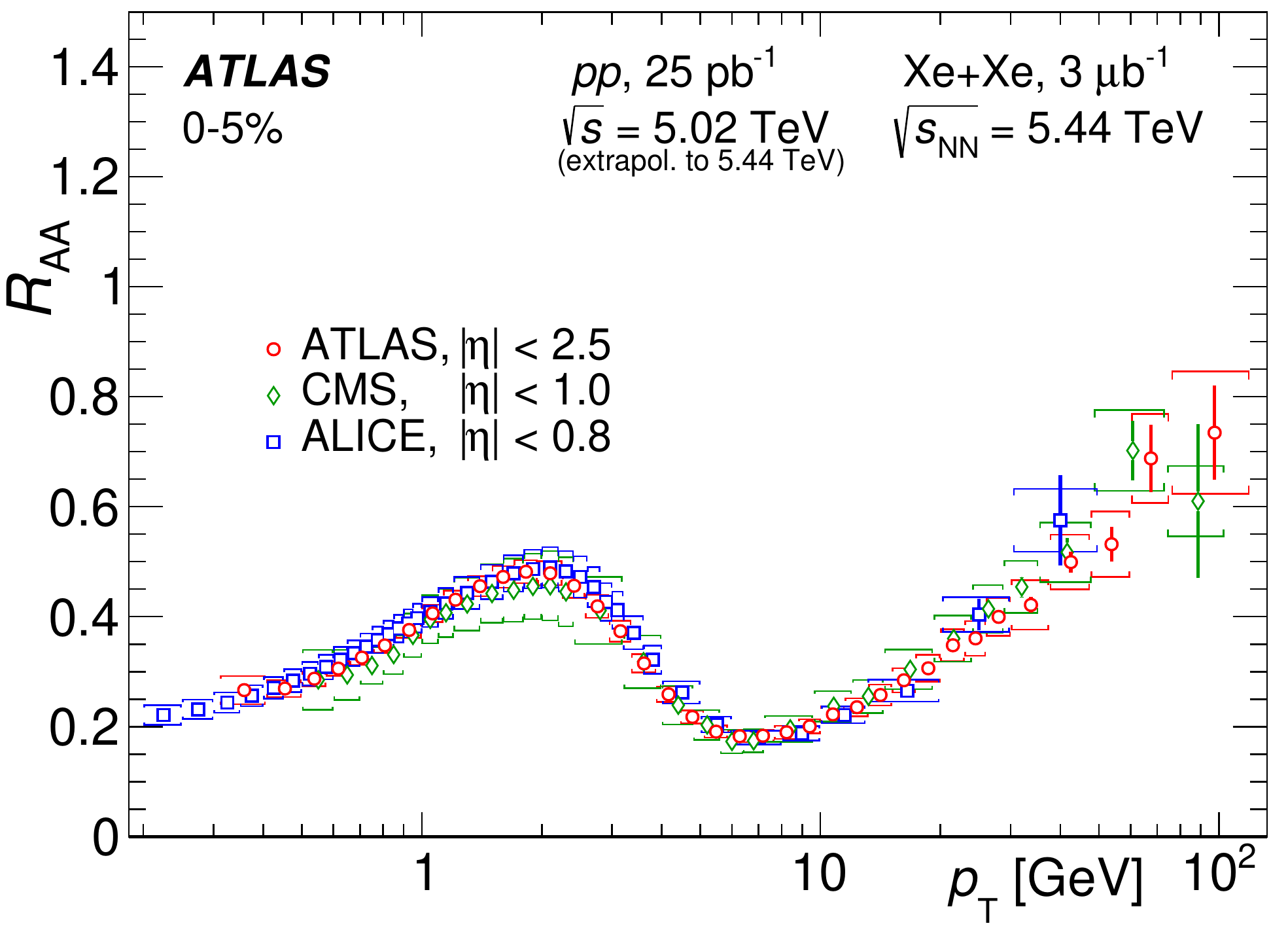}
\caption{Charged-hadron nuclear modification factor $\Raa$ for \XeXe collisions in the pseudorapidity region $|\eta| < 2.5$ measured by ATLAS compared with results from CMS~\cite{CMS-HIN-18-004} and ALICE~\cite{ALICE_Raa_XeXe}. Systematic uncertainties are shown with brackets; statistical uncertainties are shown with vertical lines.}
\label{fig:raa_compareto_xexe}
\end{figure}

Comparisons of measurements from this analysis with the measurements of CMS~\cite{CMS-HIN-15-015,CMS-HIN-18-004} and ALICE~\cite{ALICE_Raa_PbPb,ALICE_Raa_XeXe} for \pPb, \PbPb, and \XeXe collisions are shown in Figures~\ref{fig:raa_compareto_ppb}, \ref{fig:raa_compareto_pbpb}, and \ref{fig:raa_compareto_xexe}, respectively.
For \RpPb, the results of all three experiments generally agree with each other at lower $\pT$, within their uncertainties.
The ALICE result is lower than ATLAS and CMS results at $\pT\gtrsim 10$\,\GeV.
Nevertheless, the experiments do not report results in the same ranges of $\ystar$.
Considering the asymmetry of \RpPb in $\ystar$, as seen in Figure~\ref{fig:raaeta_ppb}, ALICE results would be expected to be somewhat lower than CMS results, and CMS results somewhat lower than ATLAS results\@.
The differences should diminish at $\pT\approx 10$\,\GeV.
The $\Raa$ measured in \PbPb collisions by CMS at $\pT\gtrsim 60$\,\GeV is above the one measured by ATLAS\@.
The discrepancy becomes larger at higher $\pT$.
The ALICE results are consistent with ATLAS results within the range of the ALICE measurement.
For \XeXe collisions, all three experiments are consistent.
The systematic uncertainty in $\avgTaa$ is often the largest among all systematic uncertainties.
Although estimated independently by each collaboration, it is always fully correlated across the whole $\pT$ region.
 
\afterpage{\clearpage}
 
\begin{figure}[ptbh] 
\centering
\includegraphics[width=0.49\textwidth]{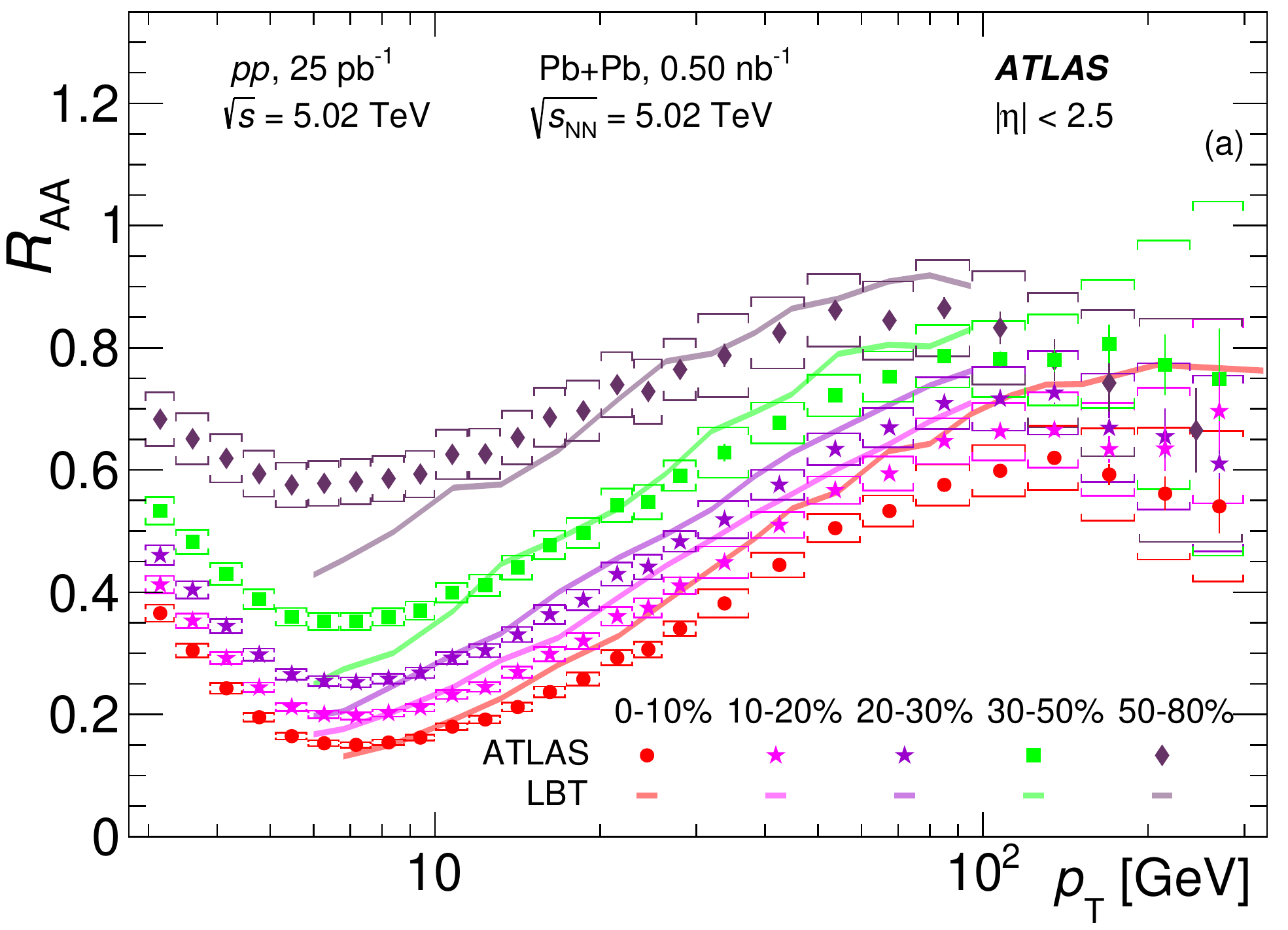} \includegraphics[width=0.49\textwidth]{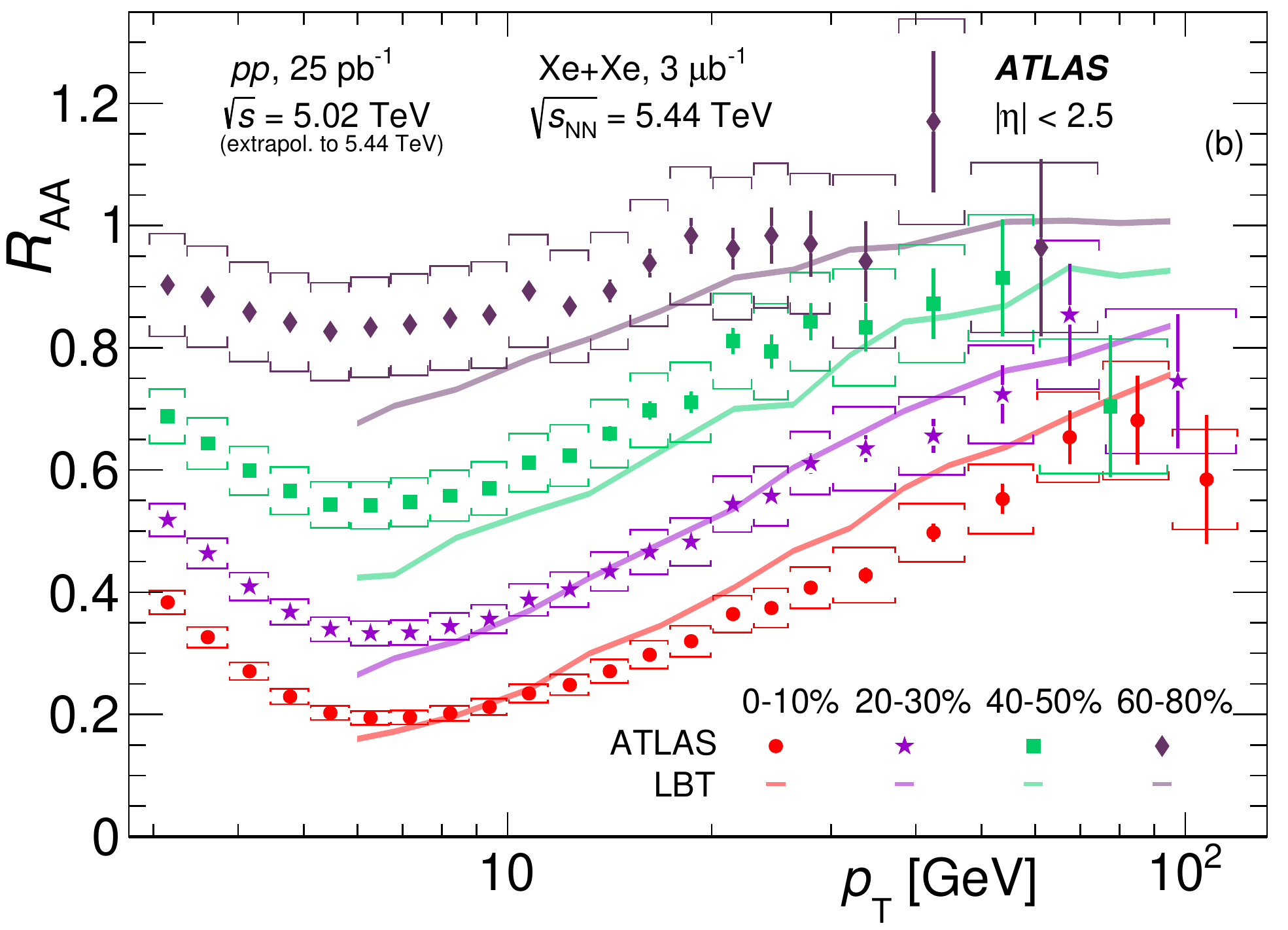}
\caption{Charged-hadron nuclear modification factor $\Raa$ for (a) \PbPb and (b) \XeXe collisions in the pseudorapidity region $|\eta| < 2.5$ measured by ATLAS compared with predictions of the LBT model~\cite{1_LBT_a,1_LBT_b}. The width of the lines representing the model is chosen only for visibility; the model was not provided with any uncertainty.}
\label{fig:raa_compareto_1}
\end{figure}

The measured distributions are also compared with several theoretical calculations. The models cover different dynamical ranges; the comparison is performed in the ranges provided by the authors of the models.
 
The Linear Boltzmann Transport (LBT) model~\cite{1_LBT_a,1_LBT_b} uses linear Boltzmann equations to describe parton transport in the QGP while considering 2$\rightarrow$2 parton scattering. The rates of scattering in a medium follow relativistic hydrodynamic equations.
A comparison of charged-hadron nuclear modification factors $\Raa$ measured by ATLAS in \PbPb and \XeXe collisions with the LBT model predictions is shown in Figure~\ref{fig:raa_compareto_1}.
The model correctly describes the trends measured in the data, but has a somewhat shorter variational range, it overestimates the \Raa in the most central collisions in both the \PbPb and \XeXe systems, and underestimates it in the most peripheral \XeXe collisions.
 
\begin{figure}[ptbh]
\centering
\includegraphics[width=0.49\textwidth]{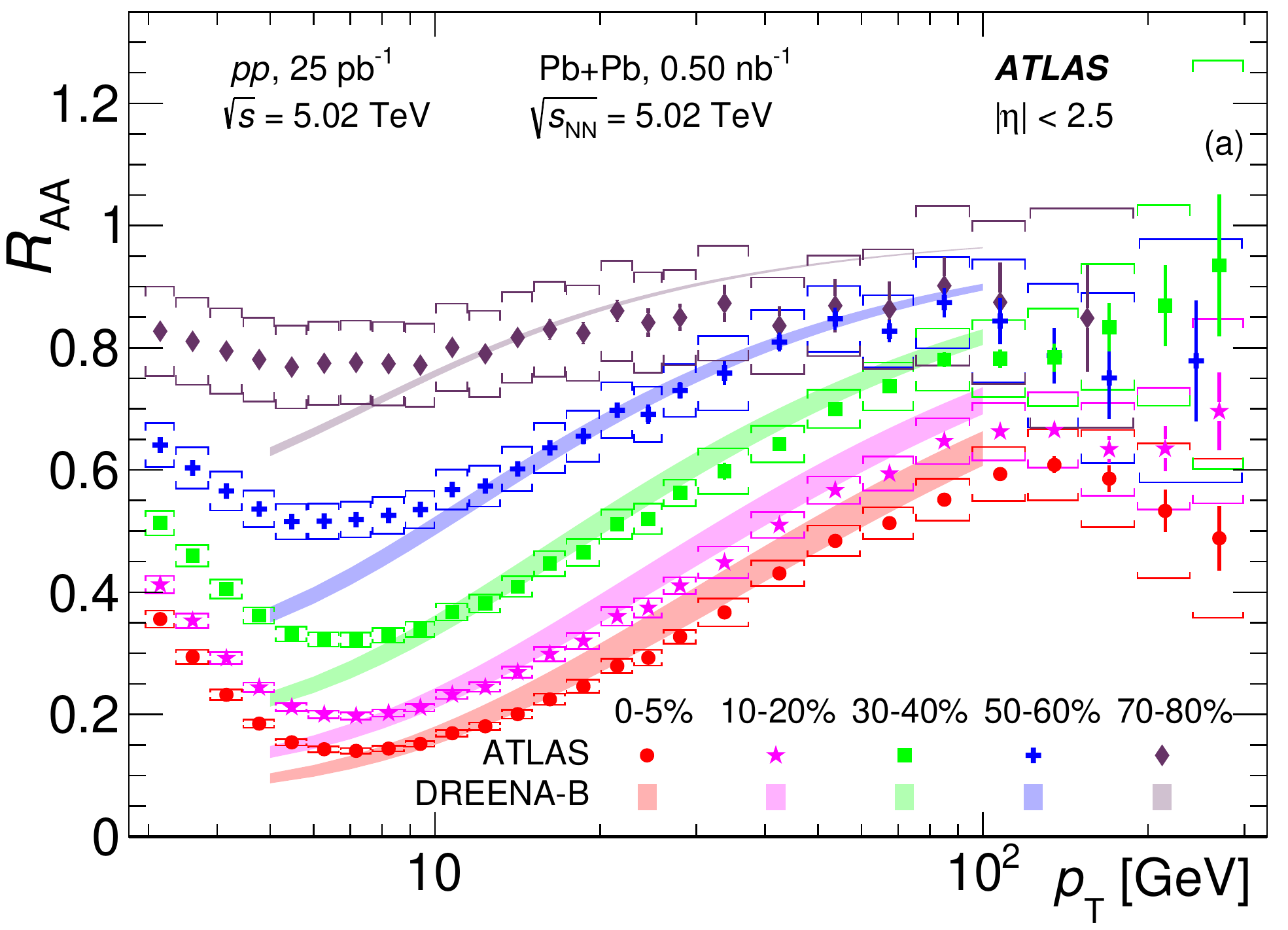} \includegraphics[width=0.49\textwidth]{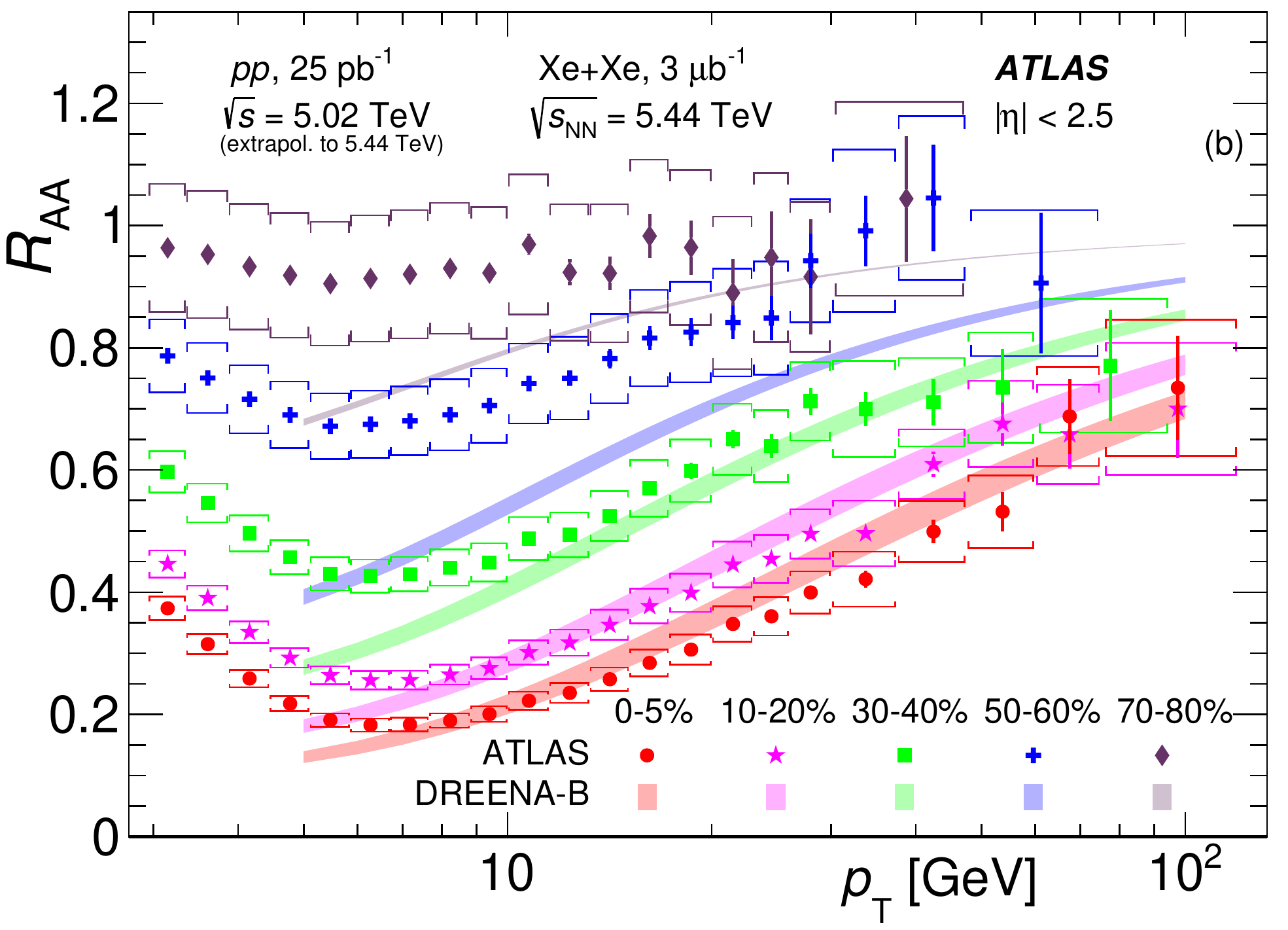}
\caption{Charged-hadron nuclear modification factor $\Raa$ for (a) \PbPb and (b) \XeXe collisions in the pseudorapidity region $|\eta| < 2.5$ measured by ATLAS compared with predictions of the DREENA-B model~\cite{2_DREENA,2_DREENA_b}. The width of the model band represents the theoretical systematic uncertainty.}
\label{fig:raa_compareto_2}
\end{figure}

The DREENA-B framework~\cite{2_DREENA,2_DREENA_b} is based on Bjorken 1+1D medium evolution~\cite{Bjorken11D} that introduces different evolution scenarios before thermalization of the QGP and the same evolution after thermalization. It is also capable of describing charged-hadron $\Raa$ and elliptic flow $v_2$~\cite{Heinz:2013th} at the same time.
The model is compared with data in Figure~\ref{fig:raa_compareto_2}.
The model correctly describes \PbPb data for $\pT\gtrsim 8$\,\GeV, but predicts stronger suppression in peripheral \XeXe collisions than is measured in the data.
 
\begin{figure}[ptbh]
\centering
\includegraphics[width=0.49\textwidth]{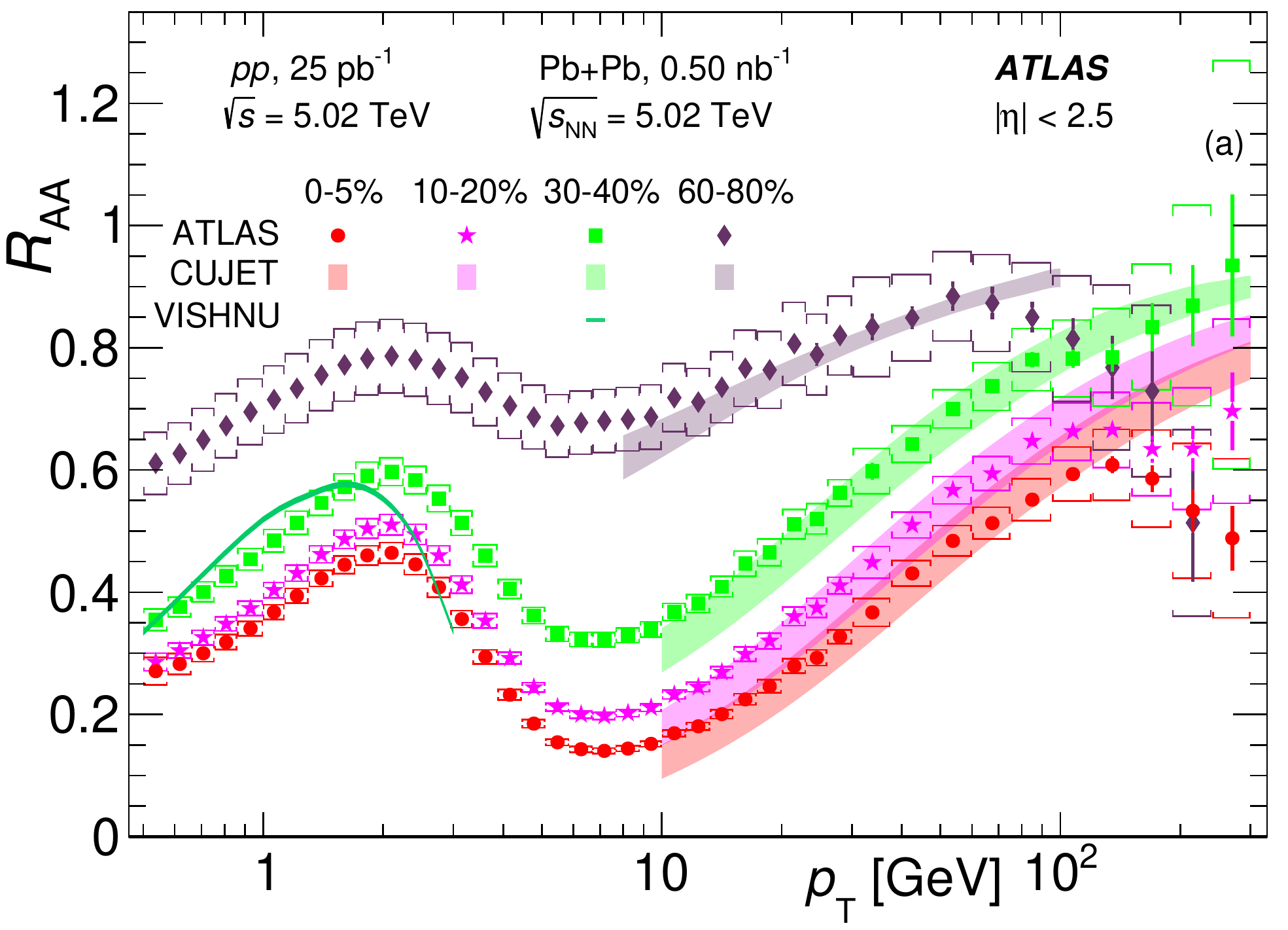} \includegraphics[width=0.49\textwidth]{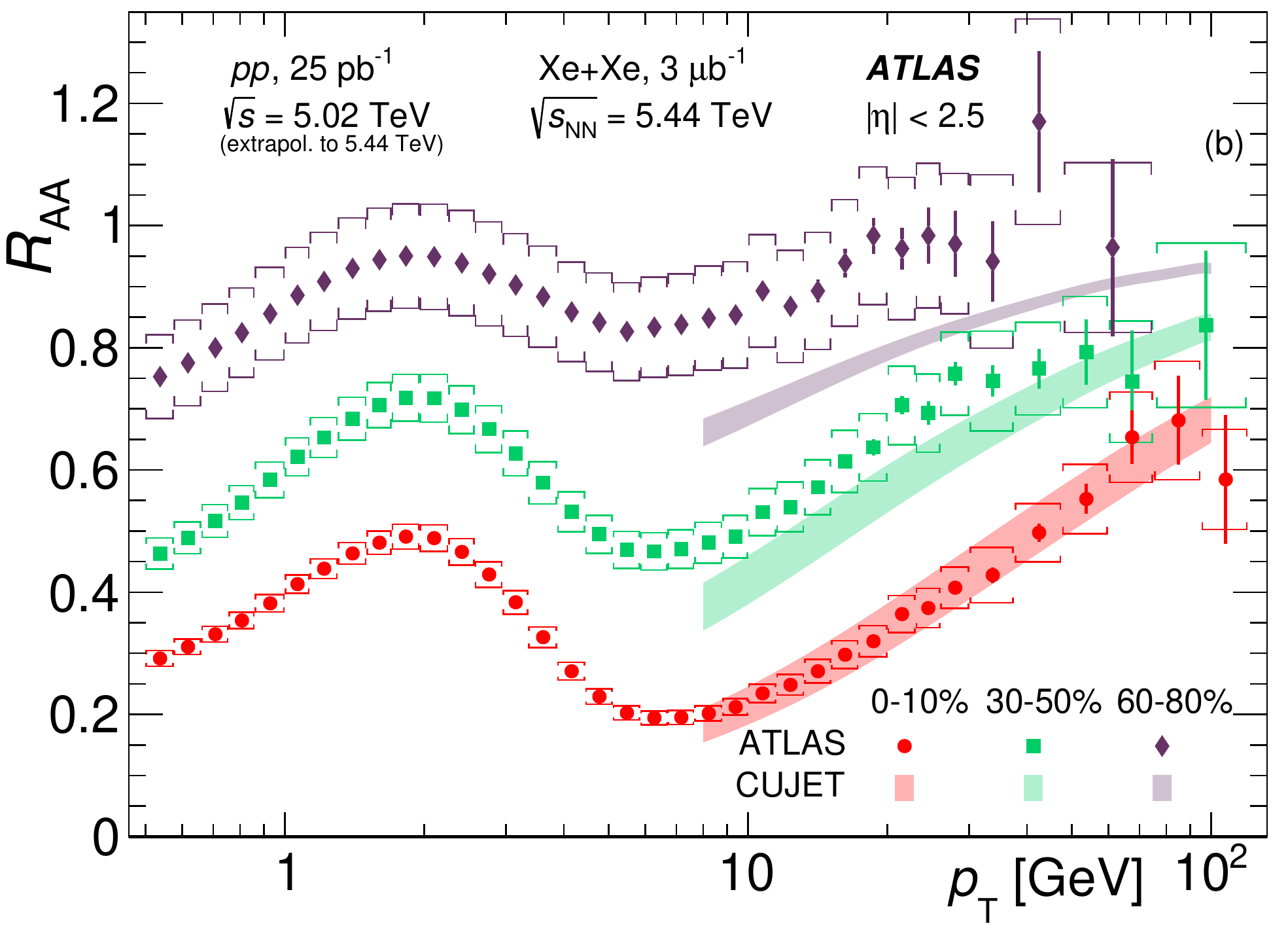}
\caption{Charged-hadron nuclear modification factor $\Raa$ for (a) \PbPb and (b) \XeXe collisions in the pseudorapidity region $|\eta| < 2.5$ measured by ATLAS compared with predictions of the CUJET~\cite{5_CUJET_a,5_CUJET_b} and VISHNU~\cite{20_VISHNU} models. The width of the CUJET band represents the theoretical systematic uncertainty while the width of the VISHNU line is chosen only for visibility; the VISHNU model was not provided with any uncertainty.}
\label{fig:raa_compareto_3}
\end{figure}

The CIBJET framework~\cite{5_CUJET_a,5_CUJET_b} combines the bulk evolution calculated by the viscous hydrodynamic simulation VISHNU~\cite{20_VISHNU} with the high-$\pT$ jet energy loss calculated in the CUJET model, which is a non-perturbative model integrating the suppression of chromo-electric degrees of freedom and emergence of the chromo-magnetic degrees of freedom.
A comparison with the measured ATLAS data is shown in Figure~\ref{fig:raa_compareto_3}.
The CUJET model successfully describes $\Raa$ in the entire \PbPb range of centrality intervals as well as in central \XeXe collisions, but it underestimates $\Raa$ in peripheral \XeXe collisions.
The VISHNU model correctly reproduces the $\Raa$ distribution shape, but calculates values lower than the data at $\pT\gtrsim2$\,\GeV.
 
\begin{figure}[ptbh]
\centering
\includegraphics[width=0.49\textwidth]{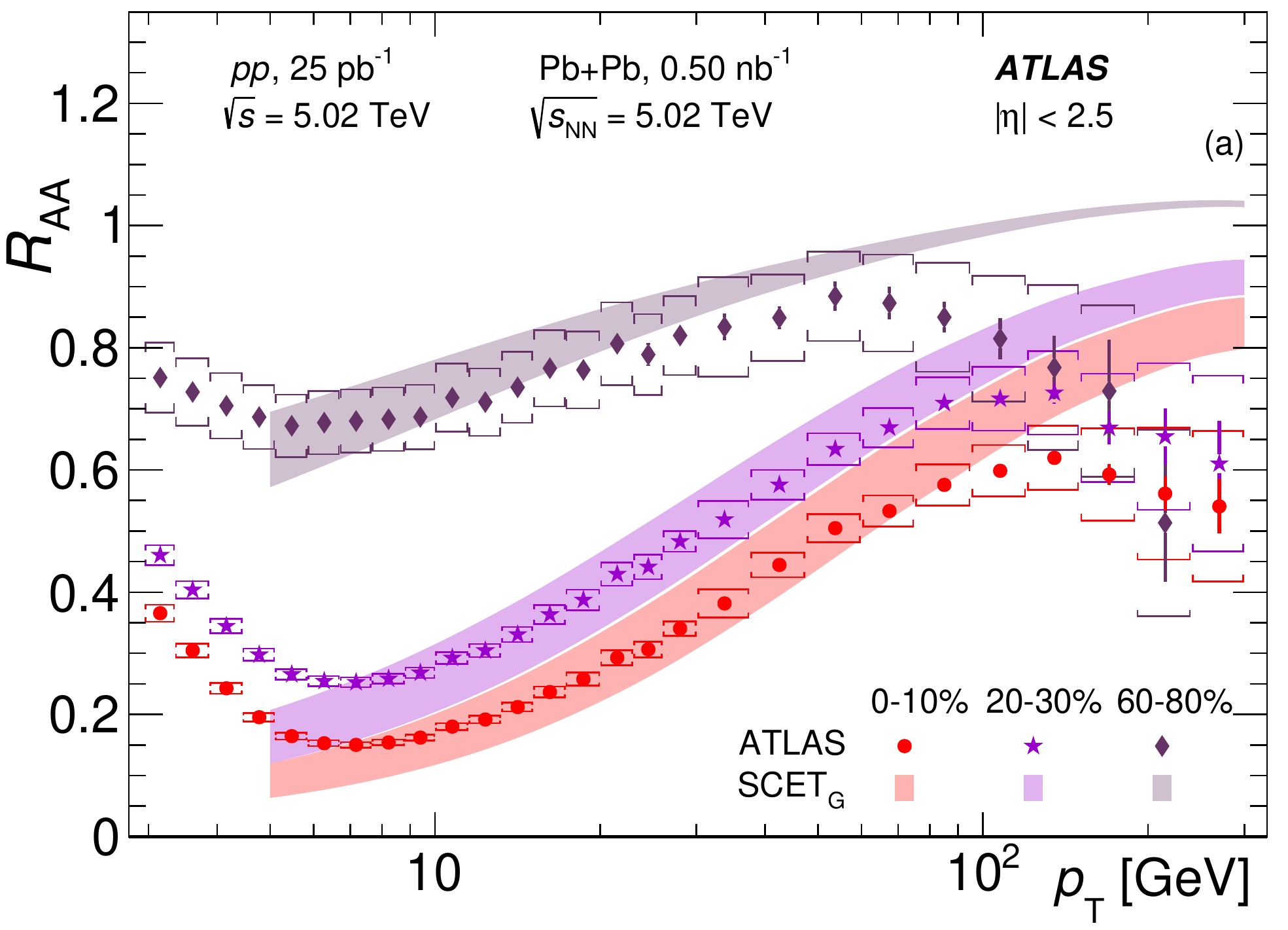} \includegraphics[width=0.49\textwidth]{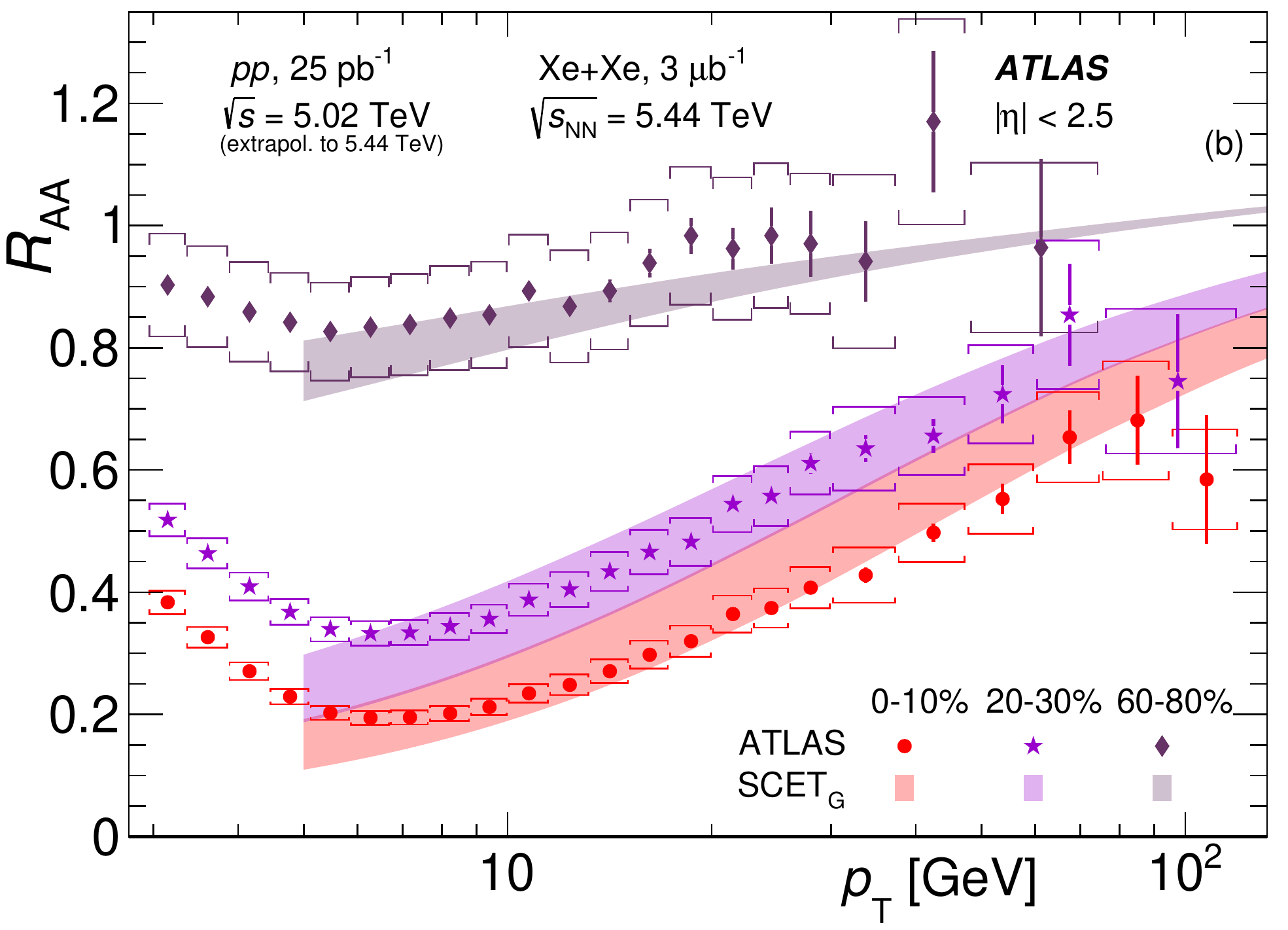}
\caption{Charged-hadron nuclear modification factor $\Raa$ for (a) \PbPb and (b) \XeXe collisions in the pseudorapidity region $|\eta| < 2.5$ measured by ATLAS compared with predictions of the SCET$_\mathrm{G}$ model~\cite{5_SCET_a,5_SCET_b,5_SCET_c}. The width of the model band represents the theoretical systematic uncertainty.}
\label{fig:raa_compareto_5}
\end{figure}

The SCET$_\mathrm{G}$ model~\cite{5_SCET_a,5_SCET_b,5_SCET_c} uses soft-collinear effective theory, extends it to also include jet propagation in the QGP, and effectively describes in-medium parton shower formation.
A comparison of the model predictions with data is shown in Figure~\ref{fig:raa_compareto_5}.
The model describes the $\Raa$ values well in both collision systems and in almost the entire range of $\pT$. At very high $\pT$, the trend calculated by the model in the \PbPb system is different from the data.
 
\begin{figure}[ptbh]
\centering
\includegraphics[width=0.49\textwidth]{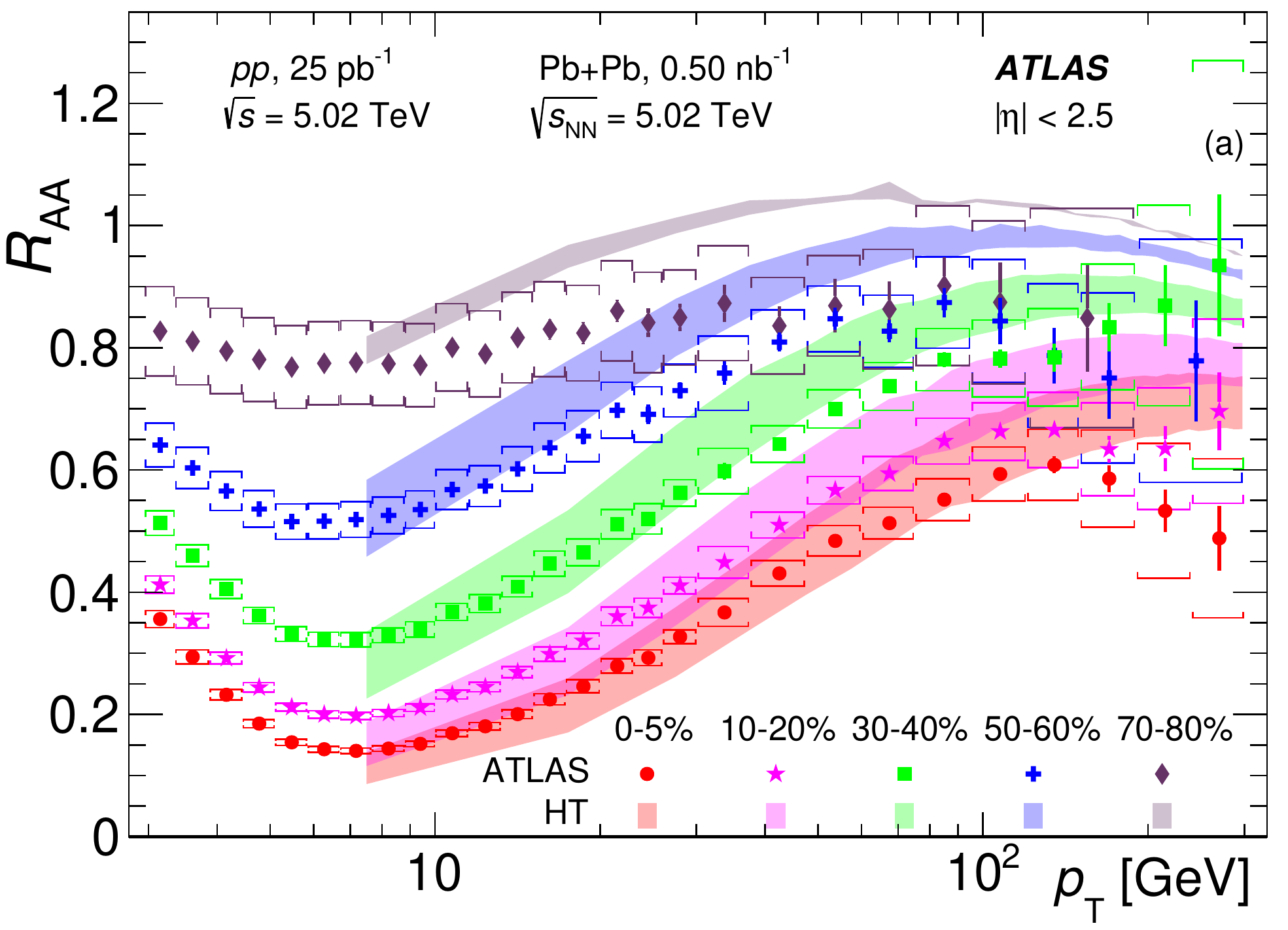} \includegraphics[width=0.49\textwidth]{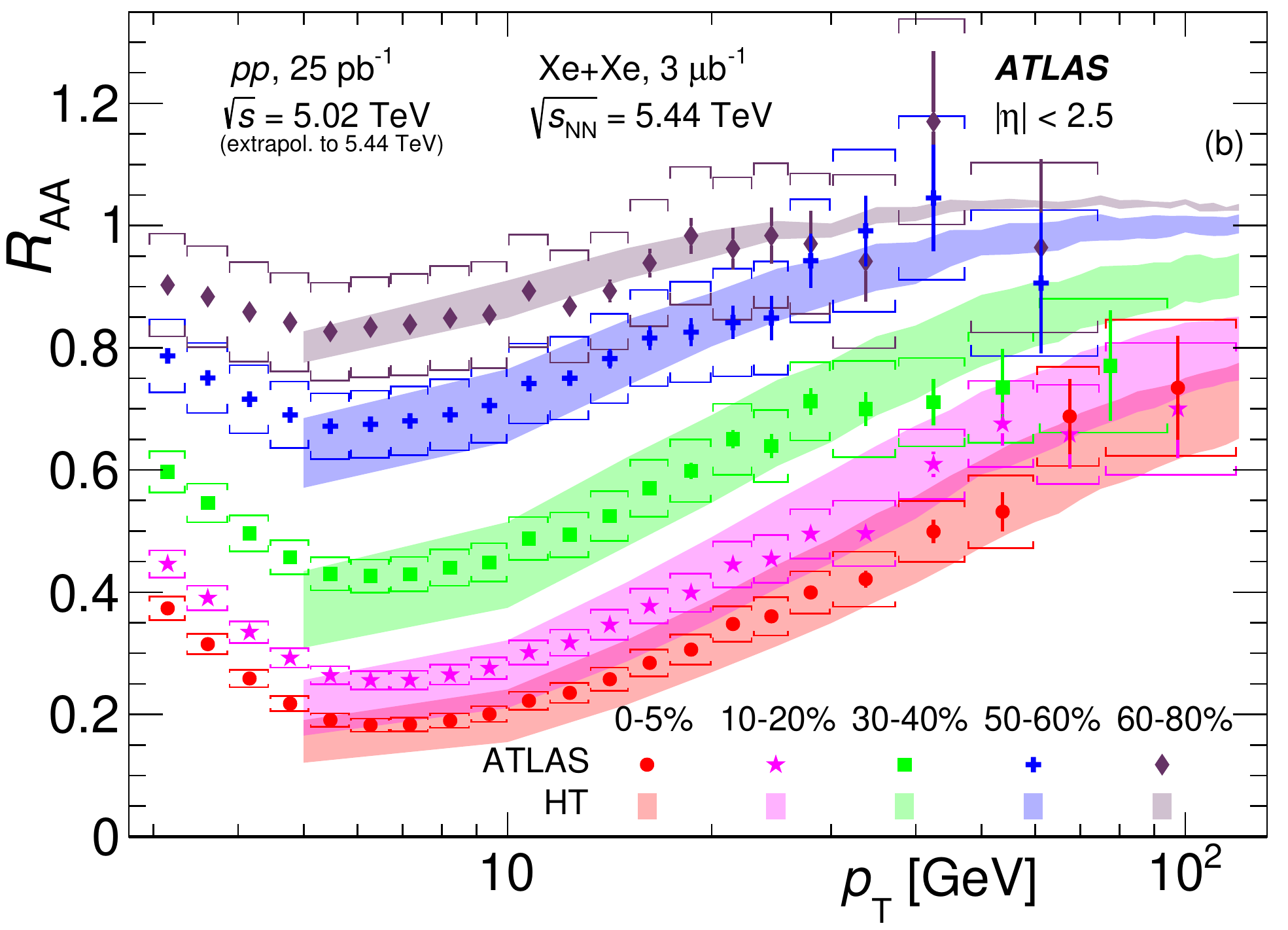}
\caption{Charged-hadron nuclear modification factor $\Raa$ for (a) \PbPb and (b) \XeXe collisions in the pseudorapidity region $|\eta| < 2.5$ measured by ATLAS compared with predictions of the HT model~\cite{19_HT_a,19_HT_b}. The width of the model band represents the theoretical systematic uncertainty.}
\label{fig:raa_compareto_19}
\end{figure}

The Higher Twist (HT) results~\cite{19_HT_a,19_HT_b} obtained using the next-to-leading-order (NLO) parton model with higher-twist energy loss are shown in Figure~\ref{fig:raa_compareto_19}.
The HT model captures the shape of the \Raa distribution well in both systems across almost the entire \pT range of the prediction. In peripheral \PbPb collisions it predicts less suppression than is measured in data.
 
\begin{figure}[ptbh]
\centering
\includegraphics[width=0.49\textwidth]{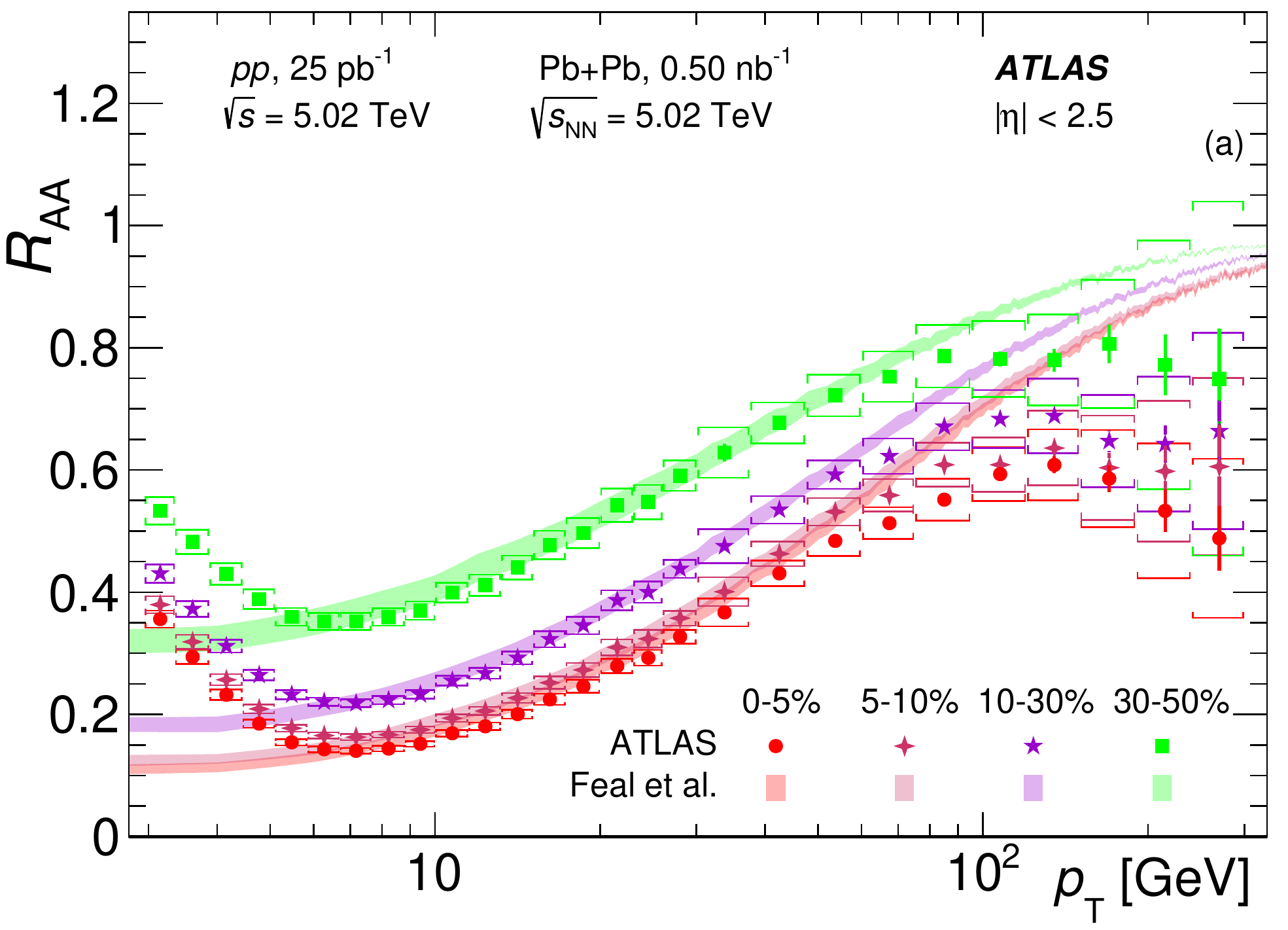} \includegraphics[width=0.49\textwidth]{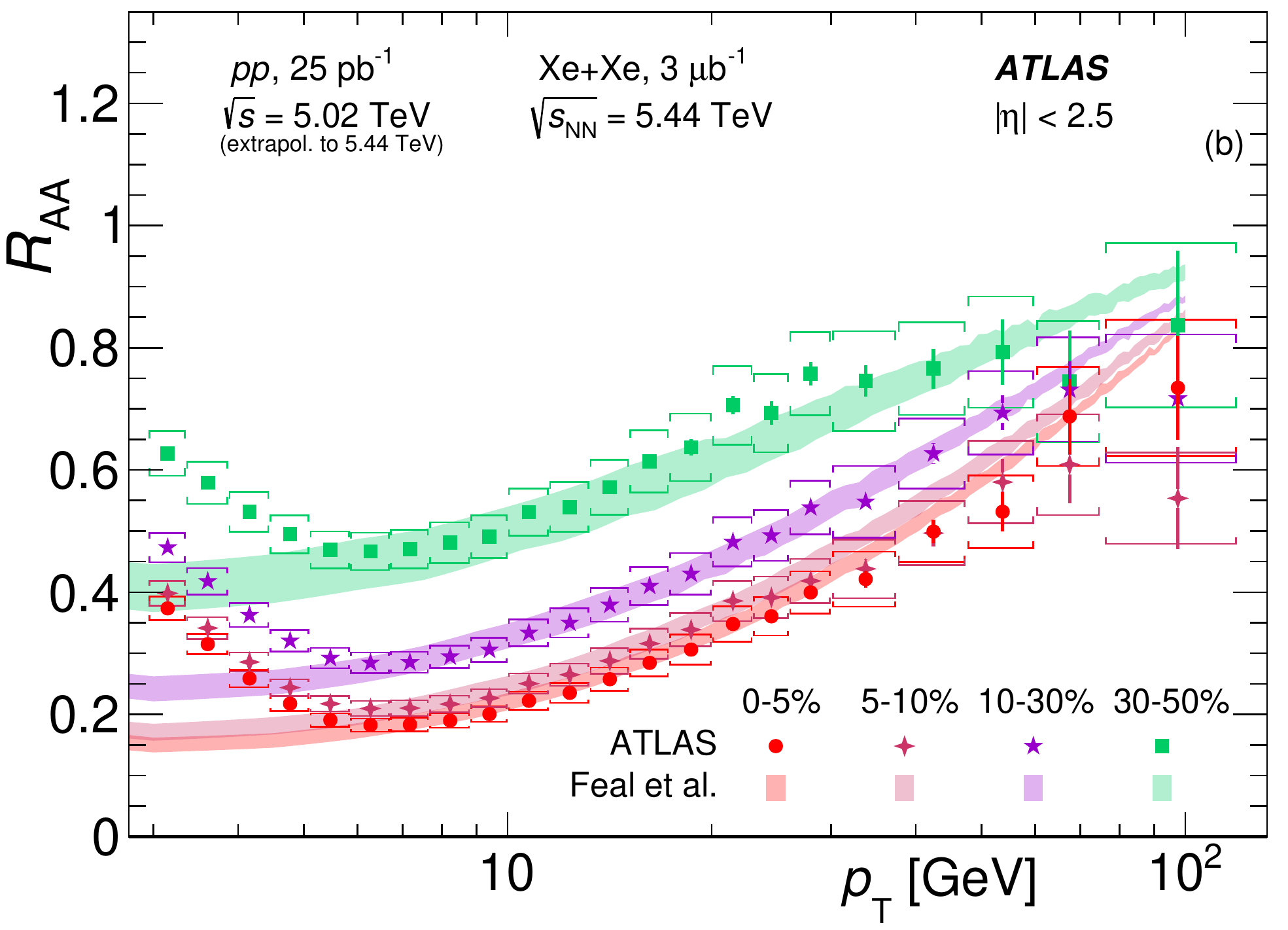}
\caption{Charged-hadron nuclear modification factor $\Raa$ for (a) \PbPb and (b) \XeXe collisions in the pseudorapidity region $|\eta| < 2.5$ measured by ATLAS compared with predictions of the model developed by Feat et al.~\cite{16_Feal}. The width of the model band represents the theoretical systematic uncertainty.}
\label{fig:raa_compareto_16}
\end{figure}

The model developed by Feal et al.~\cite{16_Feal} extracts the medium's parameter values with full resummation of scattering centres including the expected perturbative tails. The density of scattering centres qualitatively agrees with the QGP equation of state computed in lattice QCD\@.
The predictions of the model are shown in Figure~\ref{fig:raa_compareto_16}.
It is not expected to describe $\Raa$ distributions at lower $\pT$. At $8\lesssim \pT \lesssim 80$\,\GeV, the model describes the data fairly accurately, although no predictions are available for peripheral collisions.
 
\begin{figure}[ptbh]
\centering
\includegraphics[width=0.49\textwidth]{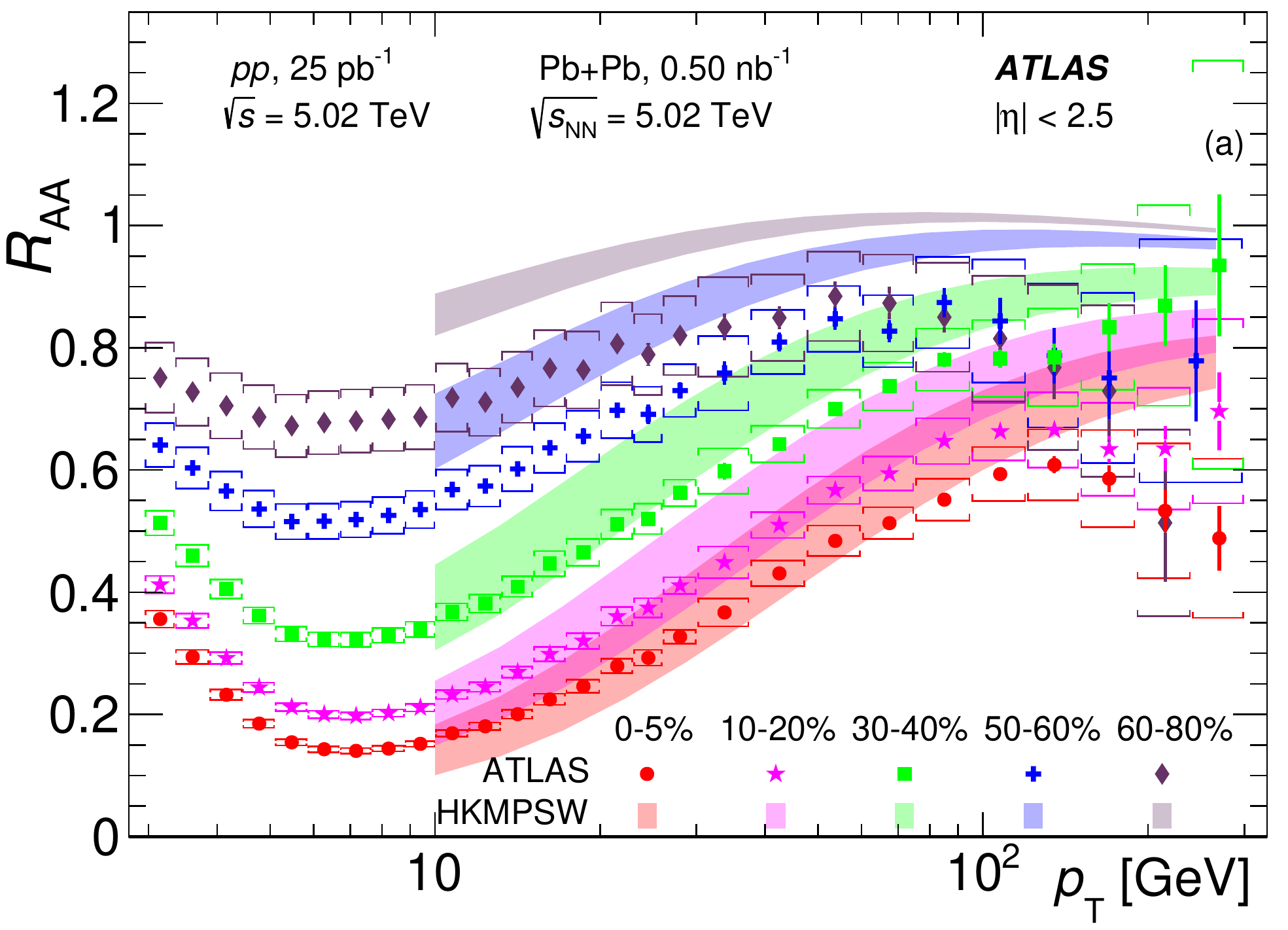} \includegraphics[width=0.49\textwidth]{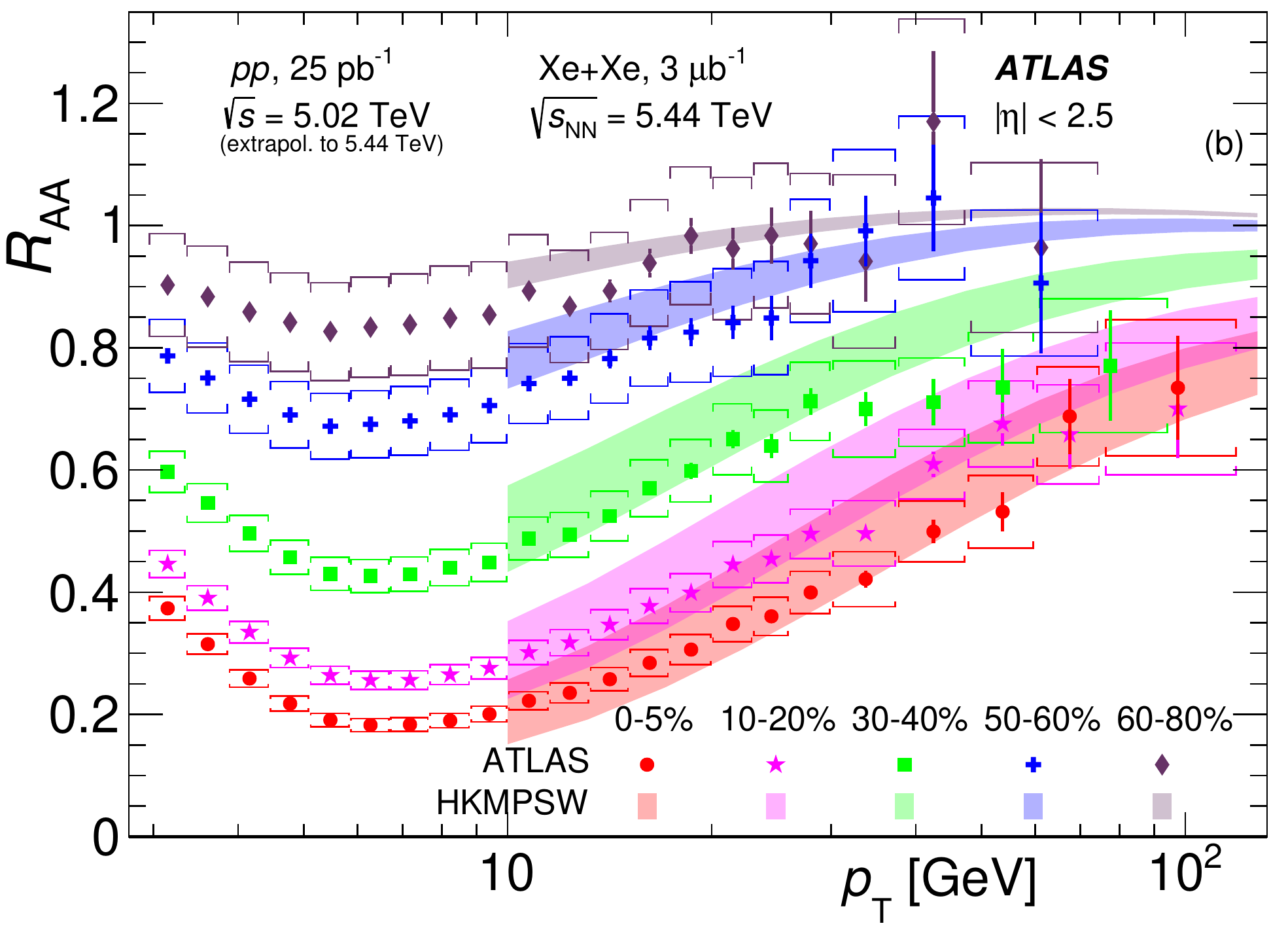} \\
\includegraphics[width=0.49\textwidth]{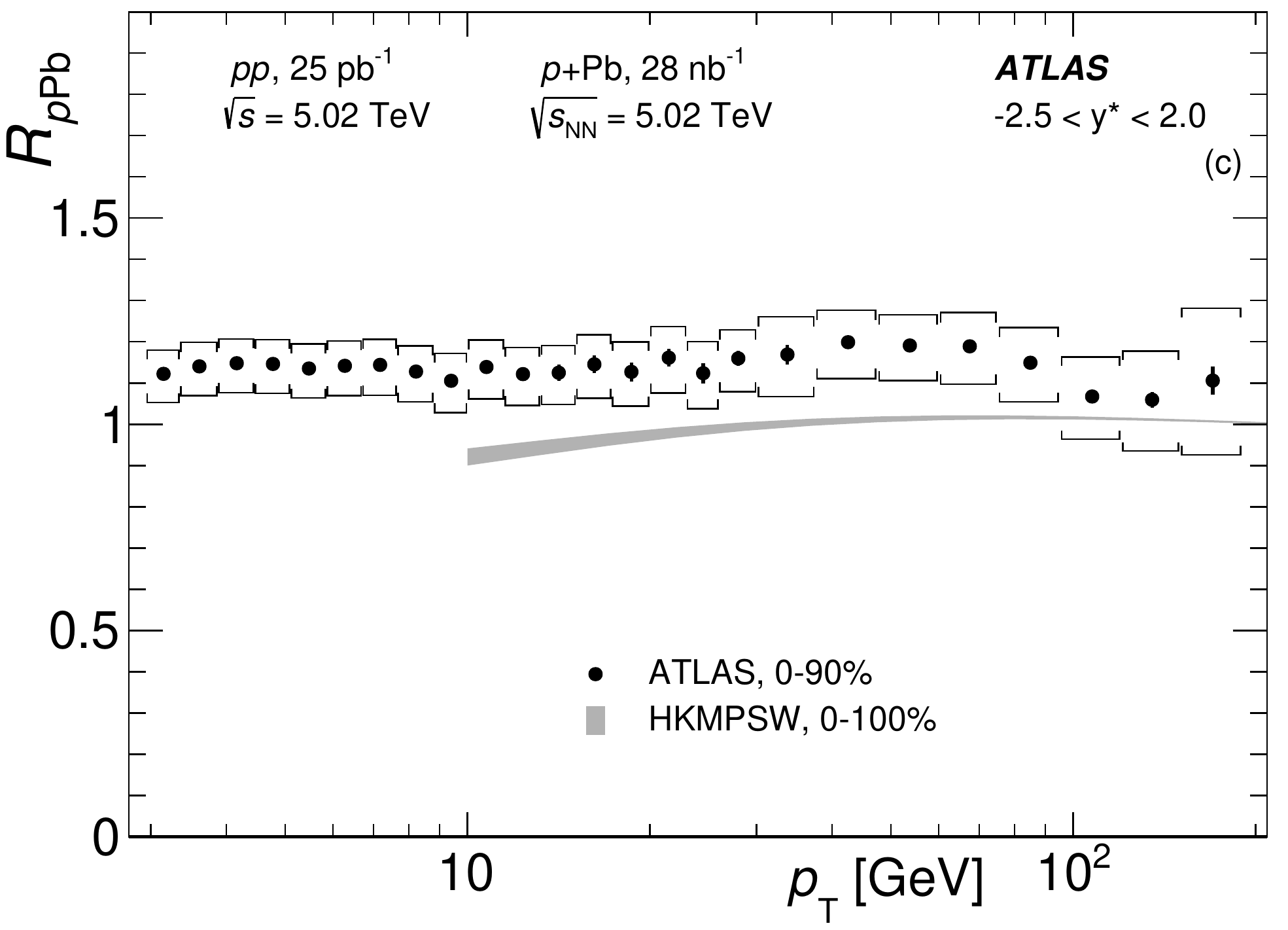}
\caption{Charged-hadron nuclear modification factor $\Raa$ for (a) \PbPb, (b) \XeXe, and (c) \pPb collisions in the pseudorapidity region $|\eta| < 2.5$ measured by ATLAS compared with predictions of the HKMPSW model~\cite{18_Wiedemann}. The width of the model band represents the theoretical systematic uncertainty.}
\label{fig:raa_compareto_18}
\end{figure}

The HKMPSW model~\cite{18_Wiedemann} has the BDMPS-Z formalism~\cite{18_Wiedemann_b,18_Wiedemann_c,18_Wiedemann_d} as a starting point from which the probability distribution of parton energy loss in QGP is derived.
A comparison with \PbPb, \XeXe, and \pPb data is shown in Figure~\ref{fig:raa_compareto_18}.
For \PbPb and \XeXe collisions, the model describes the $\Raa$ values in central and mid-central collisions in both systems well.
In peripheral \PbPb collisions, the model predictions are above the $\Raa$ values for the data.
This model's prediction for \RpPb is also available. The shape appears nearly flat above 10\,\GeV, which is similar to the data, although lower than it.

In summary, the models generally describe the data between about 10 and 100\,\GeV well. Some difficulties arise in the most peripheral collisions, where agreement is worse for several models, although the data also has larger systematic uncertainties, driven by uncertainty associated with $\avgTaa$. Above 100\,\GeV\ the ATLAS data show saturation in \PbPb collisions at values of about 0.6--0.8, depending on the centrality interval. Some models show the same trend, but others keep rising, approaching unity. At high $\pT$, the accuracy of the measurements suffers from high systematic uncertainty and thus it is difficult to draw a definite conclusion.

\FloatBarrier
 
\section{Conclusions}
\label{sec:conclusion}

A precise measurement of inclusive charged-hadron production in all collision systems produced by the LHC at $\sqn = 5.02$ or 5.44\,\TeV\ is presented by the ATLAS experiment in wide ranges of centrality, $\pT$ and $\eta$ (or $\ystar$).
The \pp data at $\sqs=5.02$\,\TeV correspond to 28\,\ipb, the \pPb data at $\sqn=5.02$\,\TeV correspond to 25\,\inb, the \PbPb data at $\sqn=5.02$\,\TeV correspond to 0.50\,\inb, and the \XeXe data at $\sqn=5.44$\,\TeV correspond to 3\,\iub.
These results extend similar ATLAS analyses of \PbPb data at $\sqn=2.76$\,\TeV~\cite{HION-2011-03}, and of \pPb data at $\sqn=5.02$\,\TeV~\cite{HION-2012-14}.
 
In \pPb collisions, the nuclear modification factor $\RpPb$ is not symmetric in rapidity, with more particles emitted in the Pb-going direction. This effect is most pronounced at low \pT and flattens out above 15\,\GeV.
In the central collisions, a sharp local maximum at $\pT\approx 3$\,\GeV is followed by a decrease.
As the collisions become more peripheral, the local maximum becomes less pronounced and wider.
Above $\pT\approx 40$\,\GeV, the \Rppb values in different centrality intervals are consistent within their systematic uncertainties.
 
In \PbPb and \XeXe collisions, the nuclear modification factors, $\Raa$, show a characteristic shape observed previously: a local maximum at $\pT\approx 2$\,\GeV is followed by a decrease to a local minimum at $\pT\approx 7$\,\GeV and then a rise towards higher $\pT$.
In \PbPb collisions at around 100\,\GeV, the slope becomes smaller; no such conclusion is drawn for \XeXe collisions because of the limited number of events.
This shape is observed in all centrality intervals, but has its highest maximum-to-minimum ratio in the most central collisions.
The suppression of charged-hadron production is also strongest in the most central collisions.
At fixed $\pT$, the $\Raa$ values have a very weak dependence on $|\eta|$.
 
The data presented here can provide comprehensive constraints on theoretical descriptions of soft and hard processes in the QGP in a variety of collision systems.
Theoretical models typically describe $\Raa$ better in central collisions and in the $\pT$ range from 7--10\,\GeV to about 100\,\GeV. They tend to deviate from the data at higher $\pT$ or in more peripheral collisions. Most of them do not attempt to describe low-$\pT$ data.


\section*{Acknowledgements}


We thank CERN for the very successful operation of the LHC, as well as the
support staff from our institutions without whom ATLAS could not be
operated efficiently.
 
We acknowledge the support of
ANPCyT, Argentina;
YerPhI, Armenia;
ARC, Australia;
BMWFW and FWF, Austria;
ANAS, Azerbaijan;
CNPq and FAPESP, Brazil;
NSERC, NRC and CFI, Canada;
CERN;
ANID, Chile;
CAS, MOST and NSFC, China;
Minciencias, Colombia;
MEYS CR, Czech Republic;
DNRF and DNSRC, Denmark;
IN2P3-CNRS and CEA-DRF/IRFU, France;
SRNSFG, Georgia;
BMBF, HGF and MPG, Germany;
GSRI, Greece;
RGC and Hong Kong SAR, China;
ISF and Benoziyo Center, Israel;
INFN, Italy;
MEXT and JSPS, Japan;
CNRST, Morocco;
NWO, Netherlands;
RCN, Norway;
MEiN, Poland;
FCT, Portugal;
MNE/IFA, Romania;
MESTD, Serbia;
MSSR, Slovakia;
ARRS and MIZ\v{S}, Slovenia;
DSI/NRF, South Africa;
MICINN, Spain;
SRC and Wallenberg Foundation, Sweden;
SERI, SNSF and Cantons of Bern and Geneva, Switzerland;
MOST, Taiwan;
TENMAK, T\"urkiye;
STFC, United Kingdom;
DOE and NSF, United States of America.
In addition, individual groups and members have received support from
BCKDF, CANARIE, Compute Canada and CRC, Canada;
PRIMUS 21/SCI/017 and UNCE SCI/013, Czech Republic;
COST, ERC, ERDF, Horizon 2020 and Marie Sk{\l}odowska-Curie Actions, European Union;
Investissements d'Avenir Labex, Investissements d'Avenir Idex and ANR, France;
DFG and AvH Foundation, Germany;
Herakleitos, Thales and Aristeia programmes co-financed by EU-ESF and the Greek NSRF, Greece;
BSF-NSF and MINERVA, Israel;
Norwegian Financial Mechanism 2014-2021, Norway;
NCN and NAWA, Poland;
La Caixa Banking Foundation, CERCA Programme Generalitat de Catalunya and PROMETEO and GenT Programmes Generalitat Valenciana, Spain;
G\"{o}ran Gustafssons Stiftelse, Sweden;
The Royal Society and Leverhulme Trust, United Kingdom.
 
The crucial computing support from all WLCG partners is acknowledged gratefully, in particular from CERN, the ATLAS Tier-1 facilities at TRIUMF (Canada), NDGF (Denmark, Norway, Sweden), CC-IN2P3 (France), KIT/GridKA (Germany), INFN-CNAF (Italy), NL-T1 (Netherlands), PIC (Spain), ASGC (Taiwan), RAL (UK) and BNL (USA), the Tier-2 facilities worldwide and large non-WLCG resource providers. Major contributors of computing resources are listed in Ref.~\cite{ATL-SOFT-PUB-2021-003}.


\printbibliography
 
\clearpage
 
\begin{flushleft}
\hypersetup{urlcolor=black}
{\Large The ATLAS Collaboration}

\bigskip

\AtlasOrcid[0000-0002-6665-4934]{G.~Aad}$^\textrm{\scriptsize 102}$,
\AtlasOrcid[0000-0002-5888-2734]{B.~Abbott}$^\textrm{\scriptsize 120}$,
\AtlasOrcid[0000-0002-1002-1652]{K.~Abeling}$^\textrm{\scriptsize 55}$,
\AtlasOrcid[0000-0002-8496-9294]{S.H.~Abidi}$^\textrm{\scriptsize 29}$,
\AtlasOrcid[0000-0002-9987-2292]{A.~Aboulhorma}$^\textrm{\scriptsize 35e}$,
\AtlasOrcid[0000-0001-5329-6640]{H.~Abramowicz}$^\textrm{\scriptsize 151}$,
\AtlasOrcid[0000-0002-1599-2896]{H.~Abreu}$^\textrm{\scriptsize 150}$,
\AtlasOrcid[0000-0003-0403-3697]{Y.~Abulaiti}$^\textrm{\scriptsize 117}$,
\AtlasOrcid[0000-0003-0762-7204]{A.C.~Abusleme~Hoffman}$^\textrm{\scriptsize 137a}$,
\AtlasOrcid[0000-0002-8588-9157]{B.S.~Acharya}$^\textrm{\scriptsize 69a,69b,p}$
% ...
\bigskip
\\

% ...
$^{29}$Physics Department, Brookhaven National Laboratory, Upton NY; United States of America.\\
% ...
$^{35}$$^{(a)}$Facult\'e des Sciences Ain Chock, R\'eseau Universitaire de Physique des Hautes Energies - Universit\'e Hassan II, Casablanca;$^{(b)}$Facult\'{e} des Sciences, Universit\'{e} Ibn-Tofail, K\'{e}nitra;$^{(c)}$Facult\'e des Sciences Semlalia, Universit\'e Cadi Ayyad, LPHEA-Marrakech;$^{(d)}$LPMR, Facult\'e des Sciences, Universit\'e Mohamed Premier, Oujda;$^{(e)}$Facult\'e des sciences, Universit\'e Mohammed V, Rabat;$^{(f)}$Institute of Applied Physics, Mohammed VI Polytechnic University, Ben Guerir; Morocco.\\
% ...
$^{55}$II. Physikalisches Institut, Georg-August-Universit\"{a}t G\"ottingen, G\"ottingen; Germany.\\
% ...
$^{69}$$^{(a)}$INFN Gruppo Collegato di Udine, Sezione di Trieste, Udine;$^{(b)}$ICTP, Trieste;$^{(c)}$Dipartimento Politecnico di Ingegneria e Architettura, Universit\`a di Udine, Udine; Italy.\\
% ...
$^{102}$CPPM, Aix-Marseille Universit\'e, CNRS/IN2P3, Marseille; France.\\
% ...
$^{117}$Department of Physics, New York University, New York NY; United States of America.\\
% ...
$^{120}$Homer L. Dodge Department of Physics and Astronomy, University of Oklahoma, Norman OK; United States of America.\\
% ...
$^{137}$$^{(a)}$Departamento de F\'isica, Pontificia Universidad Cat\'olica de Chile, Santiago;$^{(b)}$Millennium Institute for Subatomic physics at high energy frontier (SAPHIR), Santiago;$^{(c)}$Instituto de Investigaci\'on Multidisciplinario en Ciencia y Tecnolog\'ia, y Departamento de F\'isica, Universidad de La Serena;$^{(d)}$Universidad Andres Bello, Department of Physics, Santiago;$^{(e)}$Instituto de Alta Investigaci\'on, Universidad de Tarapac\'a, Arica;$^{(f)}$Departamento de F\'isica, Universidad T\'ecnica Federico Santa Mar\'ia, Valpara\'iso; Chile.\\
% ...
$^{150}$Department of Physics, Technion, Israel Institute of Technology, Haifa; Israel.\\
$^{151}$Raymond and Beverly Sackler School of Physics and Astronomy, Tel Aviv University, Tel Aviv; Israel.\\
% ...
$^{a}$ Also Affiliated with an institute covered by a cooperation agreement with CERN.\\
$^{b}$ Also at An-Najah National University, Nablus; Palestine.\\
$^{c}$ Also at Borough of Manhattan Community College, City University of New York, New York NY; United States of America.\\
$^{d}$ Also at Bruno Kessler Foundation, Trento; Italy.\\
$^{e}$ Also at Center for High Energy Physics, Peking University; China.\\
$^{f}$ Also at Center for Interdisciplinary Research and Innovation (CIRI-AUTH), Thessaloniki ; Greece.\\
% ...
$^{p}$ Also at Department of Physics, King's College London, London; United Kingdom.\\
% ...

\end{flushleft}


\end{document}